\documentclass[11pt,a4paper]{article}
\usepackage{jheppub}
\usepackage{hyperref}                            
\usepackage{graphicx}                            
\usepackage{soul}                                
\usepackage{microtype}                           
\usepackage[sharp]{easylist}                     
\usepackage{amsmath}                             
\usepackage{amssymb}                             
\usepackage{tensor}                              
\usepackage{booktabs}                            
\usepackage{wrapfig}                             
\usepackage{subcaption}                          
\usepackage[T1]{fontenc}                         
\usepackage{titlesec}                            
\usepackage[font=small]{caption}                 

\usepackage{tikz}

\usepackage[section]{placeins}

\graphicspath{{figures/}} 

\usepackage{todonotes}
\presetkeys{todonotes}{color=white, linecolor=gray, size=tiny}{}

\makeatletter
\if@todonotes@disabled
  
\else
  
\fi
\makeatother

\usepackage{color}
    \definecolor{darkgreen}{rgb}{0,0.5,0}
    \definecolor{darkred}{rgb}{0.5,0,0}
    \definecolor{darkblue}{rgb}{0,0,0.6}
    \definecolor{purple}{rgb}{0.4,.2,0.7}

\newcommand{\eg}{{\it e.g.\,}\ }
\newcommand{\ie}{{\it i.e.\,}\ }

\def\yp2{{y_+}^2}
\def\ym2{{y_-}^2}
\def\calr{\mathcal{R}}

\def\MAE{\hbox{\tiny MAE}}

\title{Strong Cosmic Censorship in Kerr-Newman-de Sitter}
\author[a]{Alex Davey,}
\author[a]{\'Oscar J.~C.~Dias,}
\author[a]{David Sola Gil}

\emailAdd{amd1g13@soton.ac.uk}
\emailAdd{ojcd1r13@soton.ac.uk}
\emailAdd{d.sola-gil@soton.ac.uk}

\affiliation[a]{STAG research centre and Mathematical Sciences, University of Southampton, University Road, Southampton, SO17 1BJ, United Kingdom}

\abstract{Christodoulou's formulation of Strong Cosmic Censorship (SCC) holds true for Kerr-de Sitter black holes. On the other hand,  Reissner–Nordström-de Sitter black holes violate SCC. We do a detailed scan of the parameter space of Kerr-Newman-de Sitter black holes between these  two limiting families, to identify the boundary that marks the transition between solutions that respect and violate SCC. We focus our attention on linear scalar field perturbations. SCC is violated inside a (roughly) `spherical' shell of the parameter space of Kerr-Newman-de Sitter, centred at the corner that describes arbitrarily small extremal Reissner–Nordström-de Sitter solutions. Outside of this region,  including the Kerr-de Sitter limit, we identify perturbation modes that decay slow enough to enforce SCC. Additionally, we do a necessary study of the quasinormal mode spectra of Kerr-Newman-de Sitter in some detail. As established in the literature, in the  Kerr-de Sitter and Reissner–Nordström-de Sitter limits, we find three families of modes: de Sitter, photon sphere and near-horizon modes. These interact non-trivially away from the Reissner–Nordström-de Sitter limit and display eigenvalue repulsions like in Kerr-Newman black holes. 

}
\subheader{\today}

\begin{document}

\maketitle
\flushbottom

\clearpage
\section{Introduction}
Predictability is a core characteristic of physics: given a hyperbolic partial differential equation and initial data, we expect to find a unique solution. In the initial value formulation of General Relativity (GR), we prescribe initial data (satisfying the constraint equations) on some partial Cauchy hypersurface $\Sigma$, where no pair of points is connected by a causal curve. The evolution equations then uniquely 
determine (up to diffeomorphisms) the future evolution of the system. The future causal region of the Cauchy hypersurface $\Sigma$ is known as the maximal future Cauchy development of $\Sigma$. For charged and/or rotating black holes its boundaries include Cauchy horizons~\cite{waldGeneralRelativity1984}. The Christodoulou formulation of Strong Cosmic Censorship (SCC) states that, for generic smooth  initial data, the maximal Cauchy development of initial data in $\Sigma$ is inextendible beyond the Cauchy horizons as a \textit{weak solution} of the Einstein equations (eventually coupled to matter fields)~\cite{christodoulouFormationBlackHoles2008}. This is a reformulation of Penrose's original $C^0$ formulation of SCC which conjectured that the solution should be inextendible as a {\it continuous} solution of the Einstein equations~\cite{penrose1978} (and that is known to be false~\cite{McNamara121,Ori:1991zz,Dafermos:2003wr,Franzen:2014sqa,Dafermos:2017dbw}).  A failure of Christodoulou's formulation of SCC implies a loss of predictability in GR (unless we find a reformulation of SCC that holds), as events in the region beyond the Cauchy horizon do not depend causally on only the initial data on $\Sigma$. 

The familiar Reissner-Nordstr\"{o}m (RN) and Kerr solutions of the Einstein equations (with vanishing cosmological constant $\Lambda$) do respect Christodoulou's formulation of SCC~\cite{lukProofLinearInstability2017, Dafermos:2015bzz, Dafermos:2012np}. 
One can argue, as Penrose did when motivating his original $C^0$ formulation of SCC, that the mechanism that prevents a Cauchy horizon crossing is a blueshift effect of energy sourced by the late-time tails of perturbations in the exterior of the event horizon of the black hole. These late-time tails are perturbations that cross the event horizon at late times close to the timelike infinity in a Penrose diagram, and they ultimately dictate the behaviour of perturbations at the Cauchy horizon
\cite{Price:1971fb,Simpson:1973ua,mcnamara1978instability, chandrasekhar1982crossing,poisson1990internal}. Christodoulou's formulation of SCC states that this blueshift mechanism (a.k.a. mass inflation when its backreaction is included~\cite{poisson1990internal,Ori:1991zz}) implies that observers cannot cross Cauchy horizons as weak solutions of the system -- infinite tidal forces prevent them from doing so. This also implies that curvature invariants diverge at the would-be Cauchy horizon, \ie the $C^2$ formulation of SCC also holds.\footnote{For predictability it is not enough that the $C^2$ formulation of SCC holds (here,  `inextendible' means `inextendible as a twice-differentiable solution'), because the tidal distortion felt by an observer crossing the Cauchy horizon can remain finite (if Christodoulou's formulation of SCC fails), so the divergence in the curvature might not be strong enough to destroy a macroscopic observer~\cite{Ori:1991zz}.} Indeed, this is the case for asymptotically flat Kerr-Newman black holes (including Reissner-Nordstr\"om and Kerr) even though the solutions turn out to be continuous across the Cauchy horizon. In other words, Christodoulou's formulation of SCC holds for $\Lambda=0$~\cite{lukProofLinearInstability2017, Dafermos:2015bzz, Dafermos:2012np} even though Penrose's original $C^0$ formulation of SCC does not~\cite{McNamara121,Ori:1991zz,Dafermos:2003wr,Franzen:2014sqa,Dafermos:2017dbw}.

However, the situation is different in asymptotically de Sitter (dS) spacetimes. When $\Lambda > 0$, the blueshift effect competes with a redshift effect: perturbations in the exterior of the black hole also have to climb the gravitational potential well sourced by the cosmological constant (\ie by the existence of a cosmological horizon). Therefore, perturbations may decay too quickly to render the Cauchy horizons unstable and eventually evolve to a singularity that enforces SCC. 
More concretely, it has been shown that the behaviour of perturbations at the Cauchy horizon depends on the decay rate  of perturbations along the event horizon~\cite{Hintz:2015jkj,Hintz:2015koq} and for asymptotically de Sitter black holes it is established that perturbations decay exponentially along the event horizon~\cite{Barreto1997,bony2007decay,Dyatlov:2011jd,Dyatlov:2013hba,Hintz:2016gwb,Hintz:2016jak}.
It follows that  for linearized matter or gravitational perturbations in de Sitter black holes, the outcome of the blueshift/redshift competition depends on whether the ratio $\beta$ between the spectral gap (\ie the magnitude of the imaginary part of the least damped quasinormal mode (QNM) of the black hole) and the Cauchy horizon surface gravity is above or below a critical value~\cite{Hintz:2015jkj}. By doing the actual computation to find the outcome of this competition, it has been established that Christodoulou's formulation of SCC is violated in (near-extremal) Reissner-Nordstr\"om-de Sitter (RNdS)~\cite{Cardoso:2017soq, Dias:2018etb, Dias:2018ufh}, while it holds in Kerr-dS black holes~\cite{Dias:2018ynt}. 

This distinct result for the charged and rotating cases is perhaps surprising, as the causal structures of RNdS and Kerr-dS are quite similar~\cite{waldGeneralRelativity1984}. In fact, there is also solid evidence that this difference is not unique to $d=4$ and extends generically to higher dimensions~\cite{Liu:2019lon,Davey:2022vyx}. Nonetheless, when it comes to studying the perturbations, there is a crucial difference between purely charged versus purely rotating spacetimes that can help understand the difference. In Kerr-dS, SCC holds essentially due to high azimuthal $m$ modes (which are necessarily present in generic initial data) that `couple' to the background rotation~\cite{Dias:2018ynt}. 
In RNdS, the analogue of $m$ can be taken to be the charge $q$ of matter perturbations, \eg a charged scalar field that couples to the background electric potential. So, a sector of perturbations that could eventually rescue SCC in RNdS, is the charged perturbation sector, especially if we have large $q$. Such charged matter must be present in a universe where charged black holes can be formed through gravitational collapse~\cite{Hod:2018dpx,Cardoso:2018nvb,Mo:2018nnu,Dias:2018ufh}. However, unlike $m$, we have to keep in mind that $q$ is  fixed by the theory. In the end of the day, although charge tends to weaken the SCC violation, the fact is that there is always a neighbourhood of RNdS extremality in which SCC is violated by perturbations arising from smooth initial data even for arbitrarily but finite large $q$~\cite{Dias:2018ufh}.\footnote{For charged perturbations, even large $q$ leads to SCC violation close enough to extremality. This was missed in the analysis of~\cite{Cardoso:2018nvb,Mo:2018nnu} because these references did not consider sufficiently near-extremal black holes~\cite{Dias:2018ufh}. On the other hand, the conclusions of~\cite{Hod:2018dpx} do not hold perhaps because it is a WKB analysis that does not account for the non-perturbative effects that are essential to conclude that SCC is still violated in RNdS for charged initial data~\cite{Dias:2018ufh}.} For neutral  field perturbations, \ie with $q=0$, the violation of SCC is even worse in the sense that it starts happening for RNdS black holes that are further away from extremality.

From an astrophysical point of view, we expect that environmental plasma quickly neutralizes purely charged black holes, hence one could even question the physical relevance of the RNdS SCC violation~\cite{Cardoso:2018nvb,Mo:2018nnu,Dias:2018ufh}. On the other hand, as charged matter is indeed observed in our Universe, charged rotating black holes with small charge may still form even if they are then short-lived. In particular, rapidly spinning weakly charged black holes may endure~\cite{Cardoso:2018nvb, Cardoso:2016olt}. Rotating black holes in an external magnetic field necessarily produce a non-zero electric field that attracts electric charge \cite{WaldPRD1974,Zajacek:2019kla,Horowitz:2024dch}:  rotating black holes become  (weakly) charged rotating black holes.

When we view RNdS and Kerr-dS as particular sub-families of the Kerr-Newman-de Sitter (KNdS) black hole, there should definitely be a boundary  in the parameter space of KNdS that marks a transition in the validity of SCC.
Therefore, in this paper we scan the parameter space of the KNdS solution, dialling the rotation and charge parameters to move from the RNdS into the Kerr-dS sub-families, to accurately identify this boundary.
This problem was already partially addressed in~\cite{Casals:2020uxa} (see also~\cite{Hod:2018lmi}) but only for very specific values of the parameter space (concretely, for fixed $\Lambda M^2=0.02$ where $M$ is the mass parameter of KNdS). In particular, \cite{Casals:2020uxa} considered  
a massless charged scalar field, a neutral conformally coupled scalar and a Dirac field and found that SCC is violated near extremal KNdS black holes in the very few families  that were considered. Ref.~\cite{Casals:2020uxa} also identified the boundary line in a (dimensionless) charge versus rotation phase diagram where SCC starts being violated for a particular value of $\Lambda M^2=0.02$.

Our study complements and extends the one of~\cite{Casals:2020uxa}, in the sense that, for neutral massless scalar fields, we will do a {\it full} scan of the 3-dimensional parameter space of KNdS and identify the surface boundary where SCC starts being violated in the 3-dimensional phase diagram of KNdS. It follows that the {\it line} boundary of~\cite{Casals:2020uxa} (for the particular value of $\Lambda M^2=0.02$) is then a particular curve sitting on our boundary {\it surface}. Our results agree with~\cite{Casals:2020uxa} for this particular curve.\footnote{Recall that KNdS is described by four parameters that we can take to be the mass, charge, rotation parameters and the cosmological radius. But the system has a scaling symmetry that implies that KNdS only has 3 physical parameters, \ie only adimensional ratios of three of them in units of the fourth are physical. Ref.~\cite{Casals:2020uxa} measures the charge, rotation and cosmological radius in units of the mass. We will instead use a `spherical polar parametrization' where the three adimensional parameters are off-extremality, off-Nariai and off-RNdS (or off-Kerr-dS) measures. This polar parametrization (which is an extension of the one introduced for Kerr-Newman in~\cite{Davey:2023fin}) will prove very convenient to systematically scan the full 3-dimensional parameter space of KNdS.}
The main outcome of our analysis  will be best summarized by Fig.~\ref{fig:eikonal_verification3D}  where we identify the boundary between solutions that violate and respect Christodoulou's formulation of SCC in the full 3-dimensional parameter space of KNdS. In particular, we identify a `wide' range of near-extremal weakly rotating KNdS black holes where generic scalar field perturbations arising from smooth initial data have finite energy at the Cauchy horizon even though they are not continuously differentiable there. This boundary is roughly a `spherical-like' region centred at the arbitrarily small extremal RNdS corner of the KNdS parameter space.

Although we will not address reformulations of the problem that can rescue SCC in RNdS, and thus for weakly rotating KNdS configurations (inside the above `spherical-like' region) SCC is still violated, for completeness we briefly mention them here. We invite the reader to see, for instance, the introduction and conclusions of~\cite{Dias:2018ufh} for a recent detailed account of SCC studies and possible resolutions of its violation when this is the case.

A natural question to ask is whether non-linear corrections could end up saving Christodoulou's formulation of SCC for RNdS and weakly rotating KN-dS. This is a non-trivial question and the few existing non-linear studies (for RNdS) are inconclusive~\cite{Luna:2019olw,Zhang:2019nye,Luna:2019olwaddendum}. Additionally, our discussion assumes smooth initial data. However, one might also consider a `rough' formulation of the SCC, which allows for non-smooth initial data: at least for RNdS, it has been shown that the degree of regularity of the solutions at the Cauchy horizon is generically lower than the one of the non-smooth initial data, and in this sense one can say that this non-smooth version of SCC still holds~\cite{Dafermos:2018tha}. The connection between the formulation in~\cite{Dafermos:2018tha} and previous studies~\cite{Mellor:1989ac,Mellor1992,Brady:1998au} is discussed in~\cite{Dias:2018etb}.

For our study we only consider a classical discussion of SCC, but we might wonder if a reformulation of SCC that includes quantum effects might end up enforcing predictability. This was conjectured to be the case in~\cite{Dias:2018etb,diasBTZBlackHole2019} and indeed such a quantum formulation of SCC was indeed proven to hold for RNdS black holes in~\cite{hollandsQuantumInstabilityCauchy2020a,Hollands:2020qpe,Klein:2024sdd}, even when Christodoulou's classical formulation of  SCC is violated. In more detail, for any non-singular quantum state within the domain of dependence of initial data, the (quantum) expectation value of the energy-stress tensor diverges fast enough to render the Cauchy horizon unstable. When they co-exist, this divergence is even stronger than the one caused by the blueshift instability associated to the classical late-time behaviour of the black hole solution. Hence we also expect that the perturbed spacetime does not admit Cauchy horizons. In principle, this result extends to KNdS~\cite{hollandsQuantumInstabilityCauchy2020a,Hollands:2020qpe,Klein:2024sdd}, implying that predictability in dS spacetimes is restored once we include quantum effects in the late-time behaviour of black hole perturbations.\footnote{\label{foot:BTZ}As far as we are aware there is only one black hole where the quantum (and classical) {\it linear} reformulation of SCC can be violated. This is for the asymptotically anti-de Sitter BTZ  black hole (where the late-time behaviour of perturbations is also ruled by QNMs). In this case, the classical and quantum linear formulations of Christodoulou's SCC are both violated~\cite{diasBTZBlackHole2019,Balasubramanian:2019qwk,hollandsQuantumInstabilityCauchy2020a}. However, in this case it is the {non-linear} quantum reformulation of Christodoulou's SCC (\ie back-reaction of the linear quantum fields on the BTZ background) that ultimately enforces predictability in BTZ~\cite{Emparan:2020rnp,Emparan:2020znc}. As a side note, perhaps this suggests that at non-linear level even Christodoulou's classical formulation of SCC can still be rescued, although non-linear studies to establish this have been, so far, inconclusive~\cite{Luna:2019olw,Zhang:2019nye,Luna:2019olwaddendum}.} 

In this paper, we extend the SCC analysis of~\cite{Casals:2020uxa} to the full parameter space of KNdS for a neutral massless scalar field, as stated above. The Klein-Gordon equation for a scalar field is separable in the KNdS geometry~\cite{carter1968hamilton}, unlike the case of gravito-electromagnetic perturbations. The scalar field is thus a good `simple' proxy to discuss Christodoulou's  SCC in KNdS given that the validity of SCC in RNdS and Kerr-dS happens independently of the field perturbation spin (the details of the violation, \eg the degree of differentiability when the violation occurs and the boundary location in the parameter space can however depend on the spin). For the sake of completeness, we allow for non-zero mass and charge when performing some of our analytical analyses. But, to find our numerical results and when discussing their implications for SCC, we will set the charge and mass of the scalar field to zero. In section~\ref{sec:Conclusions} we will comment on how our findings for the neutral massless scalar field combined with known results in the RNdS and Kerr-dS cases (and the specific cases of~\cite{Casals:2020uxa} for KNdS) can be used to produce educated expectations for the the discussion of SCC for other perturbations. 

This manuscript is structured as follows. In Section \ref{sect:background_material} we review the main properties of KNdS black holes. We introduce a set of `\textit{spherical polar parameters}' which fully cover the parameter space of KNdS and generalize the `polar parametrization' for Kerr-Newman in~\cite{Davey:2023fin}. These conveniently allow us to continuously follow the evolution of QNMs between RNdS and Kerr-dS, while keeping the ratios between two of the horizons, \ie the `off-extremality' and `off-Nariai' measures, fixed. 
Then, we set up the scalar QNM coupled eigenvalue problem and its boundary conditions. Its solution gives the 
spectral gap, a fundamental quantity needed to discuss Christodoulou's formulation of SCC in KNdS.

In Section \ref{sect:analytical_study}, we provide analytical approximations (which are accurate only in certain small regions of the KNdS parameter space) for the three distinct families of QNMs that can be identified in KNdS: de Sitter (dS), Photon Sphere (PS) and Near-Horizon (NH) modes (this is not a sharp classification due to the presence of eigenvalue repulsions that will  be discussed in section~\ref{sect:QNM_spectrum_in_KNdS}). As part of the NH mode analytic computation, we investigate the near-horizon geometry of KNdS in subsection \ref{sect:BFboundandNHgeometry}, obtaining an expression for the corresponding 2-dimensional  Breitenl\"{o}hner-Freedman (BF) bound. Coincidentally, we observe that here, a violation of the BF bound corresponds to a qualitative change in the QNM spectra, like in the Kerr-Newman case~\cite{Davey:2023fin}.

In Section \ref{sect:QNM_spectrum_in_KNdS}, we follow the evolution of different QNM families through slices of the KNdS parameter space. This will allow us to compare our numerical results with the analytical approximations and prior findings in the literature, apart from illustrating important features and complexities of the QNM spectra. The QNM spectra of KNdS (much like in the Kerr-Newman case~\cite{Dias:2021yju,Dias:2022oqm,Davey:2022vyx,Davey:2023fin}) turns out to be plagued with the so-called {\it eigenvalue repulsions} that effectively can trade dominance (damping) between the above three QNM families. We discuss these since for SCC one must find the dominant (least damped) mode. These eigenvalue repulsions are not discussed in~\cite{Casals:2020uxa}. They are not relevant for the final discussion of the validity of SCC as long as we always identify the least damped mode for each point in the parameter space (as we believe is the case in~\cite{Casals:2020uxa}), but it can change the classification attributed to each of the three aforementioned QNM families.

Section \ref{sect:SCC_in_KNdS} contains the main results and conclusions relevant for the discussion of SCC in KNdS. In particular, in Fig.~\ref{fig:eikonal_verification3D}  we present a plot of the boundary in KNdS parameter space that signals the transition from violating SCC to enforcing it, up to linear neutral massless scalar perturbations. We also discuss in some detail how our results for a neutral massless scalar field, together with previous findings, can give educated expectations for other perturbation sectors.  In section~\ref{sec:Conclusions} we present our final conclusions and discussions.

For completeness, we compare our analytical approximations to prior work in the literature in Appendices \ref{sect:eikonal_analysis}-\ref{sect:MAEanalysis}.  In Appendix \ref{sect:MAEvalueforvanishingm}, we discuss details of the specific case of vanishing azimuthal and orbital quantum numbers ($m = \ell = 0$), and we show that NH modes do not enforce SCC at extremality if $m = \ell = 0$. Finally, numerical convergence tests can be found in Appendix~\ref{sect:numerical_convergence}.
\section{Scalar perturbations of the Kerr-Newman-de Sitter black hole}\label{sect:background_material}
\subsection{The Kerr-Newman-de Sitter black hole}
The Kerr-Newman-de Sitter (KNdS) black hole is a solution of the Einstein-Maxwell action with a cosmological constant $\Lambda > 0$, \ie a generalization of the Kerr-Newman solution to asymptotically de Sitter (dS) spacetimes~\cite{GibbonsHawkingPhysRevD.15.2738}. It is described by four dimensionful parameters: the mass $M$, the rotation $a$ and charge $Q$ parameters and  the de Sitter radius $L = \sqrt{3/\Lambda}$. Considering the limits $Q = 0$, $a = 0$, $Q = a = 0$ and $Q = a = M = 0$ we obtain the Kerr-de Sitter, Reissner-Nordstr\"{o}m-de Sitter (RNdS), Schwarzschild-de Sitter (SdS) and de Sitter (dS) spacetimes, respectively. Taking the asymptotically flat limit $\Lambda \to 0$ (or $L \to \infty$), we recover the corresponding asymptotically flat spacetimes. 

In Boyer-Lindquist coordinates $\{t, r, \theta,\phi\}$ (time, radius, polar angle, azimuthal angle), the metric can be expressed as~\cite{carter1968hamilton, GibbonsHawkingPhysRevD.15.2738, Suzuki:1999nn}
\begin{equation}
  ds^{2} = - \frac{\Delta_{r}}{\Xi^2 \, \Sigma} (dt - a \sin^{2}\theta d\phi)^{2} + \frac{\Delta_{\theta}\sin^{2}\theta}{\Xi^2 \, \Sigma}\Big(a\, dt - (r^{2}+a^{2}) d\phi\Big)^{2}  + \Sigma \left( \frac{dr^{2}}{\Delta_{r}} + \frac{d\theta^{2}}{\Delta_{\theta}} \right)
  \label{eqn:metric_KNdS}
\end{equation}
with the Maxwell potential given by
\begin{equation}\label{eqn:potential_A}
    A = - \frac{Q \, r}{\Xi \, \Sigma} \big(dt - a \sin^2 \theta d\phi \big),
\end{equation}
where we defined
\begin{equation}
\begin{aligned}\label{eqn:BL_metric_functions}
  \Delta_{r}(r) &= (r^{2} + a^{2})\left(1-\frac{r^{2}}{L^{2}}\right) - 2 M r + Q^{2}, \qquad &\Xi &= 1+ \frac{a^{2}}{L^{2}}, \\
  \Delta_{\theta}(\theta) &= 1 + \frac{a^2}{L^2} \cos^{2} \theta, \qquad &\Sigma(r, \theta) &= r^{2} + a^{2} \cos^{2}\theta.
\end{aligned}
\end{equation}
We consider $M \geq 0$ to avoid naked singularities, restrict to $a \geq 0$ using the $t - \phi$ symmetry of the metric, and fix $Q \geq 0$ using the sign freedom in the definition of $A$. The positive real roots of $\Delta_r(r)$ define the Cauchy horizon $\mathcal{C}\mathcal{H}$ at $r = r_-$, the event horizon $\mathcal{H}$ at $r = r_+$ and the cosmological horizon $\mathcal{H}_c$ at $r = r_c$, with $r_- \le r_+ \le r_c$. 

\begin{figure}[t]
    \centering
    \includegraphics[width=0.5\textwidth]{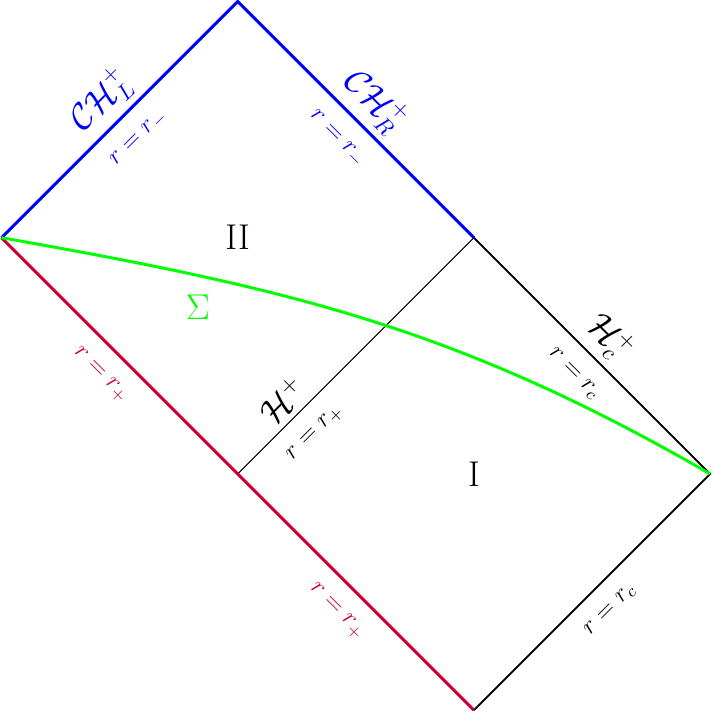}
    \caption{Section of the Penrose diagram (common to RNdS, Kerr-dS and KNdS spacetimes) that is relevant for the SCC discussion. The future event and future cosmological horizons are labelled by $\mathcal{H}^+$ and $\mathcal{H}^+_c$ and the blue lines represent the right and left future Cauchy horizons $\mathcal{C}\mathcal{H}_{L,R}$ respectively. The QNMs in this spacetime blow up along the red edge and along the past cosmological horizon (lower right black edge of region I)~\cite{Dias:2018ynt}. The green line represents an example of a Cauchy surface $\Sigma$.}
\label{fig:kerrdSspacetime}
\end{figure}

Each of the horizons $r_j$ with $j \in \{-,+, c\}$ is a Killing horizon, with the corresponding horizon generator $\xi_j$, angular velocity $\Omega_j=-g_{t\phi}/g_{\phi\phi}\big|_{r_j}$, surface gravity $\kappa_j$  and electrostatic potential $\Phi_j=-\xi^a A_a\big|_{r_j}$ given by\footnote{As a result, the first law of thermodynamics for KNdS black holes reads $T_+ dS_+ +T_c dS_c=(\Omega_+ -\Omega_c) d \mathcal{J}_+  + (\Phi_+ -\Phi_c) d \mathcal{Q}_+ $ where $S_j$ is the horizon entropy, $T_j=\kappa_j/(2\pi)$ is the temperature of the horizon $r_j$, and $\mathcal{J}_j$ and $\mathcal{Q}_j$ are the Komar angular momentum and electric charge evaluated at the horizon $r_j$~\cite{GibbonsHawkingPhysRevD.15.2738}. Here, we have used the fact that $\mathcal{J}_c=\mathcal{J}_+$ and $\mathcal{Q}_c=\mathcal{Q}_+$.} \cite{caldarelli2000thermodynamics, Chen:2010bh}
\begin{equation}\label{eqn:generatorshorizonsandpotentials}
\begin{aligned}
  \xi_j &= \partial_t + \Omega_j \partial_\phi,  \qquad\text{where}\qquad &\Omega_j \equiv \Omega(r_j) \qquad\text{with}\qquad \Omega(r) &= \frac{a}{r^2 + a^2}\,, \\
  \kappa_j &= \frac{|\Delta_r'(r_j)|}{2\, \Xi \, (r^2_j + a^2)}, \qquad &\Phi_j \equiv \Phi(r_j) \qquad\text{with}\qquad \Phi(r) &= \frac{Q\, r}{\Xi \, (r^2 + a^2)}\, .
\end{aligned}
\end{equation}
Here the prime stands for derivative with respect to the argument. Notice how we use the shorthand notation $\Omega_j \equiv  \Omega(r_j)$ and $\Phi_j \equiv  \Phi(r_j)$ when evaluating these potentials at the horizons $j \in \{-,+, c\}$.

In Fig.~\ref{fig:kerrdSspacetime} we depict the part of the causal structure of non-extremal KNdS that will be relevant for our discussion. Region I describes the region between the event and cosmological horizon with  $r_+ < r < r_c$, also referred to as the black hole exterior. Region II corresponds to the black hole interior, \ie the region enclosed by the event and Cauchy horizons with $r_- < r < r_+$. The solution \eqref{eqn:metric_KNdS}-\eqref{eqn:potential_A} in Boyer-Lindquist coordinates is appropriate to describe KNdS in the exterior region I, but is not regular at the future event horizon ${\cal H}^+$. So, within region I, we  define a tortoise coordinate $r_*$ by 
\begin{equation}
    \frac{d r_*}{dr} = \frac{\Xi \, (r^2 + a^2)}{\Delta_r(r)}.
\end{equation}
We  then define ingoing Eddington–Finkelstein (EF) coordinates $(v, r, \theta, \phi')$ as
\begin{equation}\label{eqn:ingoing_EF_transform}
    dt = dv - dr_*, \qquad d\phi = d\phi' - \Omega(r) dr_*\,,
\end{equation}
which allows us to analytically extend the KNdS solution from region I into region II. These coordinates cover regions I and II of Fig.~\ref{fig:kerrdSspacetime} and are smooth at the `left' component $\mathcal{CH}_L^+$ of the Cauchy horizon. However, these are not regular at the `right' component $\mathcal{CH}_R^+$. To have a coordinate system that is regular at $\mathcal{CH}_R^+$, we define the outgoing EF coordinates $(u, r, \theta, \phi'')$ via
\begin{equation}\label{eqn:outgoing_EF_transform}
    dt = du + dr_*, \qquad d\phi = d\phi'' + \Omega(r) dr_*\,.
\end{equation}
 These outgoing EF coordinates also allow us to extend the KNdS solution across the future cosmological horizon $\mathcal{H}_c^+$.

\subsection{Spherical polar parametrisation of KNdS}
\begin{figure}[ht]
    \centering
    \includegraphics[width=0.49\textwidth]{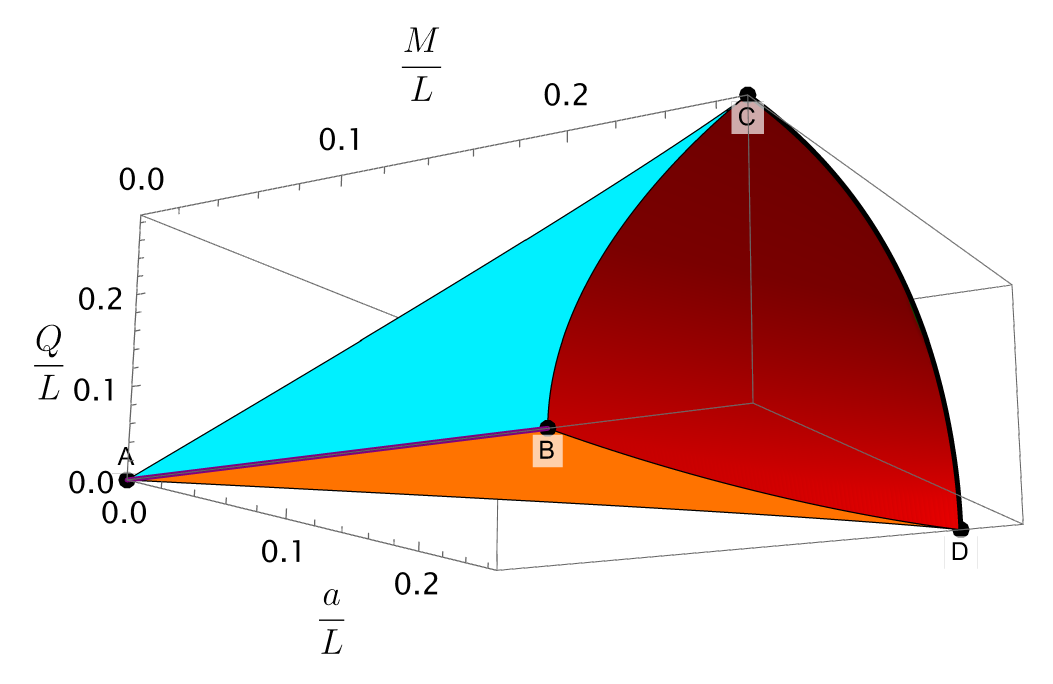}
    \includegraphics[width=0.49\textwidth]{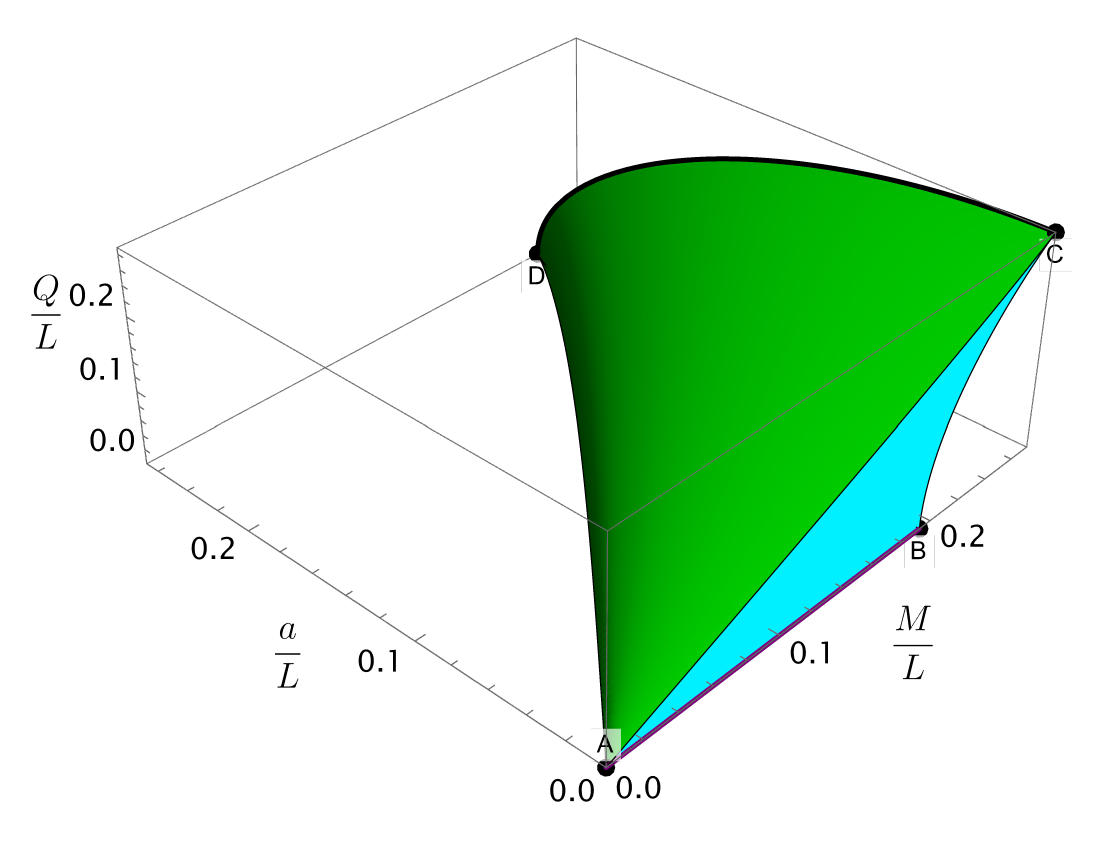}
    \caption{Region of the dimensionless parameter space $(M/L, a/L, Q/L)$ where regular KNdS black holes exist (from two distinct viewpoints in the left/right plots to clearly capture all boundary surfaces). The cyan $ABC$ surface with $a=0$ (or $\Theta=0$) describes regular RNdS black holes~\cite{Bousso:1996pn}, the orange $ABD$ surface  with $Q=0$ (or $\Theta=\pi/2$) describes Kerr-dS~\cite{Akcay:2010vt, Gregory:2021ozs}, the green $ACD$ surface describes extremal KNdS (with $r_-=r_+$ \ie $\calr = 1$) and the red $BCD$ surface describes KNdS black holes in the Nariai limit (with $r_+=r_c$ \ie  $y_+ = 1$).  The $AB$ purple line describes Schwarzschild-dS black holes, the $CD$ thick black line describes  extremal KNdS in the Nariai limit (\ie with $r_-=r_+=r_c$ or, equivalently, with $(\calr ,y_+ )= (1,1)$). Point $A=(0,0,0)$ describes the pure de Sitter solution, point $B=(\frac{1}{3\sqrt{3}},0,0)$ describes the Nariai Schwarzschild-dS black hole, point  $C=(\frac{\sqrt{2}}{3\sqrt{3}}, 0, \frac{1}{2\sqrt{3}})$ describes  extremal and Nariai RNdS (\ie with $a=0$ and $r_-=r_+=r_c$), and point  $D=(\frac{4}{3}\sqrt{26 \sqrt{3}-45}, 2- \sqrt{3},0)$ describes extremal Nariai Kerr-dS black holes  (\ie with $Q=0$ and $r_-=r_+=r_c$). \label{fig:KNdS_parameterspace}}
\end{figure}

Scanning the full parameter space of a black hole that is described by four dimensionful parameters $(M,Q,a,L)$ is not a trivial endeavour, so we should attempt to formulate a tractable setup. Factorising $\Delta_r(r)$ in terms of its roots $(r_-,r_+,r_c)$ we find 
\begin{equation}\label{eqn:Deltarinhorizons}
    \Delta_r(r) = \frac{1}{L^2} (r - r_-)(r - r_+)( r_c-r)(r + r_- + r_+ + r_c).
\end{equation}
It follows that we can express the KNdS parameters $(M, Q, a)$ in terms of the horizon radii and the cosmological radius $(r_-,r_+,r_c, L)$ via
\begin{equation}
\begin{aligned}\label{eqn:MAQToHorizonRadii}
    M &= \frac{(r_c + r_-)(r_c + r_+)(r_- + r_+)}{2 L^2}, \\
    Q^2 &= r_{c}^{2} + r_{+}^{2} + r_c (r_- + r_+) + r_- (r_- + r_+) - L^2 + \frac{r_c r_- r_+ (r_c + r_- + r_+)}{L^2}, \\
    a^2 &= L^2 - r_{c}^{2} - r_{-}^{2} - r_{+}^{2} - r_- r_+ - r_c (r_- + r_+).  
\end{aligned}
\end{equation}
Note that the absence of naked singularities and the conditions $0\leq r_-\leq r_+\leq r_c$ impose strict windows on these parameters that are not a priori intuitive.  Fortunately, these four parameters are not independent. This is because the system has a scaling symmetry\footnote{This scaling symmetry is such that, under the transformation $\{t, r, \phi, x\} \to \{\lambda t, \lambda r, \phi, x\}$ and $\{r_c, r_+, r_-\} \to \{\lambda r_c, \lambda r_+, \lambda r_-\}$, the metric is rescaled as $g \to \lambda^2 g$ and the Maxwell potential as $A \to \lambda A$, but since the Christoffel symbols, Riemann and Ricci tensors are left invariant, the Einstein-Maxwell equations of motion are left unchanged.}
that allows us to use parameters such as $M$, $L$ or $r_c$ as a unit of length. Consequently, KNdS is effectively described by `only' three dimensionless parameters. Common choices in the literature are $(\Lambda M^2, Q/M, a/M)$, $(M/L,Q/L,a/L)$ or  $(L/r_c,r_-/r_c,r_+/r_c)$. Although we do not need this information for our study, for completeness, we display the 3-dimensional parameter space  $(M/L,Q/L,a/L)$ where regular KNdS black holes exist in Fig.~\ref{fig:KNdS_parameterspace}.\footnote{For $r_-$, $r_+$ and $r_c$ to be real,  the inequality
\begin{equation}
  \Big|  54 M^2 L^4 + \left(a^2-L^2\right) \left[a^2 \left(a^2+34 L^2\right)+L^2 \left(L^2+36 Q^2\right)\right] \Big|
    \leq \left[a^2 \left(a^2-14 L^2\right)+L^2(L^2-12Q^2)\right]^{3/2}.\nonumber
\end{equation}
must be satisfied.
In the RNdS limit $a \to 0$, this reduces to the upper bound for $M/L$ given in~\cite{Bousso:1996pn}.} 

Instead of employing one of the above parameterizations of KNdS, in our study we use a simpler set of parameters. 
Motivated by the `{\it polar parametrization}' of Kerr-Newman introduced in~\cite{Davey:2023fin}, we find it very convenient to generalize it to a 3-dimensional parameter space and  introduce the following `{\it spherical polar parametrization}' of KNdS:
\begin{equation}\label{PolarCoord}
    y_+ \equiv \frac{r_+}{r_c}, \qquad\qquad \calr^2 \equiv \frac{r_-}{r_+}, \qquad\qquad \tan \Theta \equiv \frac{a}{Q}.
\end{equation}
These spherical polar parameters cover the full parameter space of KNdS and have well defined physical meaning and ranges. Namely, $y_+ \in [0, 1]$ denotes the off-Nariai measure, $\calr \in [0, 1]$ denotes the off-extremality measure, and $\Theta \in [0, \frac{\pi}{2}]$ measures the ratio of charge to angular momentum. Hence, we can we go from the RNdS limit ($a = 0$ or $\Theta = 0$) to the Kerr-dS limit ($Q = 0$ or $\Theta = \pi/2$) whilst keeping both the off-extremality and off-Nariai measures fixed. Furthermore, the extremal limit ($r_-\to r_+$ such that $T_-=T_+\to 0$) corresponds to $\calr \to 1$ whilst holding $y_+$ and $\Theta$ fixed. 
Finally,  the Nariai  limit $r_+\to r_c$ corresponds to $y_+ \to 1$ while keeping $\calr$ and $\Theta$ fixed.

Using~\eqref{eqn:MAQToHorizonRadii} and \eqref{PolarCoord}, we can express the dimensionless black hole parameters $\hat{M} \equiv M/r_c$, $\hat{a} \equiv a/r_c$ and $\hat{Q} \equiv Q/r_c$ in terms of the polar parameters $(y_+, \calr, \Theta)$ as
\begin{equation}
\begin{aligned}\label{eqn:MAQToPolar}
    \hat{M} \hat{L}^2 &= \frac{1}{2} y_+ (1 + y_+) (1 + \calr^2)(1 + y_+ \calr^2),  \\
    \hat{a} \hat{L} &= y_+ \calr\sqrt{1 + y_+ (1 + \calr^2)} \sin \Theta, \\
    \hat{Q} \hat{L} &= y_+ \calr\sqrt{1 + y_+ (1 + \calr^2)} \cos \Theta\,, 
\end{aligned}
\end{equation}
where $\hat{L} \equiv L/r_c$ is itself given in terms of $(y_+, \calr, \Theta)$ by
\begin{equation}
\begin{aligned}\label{eqn:LToPolar}
    {\hat{L}}^2 &= \frac{\hat{L}_0}{2}\left(1 + \sqrt{1 + \frac{4 \yp2 \calr^2 (1 + y_+ + y_+ \calr^2) \sin^2\Theta}{{\hat{L}_0}^2}} \,\right), \\
    {\hat{L}_0} &\equiv 1 + y_+ \Big(1 + \calr^2 + y_+ (1 + \calr^2 + \calr^4)\Big).  
\end{aligned}
\end{equation}
In these coordinates, the pure de Sitter limit corresponds to $y_+ \to 0$ and implies having $\hat{L} \to 1$.

Finally, for use in later sections, we provide explicit expressions of $\Omega_+$ and $\Phi_+$ evaluated at extremality, $\calr=1$. These are  
\begin{align}
   \Omega_+^{\rm ext}  &= \frac{1}{r_c}\frac{\hat{L}_{\rm ext} \sqrt{1+ 2 y_+} \sin \Theta }{y_+\left(\hat{L}_{\rm ext}^2+(1+2 y_+) \sin ^2\Theta \right)}, \label{eqn:omegaatext} \\
   \Phi_+^{\rm ext}&=\frac{\hat{L}_{\rm ext}^5 \sqrt{1+ 2 y_+} \cos \Theta }{\left(\hat{L}_{\rm ext}^2+(1+2 y_+) \sin ^2\Theta \right) \left(\hat{L}_{\rm ext}^4+y_+ ^2 (1+2 y_+) \sin ^2\Theta \right)} \label{eqn:Phiatext},
\end{align}
where $\hat{L}_{\rm ext}$ is given by \eqref{eqn:LToPolar} with $\calr = 1$. 
\subsection{Setup of the problem}

\subsubsection{Klein-Gordon equation on KNdS}
\label{sect:kg_setup}

The study of linearized scalar field perturbations of the KNdS black hole simply reduces to studying the Klein-Gordon equation for the scalar field on a fixed KNdS black hole background.
Although our numerical results and SCC conclusions will be presented only for a neutral massless scalar field, for the sake of completeness and because it does not compromise our presentation, we compute the analytical approximations of the QNMs for massive charged scalar fields.

The Klein-Gordon equation for a scalar field $\Psi$ with mass $\mu$ and charge $q$ on a fixed KNdS background is given by
\begin{equation}\label{eqn:KG_equation}
    (\nabla - i q A)^2 \Psi = \mu^2 \Psi
\end{equation}
where $\nabla$ is the covariant derivative with respect to the KNdS metric \eqref{eqn:metric_KNdS}, and $A$ is the Maxwell potential given in~\eqref{eqn:potential_A}.  Since $\partial_t$ and $\partial_\phi$ are Killing vector fields of KNdS, we can perform a Fourier decomposition $\Psi (t,r,x,\phi) = e^{- i \omega t + i m \phi} \chi(r) S(x)$, which introduces the frequency $\omega$ and azimuthal quantum number $m$ of the scalar perturbation and, for computational convenience, we use the new polar coordinate $x = \cos \theta$. As a result, the Klein-Gordon PDE separates into a pair of radial and angular ODEs  for $\chi(r)$ and $S(x)$ with separation constant $\lambda$:
\begin{gather}
    \frac{d}{dr}\left(\Delta_r \chi'\right) + \left[\frac{1}{\Delta_r} \left(1+ \frac{a^2}{L^2}\right)^2 (r^2 + a^2)^2\left( \omega - m\, \Omega - q \,\Phi\right)^2  - r^2 \mu^2 - \lambda \right] \chi = 0 \,,\label{eqn:KG_radial} \\
    \frac{d}{dx} \Big( (1 - x^2) \Delta_x S' \Big) - \left[a^2 \left(1 + \frac{a^2}{L^2}\right)^2 \frac{(1 - x^2)}{\Delta_x} \left( \omega - \frac{m}{a(1-x^2)} \right)^2 + a^2 x^2 \mu^2 -\lambda  \right ] S = 0 \,, \label{eqn:KG_angular}
\end{gather}
where $\Delta_r$ is defined in \eqref{eqn:BL_metric_functions}, $\Delta_x = 1 + a^2 x^2/L^2$, and the rotational $\Omega(r)$ and electromagnetic $\Phi(r)$ potentials are defined in \eqref{eqn:generatorshorizonsandpotentials}. 
We can restrict our attention to $m \geq 0$ (regularity of $S$ requires that $m$ is an integer) as long as we  study both signs of $\text{Re}(\omega)$. This is because of the $t - \phi$  symmetry of KNdS, which tells us that if $\{\omega_m, \lambda_m\}$ is a solution pair of QNM frequency $\omega$ and angular separation constant $\lambda$ for fixed $m$, then so will be $\{-\omega^*_{-m}, \lambda^*_m\}$~\cite{Leaver:1985ax, Berti:2005gp}. Equivalently, we can consider both signs of $m$ and then only look for modes with ${\rm Re}(\omega_m)>0$. 

\subsubsection{Boundary conditions}\label{sect:KG_equation_Frobenius_analysis}

To discuss the boundary conditions we need to impose at the poles of the sphere and at the event and Cauchy horizons, we first need to understand the behaviour of our field perturbations near these boundaries.

 The radial equation has regular singular points at each of the horizon radii. Performing a Frobenius expansion around the event horizon $r = r_+$ yields two linearly independent solutions, namely
\begin{equation}\label{eqn:ingoing_frobenius_expansion}
    \chi \sim (r-r_+)^{\pm\frac{i}{2 \kappa_+} (\omega - m \Omega_+ - q \Phi_+)}.
\end{equation}
Similarly, there are two linearly independent solutions at the cosmological horizon $r = r_c$,
\begin{equation}\label{eqn:outgoing_frobenius_expansion}
    \chi \sim (r_c - r)^{ \pm\frac{i}{2\kappa_c} (\omega - m \Omega_c - q \Phi_c)}.
\end{equation}
QNM boundary conditions require that the radial field $\chi$ is regular in ingoing EF coordinates~\eqref{eqn:ingoing_EF_transform} at the future event horizon $\mathcal{H}^+$ and regular in outgoing EF coordinates~\eqref{eqn:outgoing_EF_transform} at the future cosmological horizon $\mathcal{H}_c^+$. In Boyer-Lindquist coordinates~\eqref{eqn:metric_KNdS}, these QNM boundary conditions translate to the requirement that the radial wavefunction must behave as
\begin{equation}\label{eqn:QNM_BCs}
    \chi(r) \sim \begin{cases}
        \left(r - r_+\right)^{-\frac{i}{2 \kappa_+} (\omega - m \Omega_+ - q \Phi_+)} \chi_+(r)  \ &\textrm{as}\quad r \to r_+\,,\\
        \left(r_c - r\right)^{- \frac{i}{2\kappa_c} (\omega - m \Omega_c - q \Phi_c)} \chi_c(r) \  &\textrm{as}\quad r \to r_c\,.
    \end{cases}
\end{equation} 
Here, the functions $\chi_+(r)$ and $\chi_c(r)$ are smooth and non-zero at $r = r_+$ and $r = r_c$, respectively. On the other hand, performing a Frobenius expansion of the angular ODE (\ref{eqn:KG_angular}) around the poles $x=\pm 1$ yields the two independent solutions $S\sim (1-x^2)^{\pm \frac{|m|}{2}}$. Regularity then requires that we discard the solution with negative exponent.  

After imposing these boundary conditions, equations (\ref{eqn:KG_radial}) and (\ref{eqn:KG_angular}) describe a coupled eigenvalue problem for the QNM frequency $\omega$ and the angular separation constant $\lambda$. We must solve this problem to determine the least damped frequency and consequently the spectral gap. These will allow us to determine whether Christodoulou's formulation of SCC in KNdS is respected. 

In our numerical search of QNMs, we can impose the QNM boundary conditions~\eqref{eqn:QNM_BCs} straightforwardly by redefining the radial and angular eigenfunctions  as
\begin{equation}\label{eqn:radial_BC_redefinition}
\begin{aligned}
    &\chi \equiv (r-r_+)^{-\frac{i}{2 \kappa_+} (\omega - m \Omega_+ - q \Phi_+)} (r_c - r)^{ - \frac{i}{2\kappa_c} (\omega - m \Omega_c - q \Phi_c)} R(r)\,, \\
    &S \equiv (1-x^2)^{\frac{|m|}{2}}Y(x)\,, \\
\end{aligned}
\end{equation}
and then searching for smooth eigenfunctions $R(r)$ in $r \in [r_+,r_c]$ and $Y(x)$ in $x \in [-1,1]$.\footnote{For numerical computations we have also introduced a compact radial coordinate $\rho = (r-r_+)/(r_c-r_+)$, such that $\rho \in [0, 1]$ in the exterior region $r \in [r_+, r_c]$.}

With these considerations we are now ready to solve the coupled pair of radial and angular ODEs \eqref{eqn:KG_radial}-\eqref{eqn:KG_angular} subject to their boundary conditions, to find the coupled pair of  eigenvalues $\lambda$ and $\omega$. Numerically, we will find the QNMs using one of two methods. The first makes use of the fact that this is an eigenvalue problem which can  be solved using {\tt Mathematica}’s built-in routine {\tt Eigensystem}. More details of this method and the discretization scheme can be found in~\cite{Dias:2010maa}. The second method uses the Newton-Raphson root-finding algorithm to solve a eigenvalue problem, and is explained in~\cite{Dias:2015nua}. The advantage of the first method is that it gives many QNMs simultaneously. The second method  computes only a single mode at a time, and only when a seed is known that is sufficiently close to the true answer. However, this method is much faster when the size of the numerical grid and numerical precision increases, and can be used to push the numerics to extreme regions of the parameter space. We will use mostly the latter but the former also proves useful in our solution scanning of the RNdS case (where the angular equation is independent of $\omega$).

In  section~\ref{sect:analytical_study} we also solve our eigenvalue problem analytically using perturbation methods. These analytical analyses have strong limitations since they provide good approximations only in narrow windows of the parameter space. But they will prove useful to: 1) test our numerics,  2) provide good seeds for the Newton-Raphson method and 3) help identify from first principles the distinct families of QNMs that are present in our system.

\subsubsection{Criterion for the violation of Christodoulou's SCC in KNdS}\label{sec:SCCcriterion}

The behaviour of perturbations at the Cauchy horizon depends on the decay rate of perturbations along the event horizon~\cite{Hintz:2015jkj}. For asymptotically de Sitter black holes it has been established that perturbations decay exponentially along the event horizon~\cite{Barreto1997,bony2007decay,Dyatlov:2011jd,Dyatlov:2013hba,Hintz:2016gwb,Hintz:2016jak}.
It follows that for de Sitter black holes, for linearized matter or gravitational perturbations, the outcome of the blueshift/redshift competition depends on whether the ratio between the spectral gap and the Cauchy horizon surface gravity is above or below a critical value~\cite{Hintz:2015jkj},
where the spectral gap is defined as the magnitude of the imaginary part of the least damped QNM of the black hole. More precisely, we define
\begin{align}\label{def:beta}
    \beta \equiv -\frac{\text{Im}(\omega)}{\kappa_-}\,
\end{align}
and then whether SCC does or does not hold depends on whether the \emph{minimum} value of $\beta$, across all QNMs, is above or below a critical value.
For RNdS and Kerr-dS black holes, it was shown that this critical value is $\beta=1/2$, for perturbations of various spins~\cite{Hintz:2015jkj, Cardoso:2017soq, Dias:2018etb, Dias:2018ynt}. Since these two solutions are limiting solutions of KNdS, this suggests that the very same critical value of $\beta$ holds for scalar field perturbations in KNdS. It is however important to explicitly verify this is indeed the case and we do so in this subsection.
 
 Our aim is to carefully study the behaviour of the scalar QNMs at the `right' component  $\mathcal{C}\mathcal{H}_R^+$ of the Cauchy horizon (see Fig.~\ref{fig:kerrdSspacetime}), which corresponds to $r = r_-$.\footnote{There is physical motivation to focus on $\mathcal{C}\mathcal{H}_R^+$ and not on $\mathcal{C}\mathcal{H}_L^+$. If the black hole has been formed by gravitational collapse, $\mathcal{C}\mathcal{H}_L^+$ could be partially or entirely blocked by the effective radius of the body, as it happens to the event horizon in the Penrose diagram for the gravitational collapse of a spherically symmetric star~\cite{waldGeneralRelativity1984}. This will never happen to $\mathcal{C}\mathcal{H}_R^+$.} We must choose our coordinates carefully. From our boundary condition analysis of section \ref{sect:KG_equation_Frobenius_analysis}, QNMs are analytic functions of $(t, r, x, \phi)$ in region I with $r_+ < r < r_c$. To cross the event horizon, we must use ingoing EF coordinates~\eqref{eqn:ingoing_EF_transform}. Once we are in region II (\ie in $r_- < r < r_+$), notice that in these coordinates $(v, r, x, \phi')$ a QNM has time dependence $e^{- i \omega v}$, so it will diverge as $v \to \infty$, \ie at $\mathcal{CH^+_R}$. Hence, we convert back into Boyer-Lindquist coordinates $(t,r,\theta,\phi)$ and then to outgoing EF coordinates $(u, r, x, \phi'')$ as defined in~\eqref{eqn:outgoing_EF_transform}. After all the required coordinate changes, in $(u, r, x, \phi'')$ coordinates the scalar field in region II is described by
\begin{align}
    \Psi = e^{-i \omega u}e^{i m \phi''}S(x) \tilde{\chi}(r),
\end{align}
for some function $\tilde{\chi}(r)$. As in section \ref{sect:KG_equation_Frobenius_analysis}, we perform a Frobenius analysis about the `right' Cauchy horizon $\mathcal{C}\mathcal{H}_R^+$ to find the two independent behaviours 
\begin{align}\label{Twobehaviours}
    &\Psi^{(1)} = e^{-i \omega u} e^{i m \phi''}S(x)R^{(1)}(r), \\
    &\Psi^{(2)} = e^{-i \omega u} e^{i m \phi''}S(x)(r-r_-)^{i(\omega - m\Omega_- - q \Phi_-)/\kappa_-}R^{(2)}(r),
\end{align}
with $R^{(1),(2)}$ denoting smooth non-vanishing functions at $r = r_-$. For a general complex frequency $\omega$, notice that $\Psi^{(2)}$ is not smooth at $r = r_-$. In addition to that, there is no additional condition that forces either of the behaviours to vanish. Thus, as any scalar QNM solution is given by a linear combination of these two behaviours near $\mathcal{C}\mathcal{H}_R^+$, its regularity will be dictated by $\Phi^{(2)}$. 

From \eqref{Twobehaviours} one sees that $\Psi^{(2)} \sim (r - r_-)^p$, for $p = i(\omega - m\Omega_- - q\Phi_-)/\kappa_-$. Taking the derivative, we have that $\partial_r\Psi^{(2)} \sim (r - r_-)^{p-1}$. This is locally square integrable if and only if $2 (\beta - 1) > -1$, with $\beta = \text{Re}(p)$. But both $i 
 m \Omega_-/\kappa_-$ and $i q \Phi_-/\kappa_-$ are purely imaginary. Therefore, for the scalar QNM solution to be locally square integrable (\ie a weak solution) at the Cauchy horizon, we must have
\begin{align}
    \beta > \frac{1}{2}\,,
\end{align}
as we wished to show. 

If we can find a QNM with $\beta \le 1/2$ then the scalar field cannot be extended across the Cauchy horizon as a weak solution and so Christodoulou's formulation of SCC is respected. One just needs one such QNM since generic initial data will contain it.
On the other hand, if for a particular KNdS solution all of its QNMs have $\beta  > 1/2$ then SCC is violated, indicating a breakdown of predictability. Since the behaviour of perturbations at the Cauchy horizon is determined by the slowest decaying QNM, 
in this case, any scalar perturbation arising from smooth initial data on $\Sigma$ can be extended across $\mathcal{CH}^+_R$ as a weak solution of the equation of motion, so the Christodoulou's  SCC is violated for smooth initial data.

In the following sections, our aim is to scan the full 3-dimensional KNdS parameter space and for each KNdS black hole determine its least damped QNM frequency, and thus $\beta$. Comparing it with the critical value of $1/2$ will determine whether SCC holds for such a black hole.

\section{Families of quasinormal modes and their analytical approximations}\label{sect:analytical_study}
The QNMs for RNdS and weakly charged KNdS can be grouped into three distinct families: the \textit{de Sitter (dS)} modes, the \textit{photon sphere (PS)} modes and the \textit{near-horizon (NH)} modes\footnote{A fourth family of modes known as the \textit{near-Nariai} modes, associated to the Nariai limit $r_+ \to r_c$, is often brought up in the literature. Our numerical computations show that these turn out to be a subset of PS modes, in agreement with~\cite{Dias:2018ynt}. Therefore, we do not consider them to be a separate family.}~\cite{Yang:2013uba, Yang:2012he, Zimmerman:2015trm, Cardoso:2017soq, Dias:2018etb, Dias:2018ynt, Detweiler:1980gk, teukolsky1974perturbationsIII, Sasaki:1989ca, Andersson:1999wj, Glampedakis:2001js, Hod:2008zz, Hod:2014uqa, Hod:2015xlh, Kokkotas:2010zd, Dias:2021yju, Dias:2022oqm, Davey:2022vyx, Casals:2020uxa, Konoplya:2007zx, Churilova:2021nnc}. The identification of each family can be established by solving the QNM problem in specific regions (neighbourhood of `boundaries/corners') of the parameter space, where we can find analytic expressions for the QNM frequencies in some approximation. We then try to follow this family as we change the parameters of the solution to extend their classification to the full parameter space. Especially near-extremality and for large values of $\Theta$, this task is often less trivial due to a phenomenon known as {\it eigenvalue repulsion}~\cite{Dias:2021yju,Dias:2022oqm,Davey:2022vyx,Davey:2023fin}.  
The classification is sometimes also harder for very small black holes, because dS mode curves tend to split into two branches and then each branch can merge with a third branch. 

In subsection \ref{sect:dSmodes}, we provide the analytic expression for the dS modes, which correspond to the QNMs found analytically in the dS spacetime boundary ($y_+\to 0, \calr \to 0,\Theta\to 0$) of the KNdS parameter space~\cite{Berti:2005gp}. In subsection \ref{sect:PSmodes}, we analytically capture the PS modes in the eikonal limit $m = \ell \to \infty$, a.k.a. the geometric optics approximation (where $\ell$ is the wave quantum number that gives the number of nodes along the angular eigenfunction of the problem), by studying the properties of unstable circular null geodesics in the KNdS background (this is effectively a leading order WKB expansion in $1/m$). Finally, the NH modes can be found in the near-extremal limit $\calr \to 1$ after a careful study of the Klein-Gordon equation in the near-horizon geometry of KNdS, which we carry out in subsection~\ref{sect:NHmodesandNHgeometry}.

Away from these `boundaries/corners' of the parameter space, \ie at an arbitrary point of parameter space, the task of classifying a given QNM unambiguously as a dS, PS or NH mode is far from trivial. It is particularly difficult for large $\mathcal{R}$ and large $\Theta$, but not exclusively. As we will see in later subsections, this is because the imaginary part of the three families QNM can intersect, repulse and interact in non-trivial ways throughout the parameter space of KNdS. This had been observed and well documented previously in asymptotically flat black holes~\cite{Dias:2021yju,Dias:2022oqm,Davey:2022vyx,Davey:2023fin} and in rotating de Sitter black holes~\cite{Davey:2022vyx} (where repulsions can also occur with the dS family of modes). To classify the three families of QNMs, in practice we start by doing it in the RNdS limit where the three families can be clearly distinguished~\cite{Cardoso:2017soq, Dias:2018ynt}. 
Then, as we move in the KNdS parameter space away from its RNdS limit we carefully follow the evolution of each family while being extra cautious with the eigenvalue repulsions (or even with branch splitting/merging) that can occur. Ultimately, for the SCC problem, this classification is not fundamental as long as we always identify the least damped QNM for each KNdS black hole, which is an easier task than the classification problem.

\subsection{de Sitter modes}\label{sect:dSmodes}
In the limit $(M \to 0, Q \to 0, a \to 0)$ or $(y_+\to 0, \calr \to 0,\Theta\to 0)$, KNdS reduces to the pure dS spacetime. Then, we can analytically solve (\ref{eqn:KG_radial}) and impose  outgoing conditions at the cosmological horizon and regularity conditions at the origin, as pure dS does not have an event horizon. As a result, in $d = 4$ we obtain the two sets of dS mode frequencies~\cite{Lopez-Ortega:2006aal}
\begin{align}\label{eqn:dSmodesfrequencies}
\omega_{dS} r_c = -i \left[\ell + 2n + \frac{3}{2}\left( 1 \pm \sqrt{1 - \frac{4 \mu^2}{9}}\right)\right], \quad \text{with} \quad n = 0, 1, \dots
\end{align}
Here the non-negative integer $\ell$ is the orbital quantum number that effectively counts the number of zeroes of the angular wavefunction and $n$  is the radial overtone quantum number, which counts the zeroes of the radial wavefunction (modes with $\mu=0$ and $\ell=n=0$ and thus $\omega_{dS}=0$ are pure gauge).  Turning on the black hole parameters, we typically find that for $|m| = \ell > 0$ the dS frequencies depend weakly on the black hole parameters (while for $m = \ell = 0$ there is the potential for complicated interactions due to eigenvalue repulsions, discussed later). More importantly for our SCC study, from our analyses we conclude that dS family of QNMs is subdominant except for KNdS black holes with small $y_+$. Hence, we will mostly focus on the other two families of QNMs. 

\subsection{Photon sphere modes}\label{sect:PSmodes}
In the eikonal limit $|m| = \ell \to \infty$, with $m$ and $\ell$ the azimuthal and orbital quantum numbers, there is a family of QNMs known as \emph{photon sphere} (PS) modes. These can be obtained by either performing a leading order WKB analysis of the QNM problem, or as we do here, using the \textit{photon sphere correspondence}. This correspondence has been carefully studied in $d = 4$ both numerically and analytically~\cite{Ferrari:1984zz, Mashhoon:1985cya, Hod:2009td, Dolan:2010wr, Dias:2018etb, Cardoso:2017soq, Casals:2020uxa} and, at the leading WKB order, it is independent of the spin of the perturbation.  

The photon sphere correspondence relates the PS frequencies to the dynamics of unstable circular null geodesics in the equatorial plane via
\begin{equation}\label{eqn:PS_correspondence}
    \omega_{\hbox{\tiny eik}} \simeq m \Omega_{\rm K} - i \left(n + \frac{1}{2}\right) \left|\lambda_{\rm L}\right|, \qquad n = 0, 1, 2, \dots
\end{equation}
where $\Omega_{\rm K}$ is the Kepler frequency of the geodesics, $\lambda_{\rm L}$ is the principal Lyapunov exponent and $n$ the overtone number. The principal Lyapunov exponent $\lambda_{\rm L}$ is the rate at which a null geodesic congruence increases its cross-section under infinitesimal radial deformations. The solutions to (\ref{eqn:PS_correspondence}) come in pairs, constituted by a \textit{co-rotating} mode with $\Omega_{\rm K} > 0$ and a \textit{counter-rotating} mode with $\Omega_{\rm K} < 0$. In the RNdS limit, the co-rotating and counter-rotating PS modes have equal imaginary parts and their real parts only differ by a sign. However, away from RNdS ($\Theta \neq 0$), the co-rotating PS modes dominate within the family, \ie they have a smaller $|\text{Im}(\omega_{\hbox{\tiny eik}})|$. In this subsection we apply (\ref{eqn:PS_correspondence}) to provide an analytic expression for the PS modes in the large $|m|=\ell$ limit.

First, we find the circular null geodesics in the equatorial plane. The Lagrangian describing geodesics in KNdS is given by $\mathcal{L} = - \frac{1}{2} g_{ab} \dot{x}^a \dot{x}^b$, where $x = (t, r, \theta, \phi)$ are geodesic coordinates that depend on an affine parameter $\eta$ and dots denote differentiation with respect to $\eta$. Geodesics confined to the equatorial plane $\theta(\eta) = \pi/2$ are described by the Lagrangian
\begin{equation}\label{eqn:lagrangian_equatorial}
    \mathcal{L} = - \frac{1}{2} \frac{r^2}{\Delta_r} \dot{r}^2 + \frac{\Delta_r \big( \dot{t} - a \dot{\phi} \big)^2 - \big(a \dot{t} - (r^2 + a^2) \dot{\phi}\big)^2}{2 r^2 \big(1 + \frac{a^2}{L^2}\big)^2}
\end{equation}
Since $\partial_t$ and $\partial_\phi$ are Killing vectors, the conjugate momenta associated to the $t$ and $\phi$ directions are the conserved charges
\begin{equation}
    E \equiv -\pi_t, \qquad\qquad L_\phi \equiv \pi_\phi.
\end{equation}
Substituting these into the equatorial Lagrangian~\eqref{eqn:lagrangian_equatorial}, the equation of motion for null equatorial geodesics ($\mathcal{L} = 0$) takes the form of a one-dimensional radial motion in an effective potential $V_{\rm eff}(r, b)$:
\begin{equation}
    \dot{r}(\eta)^2 = V_{\rm eff}(r, b), \qquad V_{\rm eff}(r,b) \equiv \frac{E^2}{r^4} \left( 1 + \frac{a^2}{L^2}\right)^2\bigg[ (a(a - b) + r^2)^2 - (a - b)^2 \Delta_r(r) \bigg]
\end{equation}
where we have defined the impact parameter $b = L_\phi/E$. The circular geodesics have orbit radius $r = r_s$ and impact parameter $b = b_s$ such that
\begin{equation}\label{eqn:circular_conditions}
    V_{\rm eff}(r_s,b_s) = 0, \qquad\qquad \partial_r \left.{V_{\rm eff}}(r,b)\right|_{r = r_s, b = b_s} = 0.
\end{equation}
Solving the first equation in~\eqref{eqn:circular_conditions} gives the two solutions
\begin{equation}\label{eqn:impact_parameter_solns}
    b_s^{\pm} = a + \frac{r_s^2}{a \pm \sqrt{\Delta_r(r_s)}}.
\end{equation}
Inserting this into the second equation in~\eqref{eqn:circular_conditions} yields a condition for $r_s$, given by
\begin{equation}\label{eqn:Deltar_equality}
    \pm 4 a \sqrt{\Delta_r(r_s^\pm)} = r_s \Delta_r'(r_s^\pm) - 4 \Delta_r(r_s^\pm),
\end{equation}
where $r_s^{\pm}$ corresponds to the solution with $b_s^{\pm}$, respectively. Using that $b = L_{\phi}/E$ and equations \eqref{eqn:circular_conditions}-\eqref{eqn:Deltar_equality}, the two solutions for the Kepler angular velocity $\Omega_{\rm K}$ are 
\begin{align}\label{eqn:Omega_keplerian_solns}
    \Omega_{\rm K}^\pm = \frac{\dot{\phi}}{\dot{t}} = \frac{1}{b_s^{\pm}},
\end{align}
with $\Omega_{\rm K}^+$ and $\Omega_{\rm K}^-$ corresponding the co-rotating and counter-rotating modes, respectively. Finally, substituting equations \eqref{eqn:circular_conditions}-\eqref{eqn:Omega_keplerian_solns} into \eqref{eqn:PS_correspondence}, the eikonal approximation $|m| = \ell \to \infty$ of the photon sphere modes is given by
\begin{equation}
\begin{aligned}\label{eqn:omega_eik}
    \omega_{\hbox{\tiny eik}}^\pm &= \frac{m}{b_s^\pm} 
    - i \left(\frac{1}{2} + n\right) \frac{\left|a^2 - a b_s^\pm + (r_s^\pm)^2\right|}{L \left(1 + \frac{a^2}{L^2}\right) (r_s^\pm)^2 |b_s^\pm (a - b_s^\pm)|} \\
    & \hspace{1.8cm} \times \sqrt{L^2\big[a^2 - (b_s^\pm)^2 + 6 (r_s^\pm)^2\big] + (a - b_s^\pm)^2 \big[a^2 + 6 (r_s^\pm)^2\big]} \,.
\end{aligned}
\end{equation}
where $n = 0, 1, 2, \dots$ again denotes the overtone.

We find that this eikonal frequency (derived using a first principles analysis that is effectively the leading order contribution of a $1/m$ WKB analysis) provides a very good approximation for the imaginary part of the frequency (relevant for the discussion of SCC) even for values as small as $m=\ell=10$ and, in the Kerr-dS limit, that  its accuracy keeps increasing to excellent levels as $m=\ell$ grows up to 50~\cite{Dias:2018ynt}. There is a phenomenological analysis that provides an even better approximation for the real part (not relevant for SCC) of the frequency~\cite{Casals:2020uxa}. In Appendix~\ref{sect:eikonal_analysis} we describe this distinction and also test the validity of both approximations.

The eikonal approximation \eqref{eqn:omega_eik}  is independent of the spin of the perturbation (higher order WKB corrections do depend on the spin). Consequently, the approximation \eqref{eqn:omega_eik}  is also a good approximation to eikonal limit of linear gravito-electromagnetic perturbations. It also turns out to describe well the eikonal limit of massive scalar field modes and~\cite{Casals:2020uxa} reports it is even a good eikonal approximation for charged scalars if one does the replacement  $ m \Omega_{\rm K}\to  m \Omega_{\rm K} + q\Phi(r_{s})$ in the real part of \eqref{eqn:PS_correspondence} (where $r_s$ is the radius of the scalar orbit). These observations will be useful later when discussing our SCC results for massless scalars and educated extrapolations for other perturbation sectors. 

\subsection{Near-horizon modes and the near-horizon geometry}\label{sect:NHmodesandNHgeometry}
Taking the near-extremal limit $r_- \to r_+$ (\ie $\calr \to 1$), we can utilise analytical methods to compute another family of QNMs known as the \textit{near-horizon} (NH) modes. The frequencies $\omega$ of these modes saturate the superradiant bound as we approach extremality: $\omega \to m \Omega_+ + q \Phi_+$ as $r_- \to r_+$. Their nomenclature stems from the fact that wavefunctions are, near-extremality, highly peaked near the event horizon $r = r_+$. We confirmed this was the case in KNdS when analysing the associated numerical wavefunctions. Although in the following sections we will present numerical QNM results and discuss SCC only for neutral massless scalar fields, for the sake of completeness, the analysis of this subsection will consider scalar fields with mass $\mu$ and charge $q$ that are not necessarily zero (this will be useful in the Conclusions).   

In subsection \ref{sect:BFboundandNHgeometry}, we study the near-horizon limit of near-extremal KNdS (NHEKNdS), to set up the analytical computation of the NH modes. As we will cover, the associated $\text{AdS}_2$ Breitenl\"{o}hner-Freedman (BF) bound appears naturally in the expressions for the NH frequencies and turns out to play a fundamental role in determining whether SCC is respected or not. Furthermore, the BF bound also establishes a boundary between a region of the parameter space where we can clearly distinguish the PS and NH families and another where we cannot (due to eigenvalue repulsions), as it has been previously observed in rotating black holes~\cite{Davey:2022vyx, Davey:2023fin}. 

In pure $\text{AdS}_2$ (with radius $L_{\text{AdS}}$), it is well-known that scalar field perturbations are normalizable even if their squared mass $\xi^2$ is negative, provided that it obeys the BF bound $\xi^2 L^2_{\text{AdS}} \geq -1/4$~\cite{Breitenlohner:1982jf, Mezincescu:1984ev}. Conversely, if $\xi^2$ does not obey the BF bound, then the scalar field on $\text{AdS}_2$ is not stable. The near-horizon geometry of KNdS can be locally expressed as a product of $\text{AdS}_2$ times a compact space~\cite{durkeePerturbationsNearhorizonGeometries2011a}. As a result, the near-horizon limit of \eqref{eqn:KG_equation} reduces to the Klein-Gordon equation of a scalar field on pure $\text{AdS}_2$ space (with a Maxwell field) with effective mass $\mu_{\text{eff}}$ and effective charge $q_{\text{eff}}$, which can give important information about the system~\cite{durkeePerturbationsNearhorizonGeometries2011a}. In particular, we apply this analysis to NHEKNdS. 

After having laid out the groundwork, we analytically capture the NH family of modes in subsection \ref{sect:MAEexpansion} by performing a matched asymptotic expansion. To achieve so, we solve the Klein-Gordon equation in the near-horizon region of near-extremal KNdS and then match it with a vanishing far-region (since the wavefunctions are very peaked near the horizon close to extremality). The validity of this poor-man matching asymptotic expansion will be assessed via a comparison with the exact numerical data. 

\subsubsection{Near-horizon geometry}\label{sect:BFboundandNHgeometry}
To find the near-horizon geometry, we start with the KNdS metric~\eqref{eqn:metric_KNdS} at extremality ${r_{-} \to r_{+}}$. Then, we zoom into the event horizon by applying the following coordinate and gauge transformations
\begin{gather}\label{eqn:nh_transformations}
  r \to r_{+} + \epsilon \tilde{R}, \quad\qquad t \to \frac{T}{\epsilon}, \quad\qquad \phi \to \tilde{\phi} + \Omega_+^{\rm ext} \, \frac{T}{\epsilon}, \quad\qquad A \to \tilde{A} - \Phi_+^{\rm ext} \frac{dT}{\epsilon},
\end{gather}
and then taking the near-horizon limit $\epsilon \to 0$.
Here $\Omega_+^{\rm ext}$ and $\Phi_+^{\rm ext}$ are defined in (\ref{eqn:omegaatext}) and (\ref{eqn:Phiatext}) respectively. To ease notation, we will often drop the subscript `ext' when it is clear that the expressions should be evaluated at extremality.

Notice that the gauge transformation of $A$ is required to yield a finite Maxwell potential in the near-horizon limit. Explicitly, substituting (\ref{eqn:nh_transformations}) in (\ref{eqn:metric_KNdS}) and taking $\epsilon \to 0$ yields the \emph{near-horizon geometry} of extremal KNdS
\begin{multline}\label{eqn:nh_geometry_raw}
    {ds}_{\hbox{\tiny NH}}^2 = \frac{2\Sigma_+}{\Delta_r''(r_+)} \left[ - \left(\frac{\Delta_r''(r_+) \tilde{R} \, dT}{2(1 + \frac{a^2}{L^2})(a^2 + r_+^2)}\right)^2 + \frac{d\tilde{R}^2}{\tilde{R}^2} \right]  \\
    + \frac{\Sigma_+}{\Delta_\theta} d\theta^2 + \frac{\Delta_\theta (a^2 + r_+^2)^2 \sin^2\theta }{\Sigma_+\left(1 + \frac{a^2}{L^2}\right)^2 } \left( d\tilde{\phi} + \frac{2 r_+ \tilde{R} \Omega_+}{a^2 + r_+^2} dT \right)^2,
\end{multline}
with the shifted gauge field
\begin{align}
    \tilde{A} &= \frac{Q}{\Sigma_+ (1 + \frac{a^2}{L^2})} \left( \frac{r_+^2 - a^2 \cos^2 \theta}{a^2 + r_+^2} \tilde{R} \, dT + a r_+ \sin^2 \theta \, d\tilde{\phi} \right).
\end{align}
Here $\Sigma_+ \equiv \Sigma(r_+, \theta)$, with $\Sigma$ defined in \eqref{eqn:BL_metric_functions}. The $(T,\tilde{R})$ part of the near-horizon metric is AdS\textsubscript{2}, with Ricci scalar
\begin{equation}
    \textrm{Ricc}_{(2)} = - \frac{2}{L_{\rm AdS}^{2}}, \qquad L_{\rm AdS}^2(\theta) \equiv \frac{2 \Sigma_+}{\Delta_r''(r_+)}.
\end{equation}
Note that the AdS\textsubscript{2} length scale $L_{\rm AdS}$ is $\theta$-dependent. To simplify the near-horizon geometry~\eqref{eqn:nh_geometry_raw}, we pull out an overall factor of $L_{\rm AdS}^{2}$ in front of the AdS\textsubscript{2} component by rescaling the radial coordinate $\tilde{R}$, and make the definitions 
\begin{equation}\label{eqn:nh_r_rescaling}
  \tilde{R} \equiv \frac{(a^{2}+r_{+}^{2})^{2}}{2 a r_{+}} \widetilde{\Omega}\, R, \qquad \widetilde{\Omega} \equiv \frac{4 r_+}{\Delta_r''(r_+)} \left(1 + \frac{a^2}{L^2}\right) \Omega_+.
\end{equation}
These make the AdS\textsubscript{2} structure of the near-horizon metric~\eqref{eqn:nh_geometry_raw} manifest in $(T, R, \theta, \tilde{\phi})$ coordinates:
\begin{equation}
\begin{aligned}\label{eqn:nh_geometry}
  ds_{\hbox{\tiny NH}}^{2} &= L_{\rm AdS}^{2} \left( - R^{2} dT^{2} + \frac{dR^{2}}{R^{2}} \right) + \frac{\Sigma_+}{\Delta_{\theta}} {d\theta}^{2} + \frac{(a^{2} + r_{+}^{2})^{2} \, \Delta_{\theta} \, \sin^{2}\theta}{\Sigma_+\big(1+ \frac{a^2}{L^2}\big)^{2}} \left( d\tilde{\phi} + \widetilde{\Omega} R \, dT \right)^{2}, \\
  \tilde{A} &= \frac{Q L_{\rm AdS}^2}{{\Sigma_+}^2} \left[ \left( r_+^2 - a^2 \cos^2 \theta \right) R \, dT + 2 a r_+^2 \frac{\Omega_+}{\widetilde{\rm \Omega}} \sin^2 \theta d\tilde{\phi}\right],
\end{aligned}
\end{equation}
in agreement with~\cite{Hartman:2008pb, Chen:2010bh, Chen:2010jj}. This is a solution of the $d = 4$ Einstein-Maxwell equations with $\Lambda > 0$. The geometry of the metric degenerates to precisely $\text{AdS}_2 \times S^2$ in the (extremal) RNdS $a \to 0$ limit. In the asymptotically flat limit $L \to \infty$, this coincides with the near-horizon extremal KN metric studied in~\cite{Davey:2023fin}. Its isometry group $\text{SL}(2, \mathbb{R}) \times U(1)$ has been carefully  studied in the context of the \textit{Kerr/CFT correspondence}~\cite{Compere:2012jk}. For other symmetries and properties of NHEKNdS we refer the reader to~\cite{Hartman:2008pb, Chen:2010bh, Chen:2010jj}. 

To find the Klein-Gordon equation on the near-horizon geometry~\eqref{eqn:nh_geometry}, we apply the same coordinate transformations~\eqref{eqn:nh_transformations} and~\eqref{eqn:nh_r_rescaling} to the radial and angular components of the Klein-Gordon equation~\eqref{eqn:KG_radial}-\eqref{eqn:KG_angular} on the full KNdS geometry.\footnote{Equivalently, one can compute the Klein-Gordon equation directly using the near-horizon solution~\eqref{eqn:nh_geometry}.} To eliminate divergent terms when taking the near-horizon limit $\epsilon \to 0$, the frequency must approach the superradiant frequency for a charged and rotating black hole
\begin{equation}
    \omega \to m \Omega_+^{\rm ext} + q \Phi_+^{\rm ext} + \epsilon \, \tilde{\omega},
\end{equation}
where again $\Omega_+^{\rm ext}$ and $ \Phi_+^{\rm ext}$ are  given explicitly in (\ref{eqn:omegaatext}) and (\ref{eqn:Phiatext}). 
The resulting radial equation on the near-horizon geometry is
\begin{multline}\label{eqn:near_horizon_radial}
   \frac{d}{dR} \Big(R^2 \chi'(R)\Big) - \Biggr\{ \frac{(\lambda + r_+^2 \mu^2)L^2}{(r_c - r_+)(r_c + 3 r_+)}  \\
    - \left[ \frac{\tilde{\omega}}{R} + \frac{2 m \Omega_+ r_+ (a^2 + L^2)}{(r_c - r_+)(r_c + 3 r_+)} + \frac{q (1 - 2 a \Omega_+) Q L^2 }{(r_c - r_+)(r_c + 3 r_+)} \right]^2 \Biggr\} \chi(R) =0 
\end{multline}
While it is convenient to express the near-horizon radial equation in terms of $L$, the third expression from \eqref{eqn:MAQToHorizonRadii} evaluated at extremality yields $L^2 = a^2 + r_c^2 + 2 r_c r_+ + 3 r_+^2$, \ie $L$  is not independent of $r_c,r_+,a$. Here, the separation constant $\lambda$ is defined by the near-horizon limit of the angular part of the Klein-Gordon equation~\eqref{eqn:KG_angular}, \ie
\begin{multline}\label{eqn:near_horizon_angular}
    \frac{d}{dx} \Big((1-x^2) \Delta_x S'(x)\Big) - \Biggr[ a^2 x^2 \mu^2  -\lambda\\
     + a^2 \left(1 + \frac{a^2}{L^2}\right)^2 \frac{1 - x^2}{\Delta_x} \left( m \Omega_+ + q \Phi_+ - \frac{m}{a(1-x^2)}\right)^2\Biggr] S(x) = 0.
\end{multline}
In the asymptotically flat limit, (\ref{eqn:near_horizon_radial}) and (\ref{eqn:near_horizon_angular}) match with (A.3a) and (A.3b) in~\cite{Davey:2023fin}.\footnote{For an exact match, we need the redefinition $\lambda \to \lambda - (2 r_+^2 + a^2)m^2\Omega_+^2$.} To provide a physical interpretation for (\ref{eqn:near_horizon_radial}), we rewrite it as the Klein-Gordon equation for a massive charged scalar on pure $\text{AdS}_2$
\begin{equation}\label{eqn:KG_AdS2}
    \left(\tilde{\nabla} - i q_{\text{eff}} A_{\text{eff}}(R) \right)^2 \Psi = \frac{(\lambda + r_+^2\mu^2)L^2}{(r_c - r_+)(r_c + 3 r_+)  L^{2}_{\rm AdS}} \Psi,
\end{equation}
where $\tilde{\nabla}$ is the covariant derivative on pure $\text{AdS}_2$ with metric 
\begin{equation}
     ds_{\text{AdS}_2}^{2} = L_{\rm AdS}^{2} \left(- R^{2} dT^{2} + \frac{dR^{2}}{R^{2}} \right),
\end{equation}
and we have made the identifications
\begin{equation}
    q_{\text{eff}} \equiv q \Phi_+ \frac{(L^2+ a^2 )(a^2-r_+^2)}{r_+ (r_c - r_+)(r_c + 3 r_+)} - m \tilde{\Omega}, \quad  A_{\text{eff}}(R) \equiv -R dT.
\end{equation}
 We can observe that, after the near-horizon limit procedure, the scalar field acquires an effective charge and effective mass, which are non-zero even in the case of $\mu = q = 0$. Hence (\ref{eqn:near_horizon_radial}) is the radial Klein-Gordon equation in the near-horizon background  for a scalar that has effective charge $q_{\text{eff}}$ and effective mass $\mu_{\text{eff}}$ (where we will identify the latter soon). 

As we approach the boundary of the near-horizon geometry $R \to \infty$, solutions of the near-horizon radial equation~\eqref{eqn:near_horizon_radial} decay as $\chi \sim R^{-\Delta_{\pm}}$, where the 2-dimensional conformal dimensions $\Delta_\pm$ are given by
\begin{align}
    \Delta_\pm &= \frac{1}{2} \pm \frac{1}{2}\sqrt{1 + 4 \mu^2_{\rm eff}L_{\rm AdS}^{2}},\\
    \mu^2_{\rm eff}L^2_{\rm AdS} &\equiv \frac{(\lambda + r_+^2\mu^2)L^2}{(r_c - r_+)(r_c + 3 r_+)} - \left(q \Phi_+ \frac{(L^2+a^2)(a^2-r_+^2)}{r_+ (r_c - r_+)(r_c + 3 r_+)} - m \tilde{\Omega}\right)^2.
\end{align}
For these solutions to be normalisable, \ie to have finite energy, they must not oscillate at infinity. This requirement is satisfied by demanding that $\Delta_\pm$ is real, giving the $\text{AdS}_2$ BF bound of the near-horizon geometry 
\begin{align}\label{eqn:BFbound}
\mu^2_{\text{eff}}L_{\rm AdS}^{2} \geq -\frac{1}{4}.
\end{align}
Defining the constant
\begin{align}\label{eqn:delta_definition}
  \delta \equiv \sqrt{\frac{\left(a^2+L^2\right)^2 \left[ q \Phi_+(r_+^2-a^2) +  2 m \Omega_+ r_+^2)\right]^2}{r_+^2 (r_c-r_+)^2 (r_c+3 r_+)^2}-\frac{L^2 \left(\lambda +r_+^2\mu^2 \right)}{(r_c-r_+) (r_c+3 r_+)}-\frac{1}{4}}\,, 
\end{align}
we have the important relation 
\begin{align}\label{eqn:d2andeffectivemassrelation}
    1 + 4 \mu^2_{\text{eff}}L_{\rm AdS}^{2} = - 4 \delta^2 \qquad \Leftrightarrow \qquad \delta^2 =- \frac{1}{2}\sqrt{1 + 4 \mu_{\rm eff}^2 L_{\rm AdS}^2}.
\end{align}
Thus the BF bound (\ref{eqn:BFbound}) is equivalent to the bound  $\delta^2 \leq 0$. In the asymptotically flat limit, this $\delta$ matches the ones defined in~\cite{Davey:2023fin,Zimmerman:2015trm, Hod:2009td}. 

In the coming subsection, we will see $\delta$ naturally appearing in the analytic expression of the NH modes, hence establishing a connection between the NH modes and the near-horizon BF bound. Recall that a violation of the BF bound should signal an instability of the NHEKNdS geometry, which does not necessarily imply an instability of the full KNdS geometry\footnote{For an in-depth discussion in the case of the Kerr spacetime, we refer the reader to~\cite{durkeePerturbationsNearhorizonGeometries2011a, Hollands:2014lra}.}. Nonetheless, a BF bound violation might signal a transition boundary of the physical properties of the system. Indeed, as we exemplify in subsection \ref{sect:m=l=1slice}, the analytic properties of the QNM spectrum change near $\delta^2 = 0$, going from the $\delta^2<0$ region in parameter space where we can clearly distinguish the PS and NH families, to the region $\delta^2>0$ where both families are entangled and cannot be precisely distinguished apart. 
\subsubsection{Near-horizon modes and matched asymptotic expansion}\label{sect:MAEexpansion}

The \emph{near-horizon} (NH) QNMs can be found by performing a matched asymptotic expansion (MAE). We divide the spacetime into two regions: one near the event horizon, and another that is far away from it, all the way up to the cosmological horizon. In the near-region, the Klein-Gordon equation simplifies and we can solve it explicitly, up to some integration constants. For the far-region, we cannot solve the Klein-Gordon equation analytically (for asymptotically flat RN, Kerr and KN this can be done but not for KNdS or even Kerr-dS). To make some progress we must necessarily adopt a poor-man strategy where we start with the fundamental assumption that, near extremality, the wavefunction is localized very close to the event horizon (confirmed by an inspection of the exact numerical wavefunction) and then simply take the far-region solution to be the trivially vanishing one (which also trivially satisfies the boundary condition at the cosmological horizon). 

The near and far-region solutions must agree in an overlap region and we use this to fix the integration constants that were left free in the near-horizon region (after imposing the boundary condition at the event horizon) and the frequency $\omega$. This poor-man matching yields an analytical expression for NH modes near-extremality. Although this analysis has little control on the approximations made, it turns out that yields a very good approximation  (if not excellent) for the frequency of NH modes near-extremality when we compare it with the exact numerical results. Ultimately, it is this fact that justifies performing it.
The matched asymptotic expansion technique has been used successfully to similar QNM problems in~\cite{Detweiler:1980gk,teukolsky1974perturbationsIII, Dias:2018etb, Davey:2023fin, Yang:2012pj, Yang:2013uba,Zimmerman:2015trm}. 

To start the matching asymptotic expansion, we define an extremal parameter $\sigma$ and a radial parameter $y$ by
\begin{equation}
    \sigma  \equiv 1 - \frac{r_-}{r_+}, \qquad\qquad y \equiv \frac{r}{r_+} - 1,
\end{equation}
such that $\sigma \to 0$ at extremality ($r_-\to r_+$), and $y \in [0, r_c/r_+ - 1]$ in the exterior of the black hole region. Supported by past QNM studies near-extremality (including for the RNdS and Kerr-dS limits of KNdS), we now make the assumption that the frequency of the NH modes approaches the superradiant bound at extremality, $\omega \to m \Omega_+^{\rm ext} + q \Phi_+^{\rm ext}$. Altogether, the frequency of NH modes should have the following perturbative expansion around extremality 
\begin{equation}\label{eqn:NHAnsatz}
    \omega = m \Omega_+^{\rm ext} + q \Phi_+^{\rm ext} + \sigma \delta \omega + \mathcal{O}(\sigma^2).
\end{equation}
Our mission is to find the off-extremality frequency correction $\delta\omega$. As in the previous subsection, we drop the `ext' superscript for notational simplicity. Define the \emph{near-region} to be $0\leq y \ll 1$ and the \emph{far-region} to be $\sigma \ll y \le r_c/r_+-1$. Since we take the expansion parameter to be $\sigma\ll 1$, these two regions overlap in the matching region $\sigma \ll y\ll 1$.

To find the solution in the near-region $0 \leq y \ll 1$, we have to be cautious doing the perturbative expansion in $\sigma\ll 1$ because it can be as small as the radial coordinate $y$. This is closely related to the fact that the far-region ($y\gg \sigma$) solution breaks down when $y/\sigma \sim \mathcal{O}(1)$. This suggests that, to proceed with the near-region analysis, we should work with the new radial coordinate  
\begin{equation}\label{def:z}
    z \equiv \frac{y}{\sigma},
\end{equation}
so that the near-region corresponds to $z \ll \sigma^{-1}$. This amounts to absorbing a power of $\sigma$ into our new radial coordinate $z$, a fundamental step to yield a good MAE result. In terms of the original radial variable $r$, we have $z = (r - r_+)/(r_+ - r_-)$. Inserting \eqref{eqn:NHAnsatz} and \eqref{def:z} into the radial Klein-Gordon equation~\eqref{eqn:KG_radial}, to leading order in $\sigma$ it reduces to
\begin{align}\label{eqn:near_region_eqn}
&\hspace{-0.2cm} \frac{d}{dz} \Big(z (1+z) \chi_{\rm near}'(z)\Big) - \chi_{\rm near}(z) \bigg\{\frac{L^2 (\lambda + r_+^2\mu^2)}{(r_c - r_+)(r_c + 3 r_+)} \\
& \hspace{0.2cm}   +\frac{r_+ (a^2 + L^2)}{z(1+z)(r_c - r_+)(r_c + 3 r_+)} \left[\left(2 m \Omega_+ + q \Phi_+ \left(1-\frac{a^2}{r_+^2}\right) \right)z  + \left(1 + \frac{a^2}{r_+^2}\right) \delta \omega\right]^2 \bigg\}=0. \nonumber
\end{align}
This turns out to be an hypergeometric ODE. To see this is indeed the case, we do the field redefinition
\begin{equation}
    \chi_{\rm near}(z) = z^\mathcal{A} (1+z)^\mathcal{B} F(-z)
\end{equation}
where
 \begin{equation}
\begin{aligned}
    \mathcal{A} &= - i \frac{(a^2 + L^2) (a^2 + r_+^2)}{r_+ (r_c - r_+)(r_c + 3 r_+)} \delta \omega,\\
    \mathcal{B} &= i\frac{(a^2 + L^2)(2 m \Omega_+ r_+^2 - q \Phi_+(a^2 - r_+^2) - (r_+^2 + a^2)\delta\omega)}{r_+(r_c-r_+)(r_c + 3 r_+)},
\end{aligned}
 \end{equation}
and introduce the new radial variable $\bar{z} = -z$. Altogether, this turns \eqref{eqn:near_region_eqn} into
the hypergeometric differential equation 
\begin{equation}
    \bar{z} (1 - \bar{z}) F''(\bar{z}) + \big[c - (a_+ + a_- + 1)\bar{z}\big] F'(\bar{z}) - a_+ a_- F(\bar{z}) = 0,
\end{equation}
with 
\begin{align}
    \label{eqn:apandamdefinitions} a_\pm &= \frac{1}{2} \pm i \delta + \frac{2 i (a^2 + L^2) (a^2 + r_+^2)}{r_+(r_c-r_+)(r_c+ 3 r_+)} \left(\frac{m  r_+^2}{a^2 + r_+^2}\Omega_+ - \frac{q(a^2-r_+^2)}{2(a^2+r_+^2)}\Phi_+ - \delta\omega\right),\\
    \qquad c &= 1 - \frac{2 i (a^2 + L^2)(a^2 + r_+^2)}{r_+(r_c - r_+)(r_c + 3 r_+)}\delta \omega, 
\end{align}
and $\delta$ defined in \eqref{eqn:delta_definition}. 
We can now work out the general solution to the near-region equation~\eqref{eqn:near_region_eqn}. It is
\begin{multline}
    \chi_{\rm near}(z) = \, c_1 \, z^{\frac{1}{2}(c-1)} (1+z)^{\frac{1}{2}(a_-+a_+-c)} {}_2 F_1(a_-, a_+, c, -z) \\
    + c_2 \, z^{-\frac{1}{2}(c-1)} (1+z)^{\frac{1}{2}(a_- + a_+ -c)} {}_2 F_1(a_- - c + 1, a_+ - c + 1, 2 - c, -z),
\end{multline}
where ${}_2 F_1(\alpha, \beta, \gamma, x)$ is the hypergeometric function~\cite{abramowitz1968handbook} and $c_1, c_2$ are arbitrary constants. Imposing the ingoing boundary condition from~\eqref{eqn:QNM_BCs}  at the event horizon $z = 0$ requires setting $c_2 = 0$, \ie
\begin{align}\label{eqn:near_region_soln_regular}
    \chi_{\rm near}(z) = c_1 z^{\frac{1}{2}(c - 1)} (1 + z)^{\frac{1}{2}(a_- + a_+ - c)} {}_2 F_1(a_-, a_+, c, -z).
\end{align}
The constant $c_1$ as well as the frequency correction $\delta\omega$ are at this point still not determined. They are fixed matching the large $z$ behaviour of the near-region solution \eqref{eqn:near_region_soln_regular}, namely
\begin{equation}\label{Near:smallRadius}
\begin{aligned}
    \chi_{\rm near}\big|_{z\gg 1} \sim & \:\:  c_1 z^{-\frac{1}{2} + \frac{1}{2}\sqrt{1+4 \mu_{\text{eff}}^2L_{\rm AdS}^{2}}} \frac{\Gamma(a_- - a_+)\Gamma(c)}{\Gamma(c - a_+)\Gamma(a_-)} \left(1 - \frac{a_+(1+a_+-c)}{(1-a_-+a_+)}\frac{1}{z}+\mathcal{O}(z^{-2})\right) \\
    & + c_1 z^{- \frac{1}{2} -\frac{1}{2}\sqrt{1+4 \mu_{\text{eff}}^2L_{\rm AdS}^{2}}} \frac{\Gamma(a_+ - a_-) \Gamma(c)}{\Gamma(c - a_-) \Gamma(a_+)} \left(1 - \frac{a_-(1+a_--c)}{(1+a_--a_+)}\frac{1}{z}+\mathcal{O}(z^{-2})\right),
\end{aligned}
\end{equation}
with the small $y=\sigma z$ behaviour of the far-region solution in the overlapping region  $\sigma \ll y\ll 1$.
To get \eqref{Near:smallRadius} we have used (\ref{eqn:apandamdefinitions}) and (\ref{eqn:d2andeffectivemassrelation}) to rewrite the exponents of $z$. 

Formally, at this point, we should: 1) solve the radial Klein-Gordon equation~\eqref{eqn:KG_radial} in the far-region $y \gg \sigma$ to find $\chi_{\rm far}$, 2) impose the outgoing QNM boundary condition~\eqref{eqn:QNM_BCs} at the cosmological horizon $y=r_c/r_+-1$, and finally 3) take the small radius expansion of $\chi_{\rm far}$ to match it with \eqref{Near:smallRadius} in the overlapping region  $\sigma \ll y\ll 1$ to fix $c_1$ and $\delta\omega$. 
 Such a procedure has been completed successfully in asymptotically-flat spacetimes, including in Kerr and Kerr-Newman~\cite{Zimmerman:2015trm,Dias:2018ynt, Davey:2022vyx,Davey:2023fin}. However, in asymptotically de Sitter spacetimes the far-region equation turns out to be much more difficult (if not impossible) to solve analytically. To make some progress we must thus adopt a poor-man strategy where we make use of the fact that we do have exact numerical results for the NH modes. An inspection of the numerical NH wavefunctions shows that, near extremality, they are highly peaked near the event horizon and exponentially decay to zero very quickly as we move away from the event horizon. Therefore, for a leading approximation one sets $\chi_{\rm far}\sim 0$ and matches \eqref{Near:smallRadius} with a vanishing far-region solution. This is a very `rough' matching but its validity will be approved a posteriori when we compare the MAE frequency approximation with the exact numerical frequency. 
 
 Without further delays we perform the matching of \eqref{Near:smallRadius} with a trivial far-region. 
An inspection of the exponents of $z$ in \eqref{Near:smallRadius} indicates that we have to proceed differently depending on whether $1+4 \mu_{\text{eff}}^2L_{\rm AdS}^{2} $ is positive or negative.
 
For $1+4 \mu_{\text{eff}}^2L_{\rm AdS}^{2} >0$, or equivalently $\delta^2 < 0$ as defined in \eqref{eqn:delta_definition}, we  match the near-region wavefunction \eqref{Near:smallRadius} with a vanishing far-region wavefunction (for the aforementioned reasons). It follows that to have a finite solution in the limit $z \to \infty$ the first term in \eqref{Near:smallRadius}  must vanish. The gamma function has poles at negative integers, so we find that the first term vanishes if we impose the condition $a_- = -n$, where $n \in \mathbb{N}_0$ is a radial overtone. This is effectively a condition to solve for $\delta\omega$  that quantizes the frequency correction as  
\begin{align}\label{eqn:omega_MAE_positive}
    \delta \omega^{>} \simeq
    \frac{r_+}{2(a^2 + r_+^2)} \left.\left[ 2 m \Omega_+ r_+ - q \Phi_+\frac{a^2-r_+^2}{r_+} - \frac{(r_c-r_+)(r_c+3r_+)}{a^2 + L^2} \left( i \left(\frac{1}{2} + n\right) + \delta\right) \right]\right|_{\rm ext} 
\end{align}
where $\delta$ is defined in~\eqref{eqn:delta_definition}. This expression can be written in terms of the dimensionless polar parameters $(y_+, \Theta)$ using the relations~\eqref{eqn:MAQToPolar}-\eqref{eqn:LToPolar} evaluated at extremality. For fixed $(y_+, \Theta)$, we can equivalently express the requirement $1+4 \mu_{\text{eff}}^2L_{\rm AdS}^{2} >0$ as 
\begin{align}
 \hat{q}_{c}^{-}\leq \hat{q} \leq \hat{q}_{c}^{+} 
\end{align}
where $\hat{q} = q/r_c$ and 
\begin{multline}\label{eqn:qcriticalplusandminus}
    \hat{q}_{c}^{\pm} = -\frac{2 m \left(\hat{L}^4+y_{+}^2 (1+2 y_{+}) \sin ^2 \Theta \right)}{ y_{+} \hat{L}^2\left(\hat{L}^2-(1+ 2 y_{+}) \sin ^2\Theta \right)} \tan \Theta \\
   \pm \frac{(1-y_{+}) (1+3 y_{+}) \left(\hat{L}^2+(1+2 y_{+}) \sin ^2\Theta\right)}{2  y_{+} \hat{L} \sqrt{1+ 2 y_{+}}\left(\hat{L}^2-(1+ 2 y_{+}) \sin ^2\Theta\right) \cos \Theta} \sqrt{1+\frac{4 \hat{L}^2 \left(\lambda +\hat{\mu}^2 y_{+}^2\right)}{(1-y_{+}) (1+3 y_{+})}},
\end{multline}
with $\hat{\mu} = \mu/r_c$ and $\hat{L}=L/r_c$ evaluated at extremality. As a check of our computation, in the RNdS limit $\Theta \to 0$, this coincides with the expression provided in~\cite{Dias:2018ufh}.

When $1+4 \mu_{\text{eff}}^2L_{\rm AdS}^{2} < 0$, or equivalently $\delta^2 > 0$, both terms in (\ref{eqn:near_region_soln_regular}) decay whilst also exhibiting an oscillatory behaviour. Consequently, there is no longer the physical need to impose the vanishing of either term, so we must demand a distinct $z \to \infty$ behaviour. We follow the steps of~\cite{Dias:2009ex,Dias:2012pp,Dias:2018ufh} and seek travelling waves that are purely outgoing with respect to the phase velocity. This corresponds to waves that have a positive phase velocity with respect to $\text{Re}(\omega)$. 
More concretely, for $1+4 \mu_{\text{eff}}^2L_{\rm AdS}^{2} < 0$, the two oscillatory solutions at large $r$ are: 
\begin{align}\label{eqn:nearregionoscillatorybhehaviours}
    \Psi_{\pm}(r) \big|_{r\gg r_+} \sim \exp \left(\pm\frac{1}{2} i \sqrt{|1+4 \mu_{\text{eff}}^2L_{\rm AdS}^{2}| } \log\left(\frac{r}{r_+\sigma }\right) \pm \frac{a_{\pm}(c-a_\pm-1)}{1+a_\pm - a_\mp}\frac{ r_+ \sigma}{r}\right),
\end{align}
where the upper (lower) signs correspond to the first (second) term in \eqref{Near:smallRadius}. 
The $r$-dependent phase velocity of these two travelling wave solutions are
\begin{align}
    v_{\rm ph}^\pm(r) = \frac{\omega}{ k_\pm(r)}, \quad \text{with} \quad k_\pm(r) = -i \frac{1}{\Psi_{\pm}}\frac{d \Psi_{\pm}}{dr}.
\end{align}
Using (\ref{eqn:nearregionoscillatorybhehaviours}), this yields
\begin{align}
     v_{\rm ph}^\pm(r) \sim \pm \frac{2 r \,  \text{Re}(\omega)}{\sqrt{|1+4 \mu_{\text{eff}}^2L_{\rm AdS}^{2}|}}.
\end{align}
Therefore, requiring outgoing travelling waves ($v_{\rm ph}(r)>0$) we must eliminate the second term in \eqref{Near:smallRadius} vanish. We achieve this by imposing $a_+ = -n$ for $n \in \mathbb{N}_0$, which gives the condition
\begin{align}\label{eqn:omega_MAE_negative}
      \delta \omega^{<}\simeq
    \frac{r_+}{2(a^2 + r_+^2)} \left.\left[ 2 m \Omega_+ r_+ - q \Phi_+\frac{a^2-r_+^2}{r_+} - \frac{(r_c-r_+)(r_c+3r_+)}{a^2 + L^2} \left( i \left(\frac{1}{2} + n\right) - \delta\right) \right]\right|_{\rm ext}
    \end{align}
In summary, collecting both cases, the frequencies that emerge from the MAE analysis are:
\begin{align}\label{eqn:omega_MAE}
    \omega_{\hbox{\tiny MAE}} \simeq m \Omega_+^{\rm ext} + q \Phi_+^{\rm ext} + \sigma \begin{cases}
        \delta \omega^{>}\,, & \text{for $\hat{q} \in \{q_{c}^{-}, q_{c}^{+}\} \:\:\hbox{i.e.}\:\: \delta^2 < 0$} \\
        \delta \omega^{<}\,, & \text{for $\hat{q} \notin \{q_{c}^{-}, q_{c}^{+}\}  \:\:\hbox{i.e.}\:\: \delta^2 > 0$}
    \end{cases}+\mathcal{O}(\sigma^2),
\end{align}
with $\delta\omega^{>}$ given in (\ref{eqn:omega_MAE_positive}),  $\delta\omega^{<}$ given in (\ref{eqn:omega_MAE_negative}) and $\hat{q}_c^\pm$ defined in (\ref{eqn:qcriticalplusandminus}). 

Keeping in mind the regime of validity of the MAE result and associated approximations, we can make some comments with regards to SCC. Using the fact the leading order expansion of the surface gravity at the Cauchy horizon $\kappa_-$ (and also of $\kappa_+$) is
\begin{equation}
    \kappa_- = \frac{r_+(r_c-r_+)(r_c+ 3r_+)}{2(a^2 + r_+^2)(a^2 + L^2)}\sigma + \mathcal{O}(\sigma^2)
\end{equation}
we find that the prediction for $\beta$ based on the matched asymptotic expansion $\omega_{\MAE}$ is simply
\begin{equation}\label{eqn:beta_MAE}
    \beta_{\MAE} = \begin{cases}
        \frac{1}{2} + n + \operatorname{Im} \delta + \mathcal{O}(\sigma) & \text{for $\delta^2<0$}
        \\
        \frac{1}{2} + n + \mathcal{O}(\sigma) & \text{for $\delta^2>0$}
    \end{cases}.
\end{equation}
For the dominant overtone $n = 0$, when $\delta^2 > 0$ then $\beta_{\MAE}$ exactly approaches the critical value of $1/2$. However, when $\delta^2 < 0$, $\beta_{\MAE}$ goes above $1/2$. This is the zeroth-order term for $\beta_{\MAE}$, hence even if $\beta_{\MAE} = 1/2$ at extremality, it is possible that $\beta_{\MAE}$ approaches $1/2$ from above.

In Section~\ref{sect:QNM_spectrum_in_KNdS}, we will use (\ref{eqn:omega_MAE}) to identify the NH family of QNMs in our numerical data. In addition, we carefully verify the accuracy of this MAE approximation and compare it to previous work in Appendix~\ref{sect:MAEanalysis}. Finally, in Appendix~\ref{sect:MAEvalueforvanishingm} we specialize to the case $m = \ell = 0$ and $q = \mu = 0$, and show how the corresponding NH modes do not enforce SCC at extremality. 

\section{A survey of the KNdS QNM spectrum and its features}\label{sect:QNM_spectrum_in_KNdS}

In the previous section we identified, from first principles, the three families of modes that we expect to be present in the QNM spectra of KNdS. The analytical studies of the previous section are thus very useful but have significant limitations: they are only valid in narrow windows (\ie in a neighbourhood of `boundaries/corners') of the parameter space and often provide just very crude approximate values (unless proven otherwise) once we drive away from the specific background about which we do the expansion. This is not enough because, for the discussion of Christodoulou's formulation of SCC, we need to identify accurately the least damped QNM at each point of the parameter space of KNdS. Therefore, onwards we use numerical methods to exactly solve the eigenvalue problem that identifies the QNM frequencies (and coupled angular eigenvalues) of KNdS for the three families of QNM. 

A priori, we could simply use {\tt Mathematica}’s built-in routine {\it Eigensystem} to simultaneously find the tower of eigenfrequencies for each KNdS black hole and then pick the least damped mode relevant for SCC. However, scanning the full parameter space using this method would be a slow and computationally costly procedure. Alternatively, we can use the analytical approximations of the previous section to identify the QNM of each of the 3 families on a `boundary/corner' of the parameter space and use this as a seed to find the exact numerical solution using a Newton-Raphson root-finding algorithm. Once we have this first solution for each of the three QNM families, we can then use it to follow the evolution along the KNdS parameter space of the frequencies of each QNM family, at relatively low cost. Typically, we will use the latter method although we sometimes also use the former to find initial seeds for the latter (or as sanity checks to make sure we are not missing other modes).   
Before presenting the SCC results in section~\ref{sect:SCC_in_KNdS}, we must discuss key properties of the QNM spectra that are fundamental to precisely identifying the least damped modes which govern SCC. 

Recall that to scan the full parameter space of KNdS in an efficient and systematic way it is very convenient to use the `{\it spherical polar parametrization}' $\{y_+,\calr,\Theta\}$ of KNdS introduced in \eqref{PolarCoord}. 
In this parametrization, $y_+ \in [0, 1]$ denotes the off-Nariai measure, $\calr \in [0, 1]$ denotes the off-extremality measure, and $\Theta \in [0, \frac{\pi}{2}]$ measures the ratio of charge to angular momentum. With this parametrization, we can go from the RNdS limit ($a = 0$ or $\Theta = 0$) to the Kerr-dS limit ($Q = 0$ or $\Theta = \pi/2$) while keeping both the off-extremality and off-Nariai measures fixed. Moreover, the extremal limit ($r_-\to r_+$) corresponds to $\calr \to 1$ whilst holding $y_+$ and $\Theta$ fixed and the Nariai  limit ($r_+\to r_c$) corresponds to $y_+ \to 1$ while keeping $\calr$ and $\Theta$ fixed. This perfect cubic parameter grid is easier to scan (while keeping track of our `distance' to relevant boundaries of the parameter space) than, \eg the $\{M/L,Q/L,a/L\}$ grid displayed in Fig.~\ref{fig:KNdS_parameterspace}.

Onwards, we restrict our attention to neutral ($q=0$) and massless ($\mu=0$) scalar field perturbations. With the aim of explaining the corresponding QNM spectrum of KNdS in depth and to illustrate the analyses we performed `behind the scenes', we comment on various plots in which we keep $y_+$ and $\calr$ fixed and explore how the QNM families change as $\Theta$ varies. 
In general, for a given $\ell$ (which counts the number of zeros of the angular eigenfunction), the equatorial $m = \ell$ modes turn out to be the least damped modes (see \eg~\cite{Berti:2009kk,Berti:2005ys,Davey:2022vyx}), and we expect that the same is true in KNdS. Since the $t-\phi$ symmetry (discussed in Section~\ref{sect:kg_setup}) allows us to restrict considerations to $m \ge 0$, we focus our discussion on the modes with $m = \ell \ge 0$. It is important to retain this fact in future discussions. The qualitative features of the $m = \ell > 0$ QNM spectra are similar for different non-vanishing $\ell$'s but they differ from the $m = \ell = 0$ case. Therefore, in subsection~\ref{sect:m=l=1slice} we describe key properties of the $m = \ell =1 $ QNM spectra as a representative example of the $m = \ell > 0$ class, while in subsection~\ref{sect:m=l=0slice}  we discuss the $m = \ell = 0$ case.

\subsection{QNMs with $m = \ell \neq 0$ and eigenvalue repulsion}\label{sect:m=l=1slice}

\begin{figure}[t]
    \centering
    \includegraphics[width=0.49\textwidth]{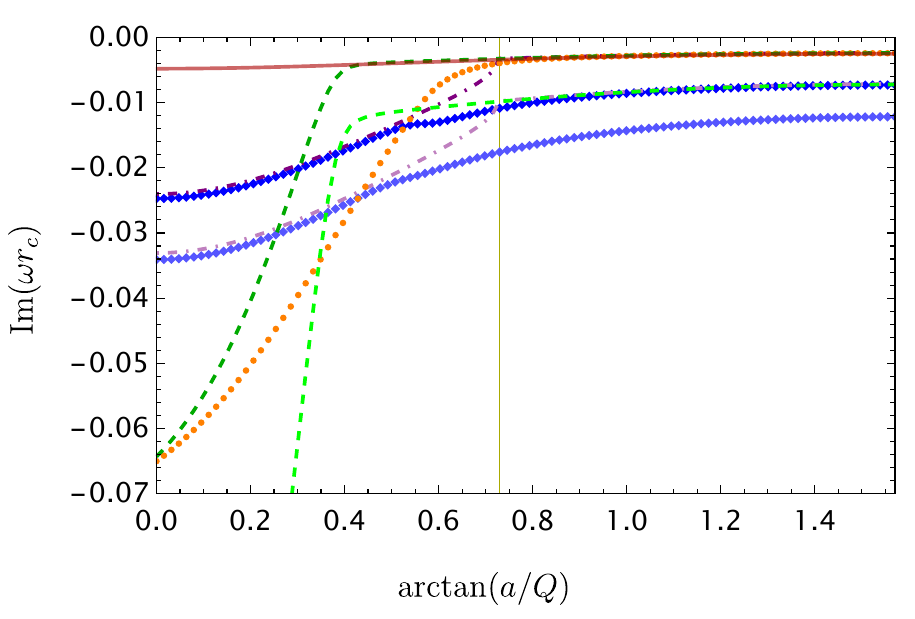}
    \includegraphics[width=0.49\textwidth]{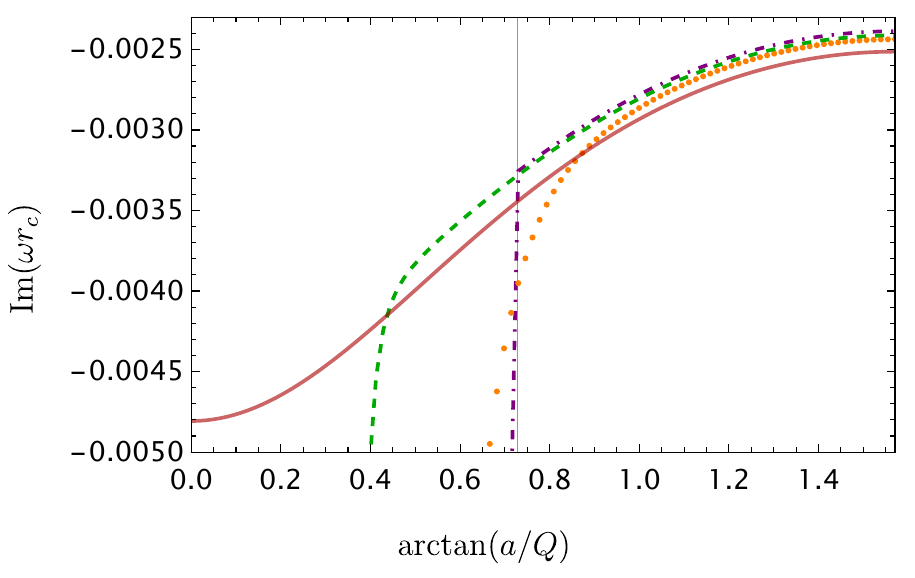}
    \caption{Imaginary part of the frequency (in units of the cosmological horizon length $r_c$) of the $n=0$ NH (blue diamonds), $n=1$ NH (light blue diamonds) and $n=0$ PS (orange disks) modes for $q =\mu = 0$ linear scalar perturbations with $m = \ell = 1$ on KNdS with $\calr = 0.99, y_+ = 1/2$  as a function of $\Theta = \arctan(a/Q)$. The $n = 0$ ($n=1$) MAE approximation $\omega_{\MAE}$ of the NH modes is described by a dot-dashed purple (light purple) curve. The $n = 0$ ($n=1$) eikonal approximation $\omega_{\hbox{\tiny eik}}$ of the PS modes are given by a green (light green) dashed curve. 
    The red curve is the critical curve ${\rm Im}(\omega r_c)=-\beta_c \kappa_-$ with $\beta_c=1/2$ relevant for the SCC discussion, and the $n=0$ PS curve is above it for for $\Theta\gtrsim 0.87$.
    The thin vertical dark yellow line at $\Theta=\Theta_c  \sim 0.73$ describes the $\Theta$ where $\delta^2 = 0$, with $\delta^2<0$ for smaller $\Theta$ values and $\delta^2>0$ for larger $\Theta$. The right panel is a zoom of the left panel, in a range of ${\rm Im}(\omega r_c)$ where we better see the orange PS mode (and green eikonal curve) crossing the red curve.
     \label{fig:PS_NH_modes_ml1_R099_yp05_im}}
\end{figure}

\begin{figure}[ht]
    \centering
    \includegraphics[width=0.48\textwidth]{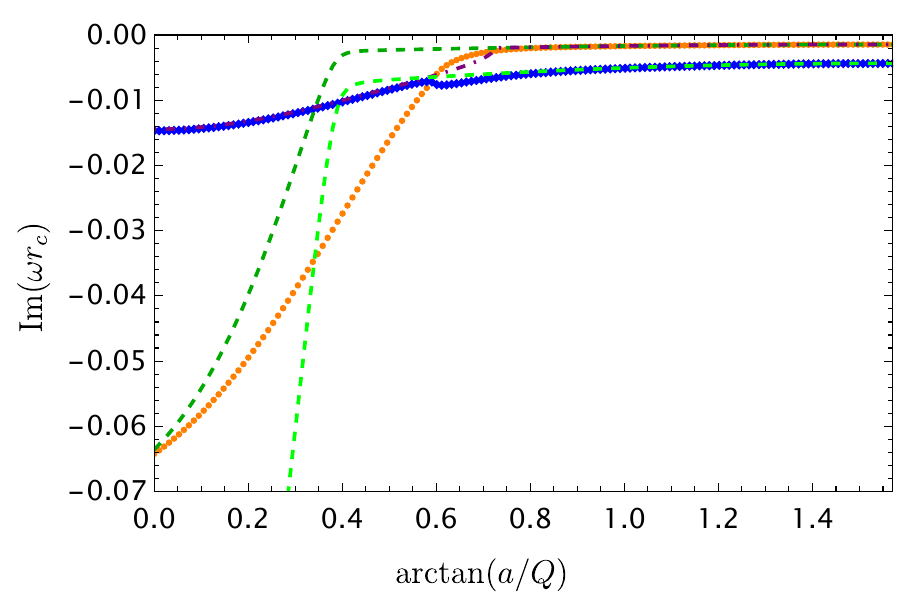}
    \includegraphics[width=0.48\textwidth]{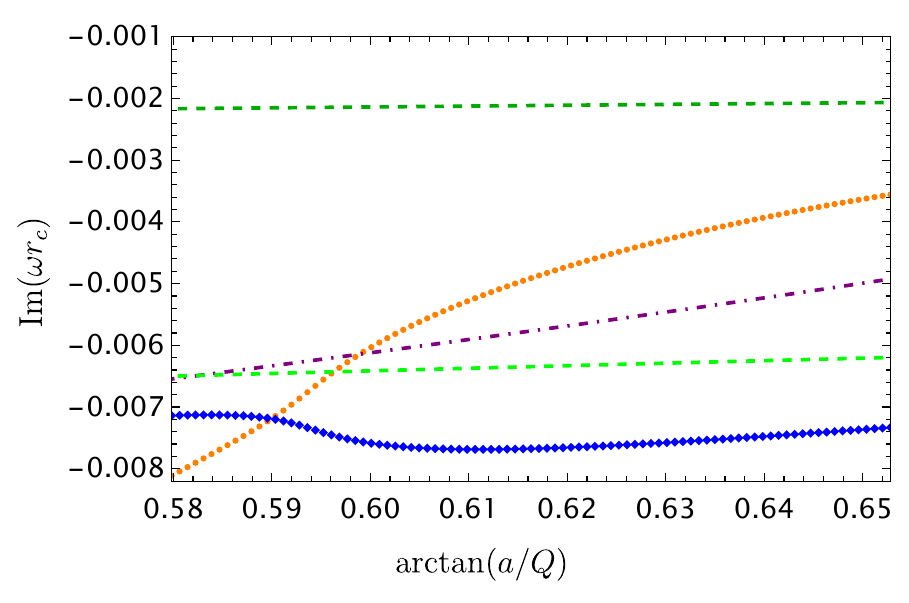}
    \includegraphics[width=0.48\textwidth]{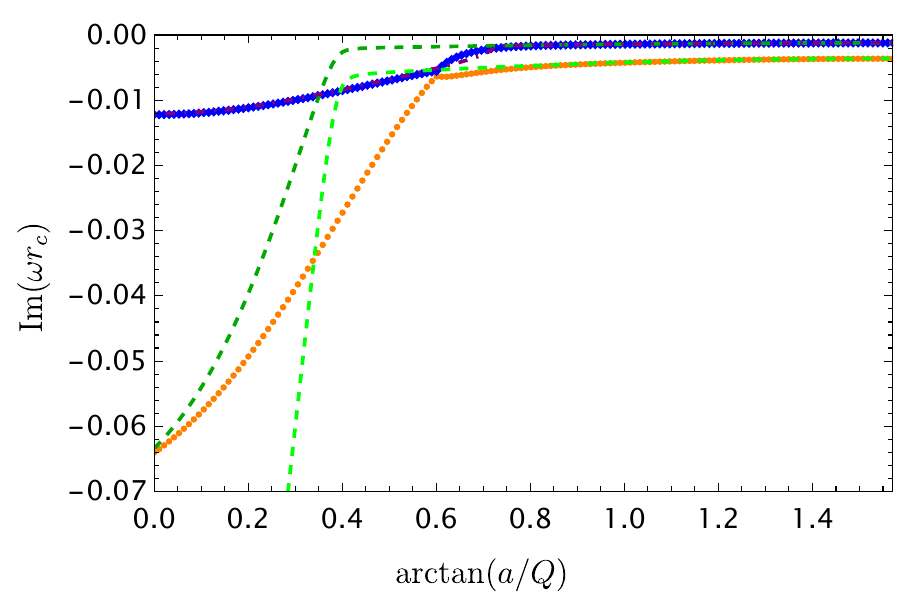}
    \includegraphics[width=0.48\textwidth]{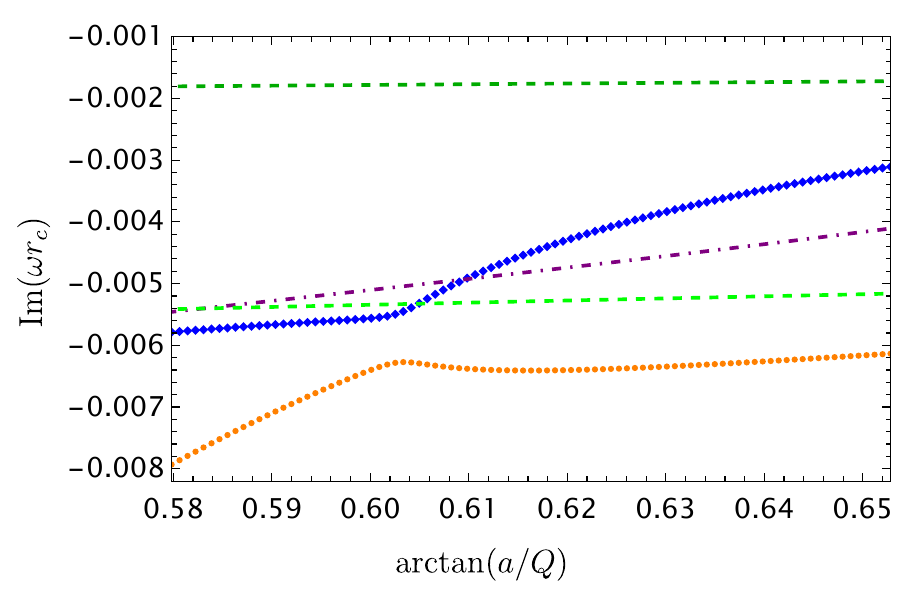}
    \includegraphics[width=0.48\textwidth]{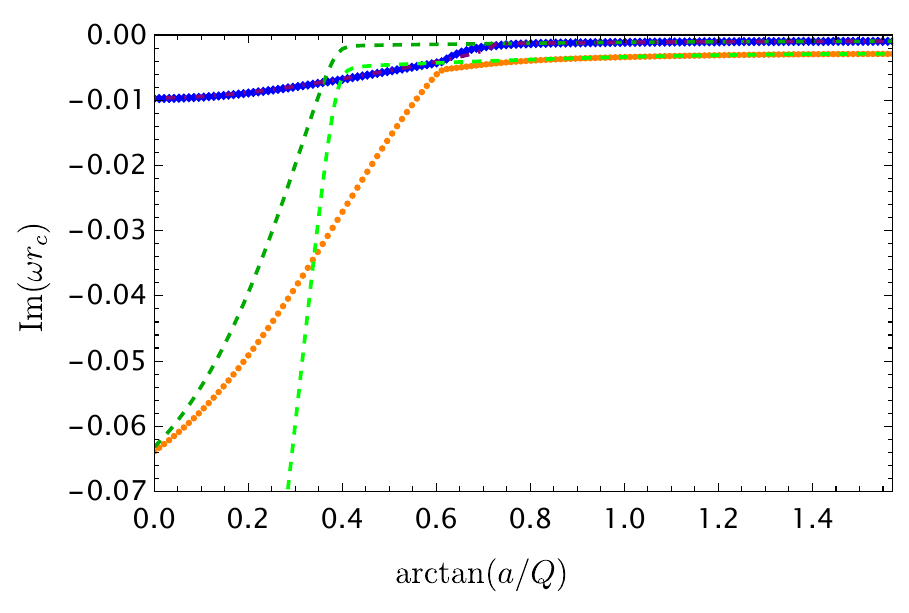}
    \includegraphics[width=0.48\textwidth]{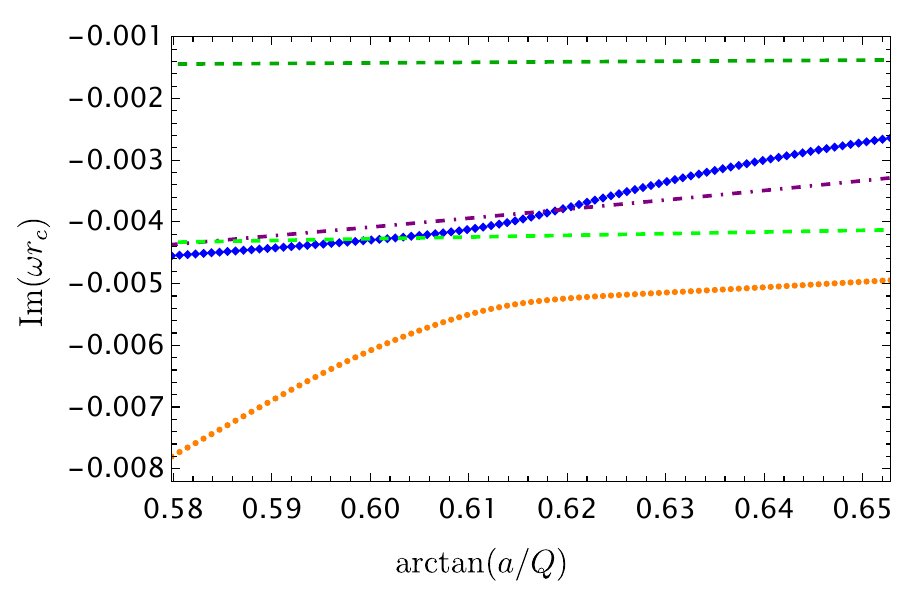}
    \caption{Imaginary part of the frequency for $q =\mu = 0$ linear scalar perturbations with $m = \ell = 1$ on KNdS with $y_+ = 1/2$  as a function of $\Theta = \arctan(a/Q)$ for three different values of $\calr$ (top to bottom).
 \textbf{Left column:} constant $\calr = 0.994$ (top), $\calr = 0.995$ (middle) and $\calr = 0.996$ (bottom). \textbf{Right column:} zoomed in plots near the `intersection' or eigenvalue repulsion regions for constant $\calr = 0.994$ (top), $\calr = 0.995$ (middle) and $\calr = 0.996$ (bottom).
 The colour code is the same as in Fig.~\ref{fig:PS_NH_modes_ml1_R099_yp05_im} but this time we just display the $n=0$ overtone modes.}
 \label{fig:eigenvalue_repulsion}
\end{figure}

We first look at the QNMs with $m = \ell= 1$. In particular, near extremality, where the most interesting and relevant behaviours are. Key properties in this regime are  illustrated in Fig.~\ref{fig:PS_NH_modes_ml1_R099_yp05_im} for fixed $y_+=0.5$ and $\calr = 0.99$ (recall that extremality occurs at $\calr = 1$). In both the RNdS limit ($\Theta=\arctan(a/Q)=0$) and the Kerr-dS limit ($\Theta=\pi/2$), we find agreement with the QNM literature \cite{Cardoso:2017soq, Dias:2018ynt}. In this and future figures, we use the following convention: modes are designated as dS, NH or PS modes according to their classification in the RNdS limit. As we increase the rotation $\Theta > 0$, these modes may not always have a unique classification. This has been discussed in great detail in Kerr-Newman~\cite{Dias:2021yju,Dias:2022oqm,Davey:2023fin} and so we do not elaborate on it further here. Close to the (near-extremal) RNdS limit ($\Theta=0$),  the near-horizon (NH; blue diamonds) modes dominate the spectra.
 At the opposite Kerr-dS limit ($\Theta=\pi/2$), the dominant mode is the photon sphere (PS; orange disks) mode.
 Here and henceforward, stating that a particular mode dominates the QNM spectra means that it is the one with smallest $|{\rm Im}(\omega r_c)|$.  In  Figs.~\ref{fig:PS_NH_modes_ml1_R099_yp05_im}~(and~\ref{fig:eigenvalue_repulsion}) we do not display the third family of QNMs, namely the de Sitter (dS) family, because it is significantly more damped (\ie it has much lower ${\rm Im}(\omega r_c)$) that the other two families in this region of KNdS black holes near extremality. 
 
 In Fig.~\ref{fig:PS_NH_modes_ml1_R099_yp05_im}, we also display the red line that looks very much horizontal (but it is not, \ie it changes slightly with $\Theta$; see right panel). This curve describes the critical SCC curve ${\rm Im}(\omega r_c)=-\beta_c \kappa_-$ with $\beta_c=1/2$ and $\kappa_-$ defined in (\ref{eqn:generatorshorizonsandpotentials}). From the discussion of subsection~\ref{sec:SCCcriterion} it follows that modes above this red curve have $\beta<1/2$ and thus they enforce Christodoulou's formulation of SCC. This is thus the case for fast rotating charged black holes with sufficiently large $\Theta$  in Fig.~\ref{fig:PS_NH_modes_ml1_R099_yp05_im}. Indeed, we see that the least damped orange disk PS curve is above this critical SCC red curve for $\Theta\gtrsim 0.87$ (at $y_+=0.5,\calr = 0.99$) and all the way up to the Kerr-dS limit ($\Theta=\pi/2$).

On the other hand, for smaller values, $\Theta\lesssim 0.87$,  Fig.~\ref{fig:PS_NH_modes_ml1_R099_yp05_im} shows that the least damped mode (be it a PS or NH mode) is now below the critical SCC red curve. This means that for black holes with $\Theta\lesssim 0.87$ (including RNdS with $\Theta=0$) and $y_+=0.5,\calr = 0.99$, the $m = \ell = 1$ modes do not enforce SCC. To determine whether SCC is respected in that region we need to confirm that there is no QNM with another value of $\{m, \ell\}$ which enforces SCC (by having $\beta \le \frac{1}{2}$). Furthermore, to discuss the validity of SCC in the full KNdS parameter space we have to repeat these procedures for all values of $y_+\in (0,1]$, $\calr\in (0,1]$ and $\{m, \ell\}$. We postpone the discussion of the outcome of this survey till section~\ref{sect:SCC_in_KNdS}.\footnote{Of course to find whether the least damped modes have $\beta$ below or above $\beta_c=1/2$ it can be more convenient to simply plot $\beta$ instead of ${\rm Im}(\omega r_c)$ for all the QNM families and see if the dominant one is below or above the critical curve $\beta=\frac{1}{2}$.}

Information relevant to the discussion of interactions between NH and PS families of modes and their classification can also be obtained from 
Fig.~\ref{fig:PS_NH_modes_ml1_R099_yp05_im}. 
In particular, such figure allows us to compare our numerical results with the analytical eikonal ($\omega_{\hbox{\tiny eik}}^+$) and matched asymptotic expansion ($\omega_{\hbox{\tiny MAE}}$) frequencies computed in sections~\ref{sect:PSmodes} and~\ref{sect:NHmodesandNHgeometry} respectively, which helps to identify their family class.

Recall that near extremality the NH modes are expected to be well described by the MAE frequency \eqref{eqn:omega_MAE} (in the present case with $q=\mu=0$): this is the purple dot-dashed line ($n=0$ overtone) and light purple dot-dashed line ($n=1$) in Fig.~\ref{fig:PS_NH_modes_ml1_R099_yp05_im}. On the other hand, the eikonal frequency \eqref{eqn:omega_eik}, which is expected to be valid only in the strict $m=\ell\to\infty$ limit, is represented by the dark green ($n=0$) and light green ($n=1$) dashed lines.
We conclude that in the RNdS limit ($\Theta=0$), the MAE (purple dot-dashed line) and eikonal (green dashed line) approximations are really accurate (even though the eikonal approximation should not be valid for such low $m=\ell=1$!), and we can clearly identify the NH and PS modes based on the MAE and eikonal approximations. 

However, as $\Theta$ increases the NH and PS modes trade dominance (around $\Theta \sim 0.52$). At the value $\Theta=\Theta_c  \sim 0.73$, which corresponds to $\delta^2 = 0$ as defined in (\ref{eqn:delta_definition}), we can observe that the (co-rotating) eikonal ($\omega_{\hbox{\tiny eik}}^+$) and MAE ($\omega_{\hbox{\tiny MAE}}$)  approximations essentially merge in the $n=0$ overtone case (and for slightly higher $\Theta$ the $n=1$ $\omega_{\hbox{\tiny eik}}^+$ and $n=1$ $\omega_{\hbox{\tiny MAE}}$ curves also merge). For $\Theta \gtrsim \Theta_c$ the dominant $n=0$ PS mode (orange disk) curve is essentially on top of both the $n=0$ eikonal and  $n=0$ MAE curves, while the $n=0$ NH mode (blue diamonds) is on top of the $n=1$ eikonal and $n=1$  MAE approximations. This illustrates why we can  precisely classify  modes as PS and NH in the RNdS limit but not necessarily away from it. Additionally, it provides an example of how the 2-dimensional BF bound violation $-$ pinpointed by  $\delta^2 = 0$  $-$  signals a change in the properties of the QNM spectra. Strictly speaking, beyond the rough guide of $\Theta_c$ where the BF bound starts being violated, we  lose the distinction between the $n=0$ PS and $n=0$ NH families and the system is effectively better described by a single PS-NH family with its two first overtones $n=0$ and $n=1$ displayed. We refrain ourselves from a more detailed description of this aspect here: the reader interested on a very exhaustive discussion of these matters is invited to read~\cite{Dias:2021yju,Dias:2022oqm,Davey:2023fin} which describe a similar system in the Kerr-Newman case (the quantitative values are of course different for $\Lambda=0$ but the properties are otherwise qualitatively similar) and~\cite{Davey:2022vyx} for rotating dS black holes.

Moving on to Fig.~\ref{fig:eigenvalue_repulsion}, the series of plots in this figure illustrate another important characteristic of the $m = \ell= 1$, $n=0$ QNM spectra of KNdS. 
Namely, we can observe how the interaction between the PS and NH families can become even more involved as we approach extremality further, $\calr \to 1$. As it occurs in KN spacetime~\cite{Dias:2021yju,Dias:2022oqm,Davey:2023fin} and dS Myers-Perry~\cite{Davey:2022vyx}, we often observe the phenomenon of \textit{eigenvalue repulsion} when scanning the QNM spectra throughout the KNdS parameter space (typically when approaching extremality). In Fig.~\ref{fig:eigenvalue_repulsion}, all plots describe KNdS black holes with $y_+ = 0.5$. From the top to the bottom we then increase the off-extremality parameter $\calr$. Concretely, the top row is for  $\calr = 0.994$, the middle row is for $\calr = 0.995$ and the bottom row is for $\calr = 0.996$. The plots on the right column are simply a zoom-in of the left column plots in regions of interest.

For $\calr = 0.994$ (top row), the PS (orange disks) and NH (blue diamonds) families `intersect'\footnote{\label{foot:intersection}Every single time we state that two QNM curves `intersect', we mean that the imaginary part of their frequencies, ${\rm Im}(\omega r_c)$,  does intersect but, it is important to emphasize, the curves describing the real part of the frequencies (not shown) do {\it not} intersect~\cite{Dias:2021yju,Dias:2022oqm,Davey:2022vyx,Davey:2023fin}. To always keep this property in mind, we will put quotation marks whenever we use the word intersect.} near $\Theta \sim 0.6$, trading dominance for higher $\Theta$ values. The top right plot is a zoom of the top left plot around the $\Theta \sim 0.6$ where the `intersection' occurs. Increasing the $\calr$ value by a small amount  to $\calr = 0.995$ (middle row), changes the `intersection'  into an {\it eigenvalue repulsion} between the PS and NH modes. Instead of `crossing past' each other, now the PS and NH curve `repel' each other around $\Theta \sim 0.6$ and no longer trade dominance  for higher $\Theta$ values. Finally, even closer to extremality, for $\calr = 0.996$ (bottom row) we can see that the repulsion between the PS and NH $n=0$ modes is not so sharp as before but, again, the two families do not `intersect' (like for the middle plot). 

We emphasize again that although we have only discussed key aspects of $m = \ell  =1$ modes (illustrating the challenges in finding them and in identifying the least damped mode, a.k.a. spectral gap), the discussion in this subsection applies qualitatively for all $m = \ell  > 0$ modes. It is also in qualitative agreement with previous findings in Kerr-Newman or Kerr-dS black holes~\cite{Dias:2021yju,Dias:2022oqm,Davey:2022vyx,Davey:2023fin} which justifies not presenting an exhaustive list of plots. 
Also note that the existence of eigenvalue repulsions are not reported in detail in the study~\cite{Casals:2020uxa} of the QNM spectra of the KNdS family with constant $\Lambda M^2$ (although a reference to their existence seems to be mentioned in footnote 3 of~\cite{Casals:2020uxa}). Eigenvalue repulsions like the ones illustrated in Fig.~\ref{fig:eigenvalue_repulsion} are typically observed if one approaches extremality sufficiently close and if $\Theta$ is larger than $\sim \Theta_c$.    

We now comment on the special case of $m = \ell= 0$ which is qualitatively different from the $m = \ell  > 0$ case. 

\subsection{QNMs with  $m = \ell= 0$ and eigenvalue splitting}\label{sect:m=l=0slice}

\begin{figure}[t]
    \centering
    \includegraphics[width=0.48\textwidth]{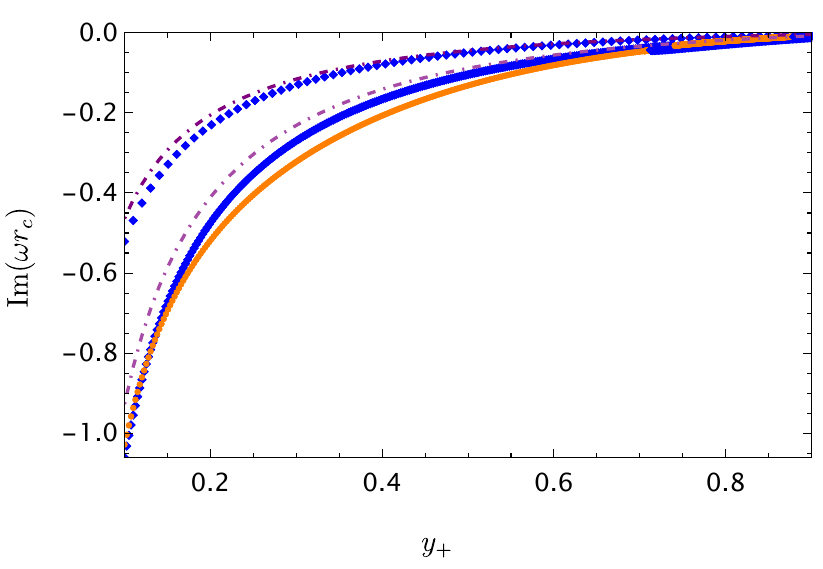}
    \includegraphics[width=0.48\textwidth]{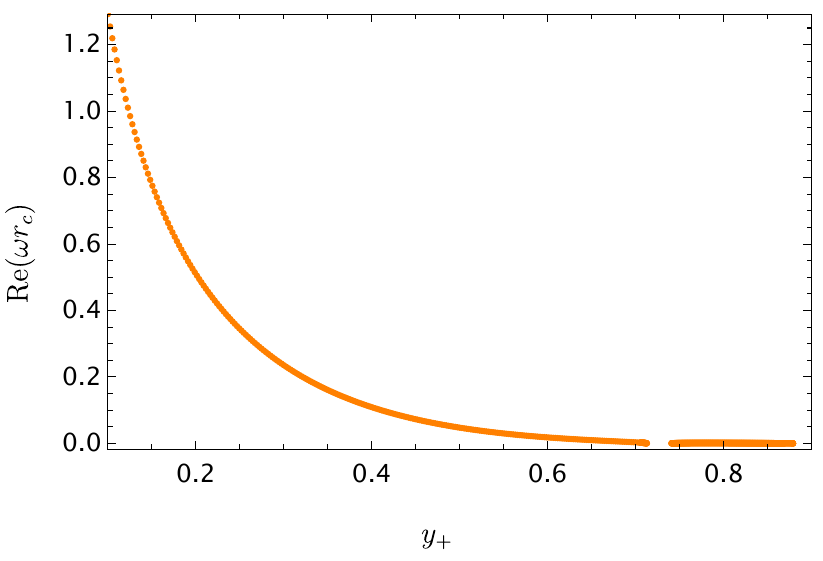}
    \includegraphics[width=0.48\textwidth]{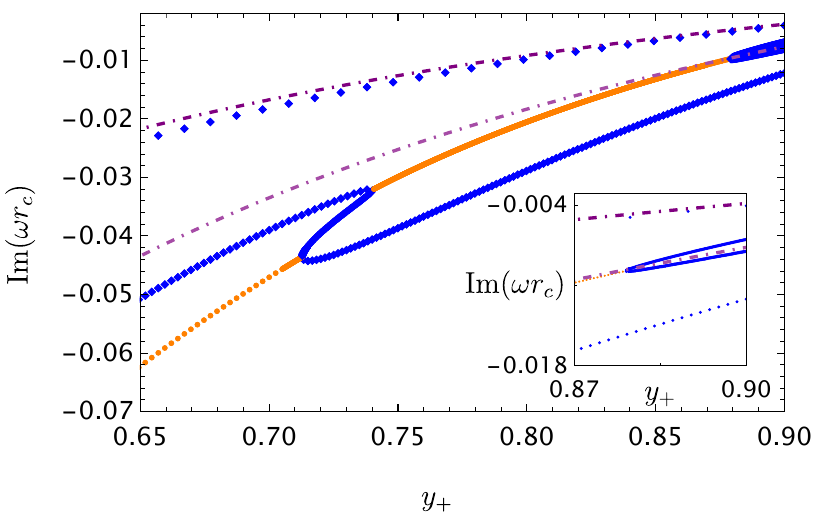}
    \includegraphics[width=0.48\textwidth]{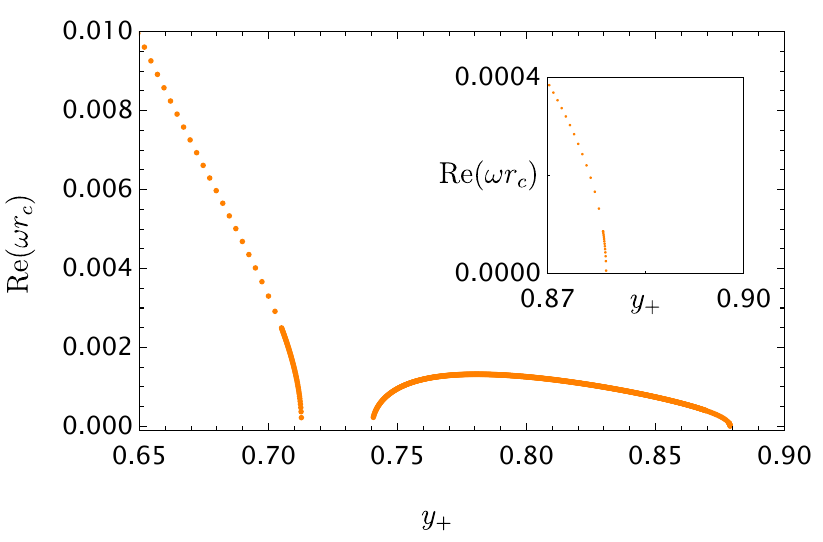}
    \caption{Imaginary (left panels) and real (right panels) parts of the frequency as a function of the off-Nariai parameter $y_+$ for ($q =\mu = 0$) $m=\ell=0$ QNMs in RNdS ($\Theta=0$) with fixed off-extremality parameter $\calr = 0.95$.
The orange disks represent $n=0$ PS modes with complex frequencies and we also display the first two purely imaginary dominant modes with blue diamonds (since they have ${\rm Re}(\omega r_c)=0$ we simply do not display them in the right panels for the sake of clarity). The purely imaginary modes are initially closely followed by the MAE frequencies with $n=0$  (purple dot-dashed curve) and $n=1$ (lighter dot-dashed lines). The second row gives zoom plots of the top panel in the regions where interesting splittings and/or mergers happen. Here, the inset plots are further zooms around the region $y_+ \in [0.87, 0.9]$.}
\label{fig:m0l0RNdSR095slice}
\end{figure}

 In this section we look at the $m = \ell= 0$ sector of QNMs, as the spectrum presents a distinct behaviour when compared to that of $m = \ell  \neq 0$ (and they also have to be considered in the SCC discussion of section~\ref{sect:SCC_in_KNdS}). To start with, the  $m = \ell= 0$  spectrum is never dominated by PS modes. Nevertheless, the $m = \ell= 0$ QNM spectrum can also be very intricate although for distinct reasons  from the $m=\ell\neq 0$ case. This intricacy in the KNdS spectrum can be understood to emerge from non-trivial features already present in the spectrum of $m=\ell= 0$ RNdS black holes ($\Theta=0$).
For this reason, we first look into the the QNM spectrum of RNdS. 
Some relevant features are visible in Fig.~\ref{fig:m0l0RNdSR095slice}, which uses RNdS black holes with $\calr=0.95$ to illustrate common properties to other elements of the RNdS family. We plot both the imaginary (left column) and real (right column) part of the frequencies as a function of $y_+$. For the sake of clarity, we do not represent the purely imaginary modes (blue diamonds) in the real part plots. The bottom plots are simply a zoom in of the top plot in a region with non-trivial features.
We plot two families of modes: the orange disks are the PS modes (with complex frequencies)\footnote{We have checked that this family is continuously connected to the $m = \ell = 1$ PS family by performing the (unphysical) marching in non-integer $m$ from $m=0$ till $m=1$. Note that the eikonal approximation is certainly not expected to hold for $m=\ell=0$.} while the blue diamonds, in this figure only, represent purely imaginary modes (\ie with ${\rm Re}(\omega r_c)=0)$ in RNdS (they acquire a real part when $\Theta>0$). The intricate interactions within the QNM spectrum only allow us to specify the distinction between NH and dS modes in some specific cases as discussed next.

In Fig.~\ref{fig:m0l0RNdSR095slice} we start by noticing that the two least damped modes for  RNdS with $y_+\lesssim 0.71$ are two blue diamond curves with purely imaginary frequencies. Both are well described by the purple and light-purple dot-dashed curves given by the $n=0,1$ MAE approximation  \eqref{eqn:omega_MAE}. Hence we could be naively tempted to classify both of them as NH modes. This is indeed the case for the $n=0$ family (upper curve).  However, if we take the $n=1$ blue diamond with  $y_+ = 0.1$ and $\calr = 0.95$ in Fig.~\ref{fig:m0l0RNdSR095slice} and march the $\calr$ parameter down to $\calr=0.1$, we conclude that here this mode is undoubtedly a dS QNM at  (discussed in section \ref{sect:dSmodes}). This fact highlights the important remark above: given a QNM with $m = \ell= 0$ in an arbitrary point of the KNdS parameter space, we typically cannot claim whether it belongs to the NH, PS or even the dS families. We will refrain from adding further discussions about intricate features that are not relevant for the ultimate aim of finding the least damped mode.

Next, starting at $y_+=0.1$ in the left panels of Fig.~\ref{fig:m0l0RNdSR095slice}, we follow the orange disk PS family (with $n=0$) to larger values of $y_+$. We see this is a single complex frequency family of modes till $y_+\sim 0.71$. But here an interesting phenomenon occurs: the $n=0$ PS families {\it splits} into two {\it purely imaginary} frequency families of modes (see zoom in plot around the splitting region in the left-bottom plot). Then, for $y_+\gtrsim 0.71$ the lower blue diamond branch remains purely imaginary until $y_+ \simeq 0.925$ (not shown, it has a splitting at this value of $y_+$), while the upper blue diamond branch merges, at $y_+ \sim 0.735$, with the purely imaginary blue curve followed closely by the $n = 1$ MAE curve  (that is above it for $y_+\lesssim 0.735$). The merged blue diamond curves with purely imaginary frequencies then continue for larger $y_+$ ($\gtrsim 0.735$) as a orange disk branch with complex frequency (which itself splits then around $y_+\sim 0.88$ into two families of purely imaginary modes as best seen in the inset plot of the bottom-left panel; not shown, they both continue as purely imaginary curves till at least $y_+=0.999$).  Notice that when a complex family splits into two purely imaginary families or two purely imaginary families merge into a complex one,  the number of degrees of freedom of the system are conserved. For a detailed discussion of this type of behaviour, we refer the reader to~\cite{Davey:2022vyx, Dias:2021yju,Dias:2022oqm}.

Given the intricate features already observed in the RNdS QNM spectra, it is with no surprise that similar, but much enhanced, intricate features are observed when we move to the KNdS black holes with $\Theta>0$. We navigated through these carefully having in mind that our mission was to always capture the least damped QNM with $m=\ell=0$ for each $(y_+,\calr,\Theta)$ point in the parameter space of KNdS.     

\begin{figure}[t]
    \centering 
    \includegraphics[width=0.495\textwidth]{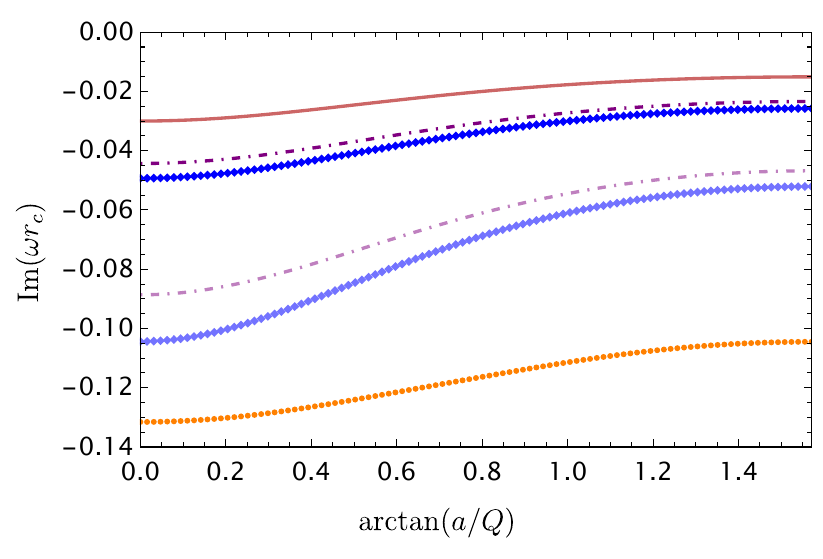}
    \includegraphics[width=0.495\textwidth]{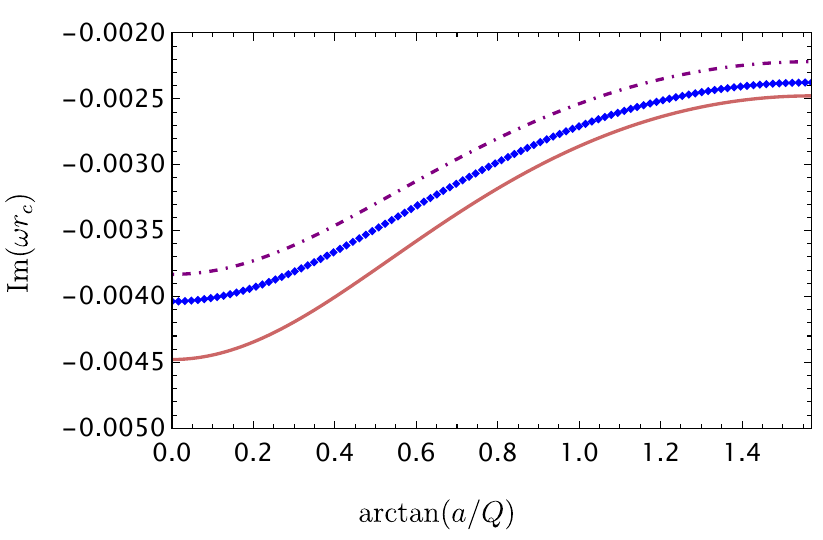}
    \caption{Imaginary part of the frequency as a function of $\Theta = \arctan(a/Q)$ for $q =\mu = 0$ linear scalar perturbations with $m = \ell = 0$ on KNdS with $\calr = 0.95$, with two different values of $y_+$: $y_+=0.5$ (left plot) and $y_+=0.9$ (right plot). \textbf{Left plot}: for $y_+ = 1/2$,  
    we show the $n=0$ NH (blue diamonds), $n=1$ NH (lighter blue diamonds) and $n=0$ PS (orange disks) modes. The $n = 0$ ($n=1$) MAE approximation $\omega_{\MAE}$ to the NH modes is given by a dot-dashed purple (lighter purple) curve. The red curve is the critical curve ${\rm Im}(\omega r_c)=-\beta_c \kappa_-$ with $\beta_c=1/2$ relevant for the SCC discussion. \textbf{Right plot}: for $y_+ = 0.9$ and with the same colour coding, we show that the dominant (blue diamond) NH mode enforces SCC for all $\Theta$ since its curve is above the critical red curve ${\rm Im}(\omega r_c)=-\frac{1}{2} \kappa_-\,$.
    } 
    \label{fig:PS_NH_ml0_R095_yp05_im}
    \end{figure}

As illustrated above for $\Theta=0$, although the QNM spectra of KNdS in the $m=\ell=0$ sector can be very involved, it is also true that in many regions of the KNdS space it can look relatively simple. This is the case for  the $y_+ = 0.5, \calr = 0.95$ KNdS displayed on the left plot of Fig.~\ref{fig:PS_NH_ml0_R095_yp05_im}.
To obtain this plot and its colour code, we have continuously traced back each of the modes to their corresponding family in RNdS ($\Theta=0$), as for $y_+ = 0.5$ and $\calr = 0.95$ there is no splitting or merging of the three dominant modes (of the type observed in Fig.~\ref{fig:m0l0RNdSR095slice}). Hence, the QNM classification is unambiguous in this particular case. On the other hand, in the right plot of Fig.~\ref{fig:PS_NH_ml0_R095_yp05_im} which is still for KNdS with $\calr = 0.95$ but this time with $y_+ = 0.9$, we only display the dominant NH mode and the corresponding $n=0$ MAE approximation, as the subdominant QNM spectrum is quite complicated. That this is the case can be understood looking again to the RNdS spectrum of Fig.~\ref{fig:m0l0RNdSR095slice} around $y_+ = 0.9$: before this $y_+$, several mergers and splittings have already occurred.

In both cases of Fig.~\ref{fig:PS_NH_ml0_R095_yp05_im}, despite being relatively far from extremality, we see that the NH modes (blue diamonds) dominate the QNM spectra, \ie they have the lowest $|{\rm Im}(\omega r_c)|$ for any $\Theta$. This is in sharp contrast with the $m = \ell = 1$  case  (c.f. Fig.~\ref{fig:PS_NH_modes_ml1_R099_yp05_im}) where away from extremality the PS mode becomes the dominant mode above a critical value of $\Theta$. In Fig.~\ref{fig:PS_NH_ml0_R095_yp05_im} we also display the red line that looks very much horizontal (but it changes slightly with $\Theta$). It describes the critical SCC curve ${\rm Im}(\omega r_c)=-\beta_c \kappa_-$ with $\beta_c=1/2$. From the discussion of subsection~\ref{sec:SCCcriterion} it follows that modes above this red curve have $\beta<1/2$ and thus they enforce SCC. The right plot of Fig.~\ref{fig:PS_NH_ml0_R095_yp05_im}  with $y_+ = 0.9$ provides an example where the dominant NH mode enforces the SCC for all  $\Theta$ values. Notice however, that in the left plot of Fig.~\ref{fig:PS_NH_ml0_R095_yp05_im} with $y_+ = 0.5$, even though the NH family dominates the spectrum, it no longer enforces SCC. These examples highlight that NH modes are particularly important in the special sector of $m = \ell = 0$ perturbations. Moreover, they also illustrate that $m = \ell = 0$ modes also have to be considered  in the SCC discussion of section~\ref{sect:SCC_in_KNdS}. In this context, in Appendix~\ref{sect:MAEvalueforvanishingm} we perform an analytic study of the NH modes at the specific limit of extremal ($\calr = 1$) KNdS with $m = \ell = 0$, and we rule out the possibility of NH modes enforcing SCC in this very precise slice.

After this survey of the properties of the $m=\ell=0$ QNMs of KNdS, without further delay we can state the main conclusion of our analysis in this subsection. We did not find any mode  with $m=\ell=0$ crossing the $\beta = 1/2$ threshold at a point of 3-dimensional parameter space closer to RNdS (\ie with smaller dimensionless rotation for a given dimensionless charge) than where the eikonal approximation for $m=\ell\to \infty$ will ultimately enforce SCC (see next section~\ref{sect:SCC_in_KNdS}). Therefore, after accounting for all sectors of perturbations and with the caveat of considering only $y_+\geq 0.1$,  the eikonal approximation boundary with $\beta=1/2$ that we will present in the next section will turn out to be the one that signals the transition boundary between regions in the KNdS parameter space that violate SCC (including near-extremal RNdS) and regions that respect it (including Kerr-dS). 

\section{Strong Cosmic Censorship in KNdS}\label{sect:SCC_in_KNdS}

Recalling the discussion of section~\ref{sec:SCCcriterion}, if for a given KNdS black hole we can find a QNM with $\beta \equiv -\text{Im}(\omega)/\kappa_- < 1/2$ then the scalar field we are probing cannot be extended across the Cauchy horizon as a weak solution of the system and so Christodoulou's formulation of SCC  is respected. We just needs one such QNM since generic initial data will contain it. On the other hand, if for a particular KNdS solution all its QNMs have $\beta  > 1/2$ then Christodoulou's formulation of SCC is violated.

In the previous two sections we have outlined the strategy and some challenges we had to face to scan the full 3-dimensional KNdS parameter space to determine the least damped QNM frequency (a.k.a. spectral gap) for each KNdS black hole, and thus compute $\beta \equiv -\text{Im}(\omega)/\kappa_-$. In this section we present our conclusions and, for each KNdS black hole, we compare $\beta$  with the critical value of $1/2$ to determine if Christodoulou's formulation of SCC does or does not hold for such a black hole.

Our general strategy to address the problem is the following. The eikonal, $m=\ell\to \infty$,  approximation~\eqref{eqn:omega_eik} of the PS modes has $\beta_{\rm eik} < 1/2$ for a large portion of the parameter space of KNdS. In such cases, we can state that $-$ if and only if we confirm this to be indeed a good approximation $-$ in these regions there is at least one QNM with $\beta<1/2$ that enforces Christodoulou's formulation of SCC (since generic initial data will include  $m=\ell\gg 1$ modes). Actually, this is precisely the family of modes that enforces SCC in the whole parameter space of the Kerr-dS limit of KNdS~\cite{Dias:2018ynt}. So this indeed seems to be our best starting point. However, to make such a claim we first need to verify that the  eikonal frequency~\eqref{eqn:omega_eik} indeed provides a good approximation for large $m=\ell$ modes. For this reason, we first verify that the exact numerical PS modes with $m = \ell = 10$ (the highest value we considered) agree well with the eikonal approximation~\eqref{eqn:omega_eik}, $\omega_{\hbox{\tiny eik}}$, and also establish the region of the KNdS parameter space where these numerical PS modes (and the eikonal approximation) enforce SCC. Then, in regions of the parameter space where SCC is not enforced by the PS modes, we need to check if there is some other family of QNMs with $\beta<1/2$.
 
\begin{figure}[t]
    \centering   
    \includegraphics[width=0.495\textwidth]{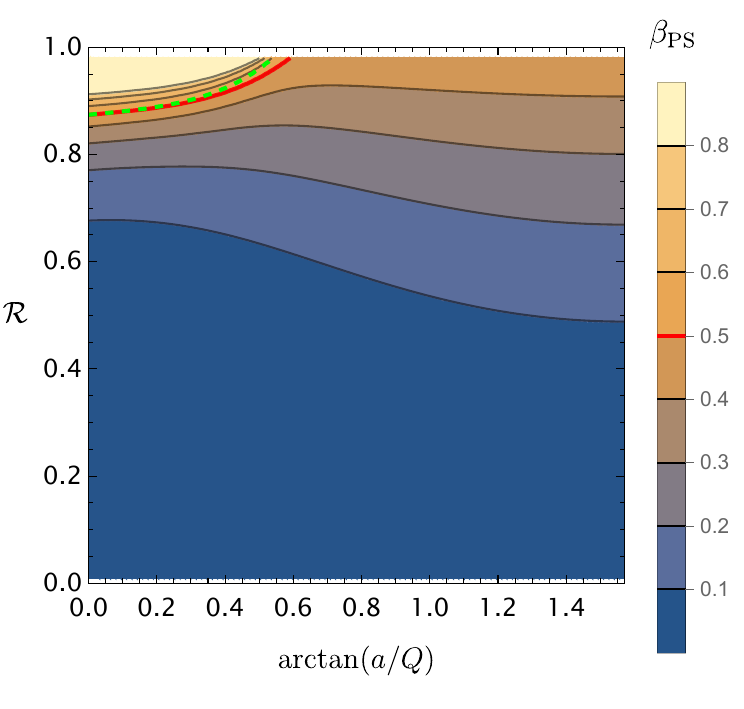}
    \includegraphics[width=0.495\textwidth]{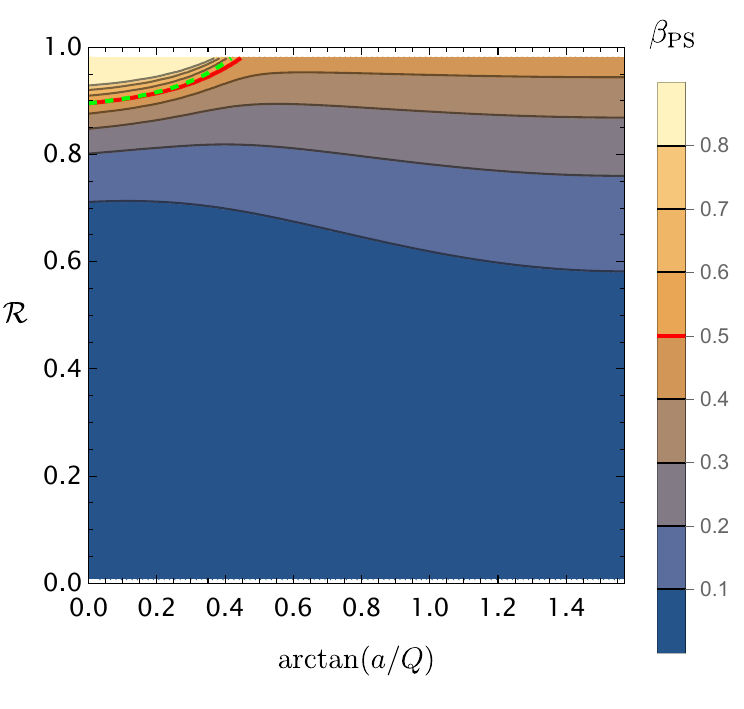}
    \includegraphics[width=0.495\textwidth]{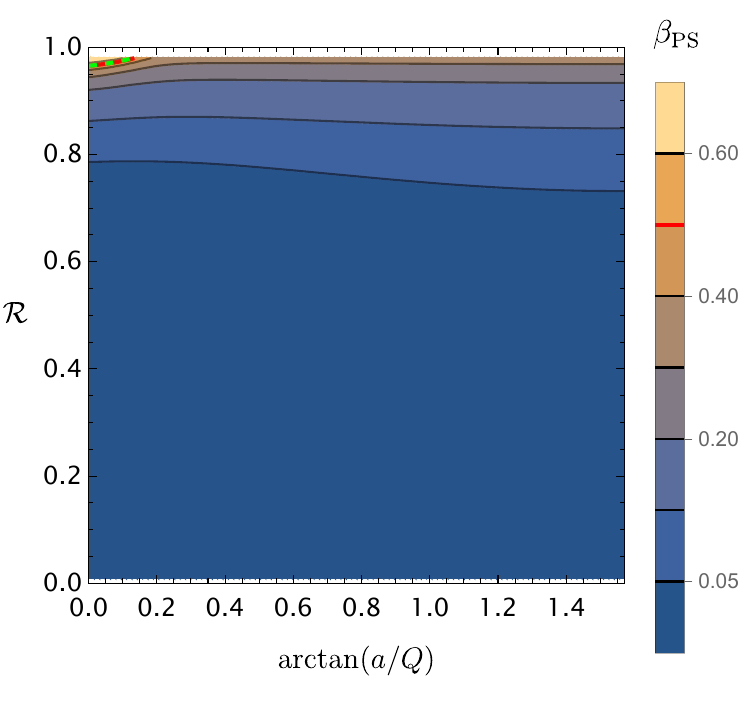}
    \caption{Contour plots of $\beta_{\rm PS}$ for exact numerical $m = \ell = 10$ PS modes as a function of   $\{\calr, \Theta=\arctan(a/Q)\}$ for KNdS families with fixed $y_+ = 0.1$ (top-left panel), $y_+ = 0.5$ (top-right panel) and $y_+ = 0.9$ (bottom panel). In each plot, the critical curve with  $\beta_{\rm PS} = \frac{1}{2}$ is  the red curve (around the top left corner), and the dashed green line identifies the eikonal prediction $\beta_{\rm eik}=\frac{1}{2}$.}\label{fig:eikonal_verification}  
\end{figure}

\begin{figure}[t]
    \centering
     \includegraphics[width=0.52\textwidth]{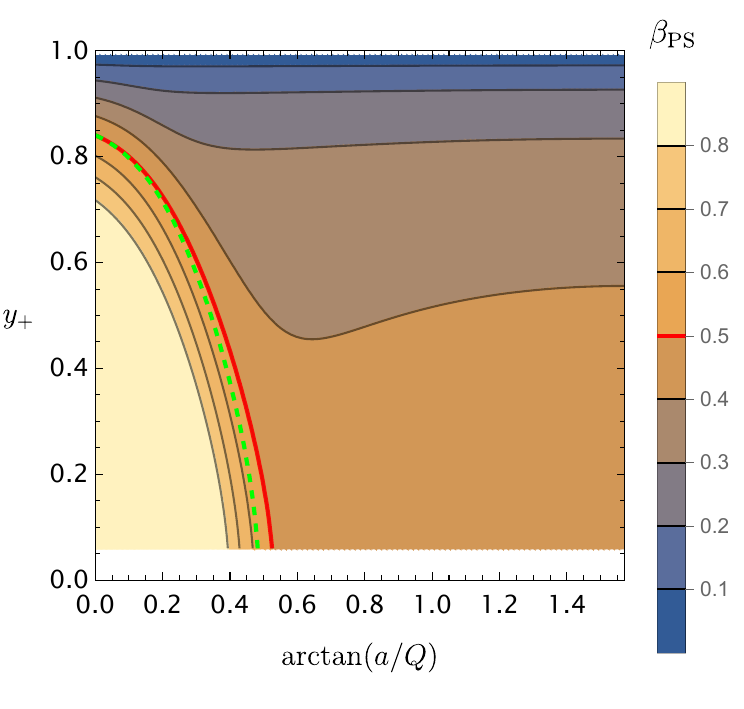}
    \includegraphics[width=0.47\textwidth]{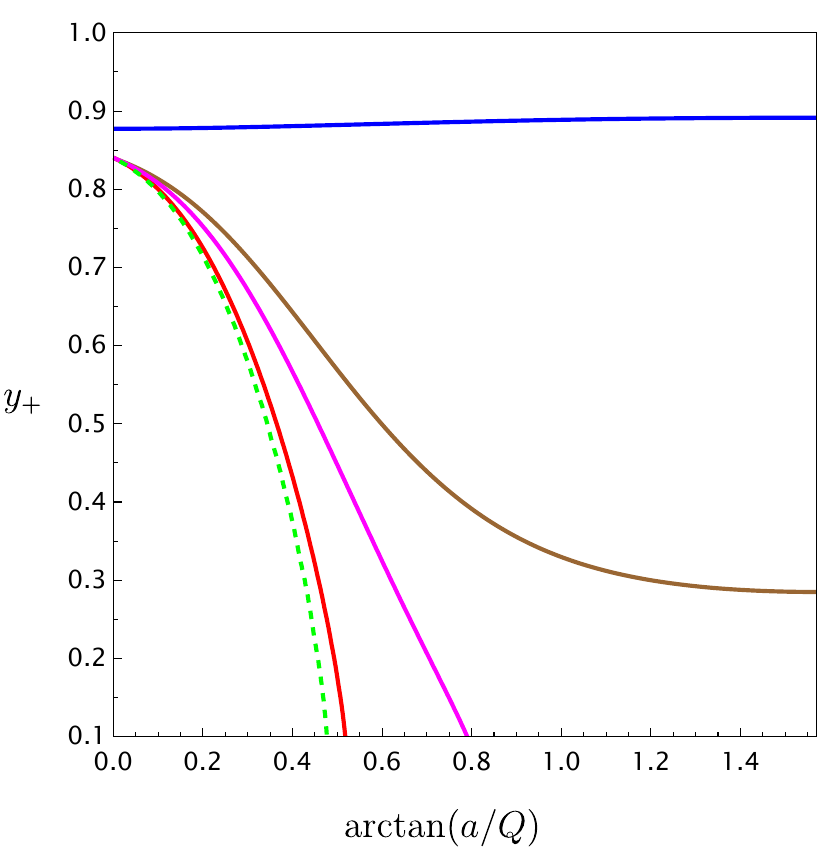}
    \caption{{\bf Left panel:}  Density plot of $\beta_{\rm PS}$ for  PS modes with $m = \ell = 10$ as a function of $\Theta$ and $y_+$ for a KNdS slice of constant $\calr = 0.95$, \ie a near-extremal KNdS family.
    {\bf Right panel:} Critical $\beta= 1/2$  curves at $\calr = 0.95$ for different values of $m = \ell $. Left to right, we have the $\beta_{\rm eik}= 1/2$ critical boundary in the eikonal approximation (green dashed line), followed by the numerical $\beta_{\rm PS} = 1/2$ curve for $m = \ell= 10$ (red solid line), $m = \ell= 2$ (magenta solid line), $m = \ell = 1$ (brown solid line) and finally the $\beta_{\rm NH} = 1/2$ curve for $m = \ell= 0$ (blue solid line) modes. As we increase the value of $m = \ell $, we observe that the $\beta_{\rm PS}  = 1/2$ boundary curve for $m=\ell$ modes increasingly converges to the $\beta_{\rm eik}= 1/2$ curve of the eikonal approximation, which corresponds to $m = \ell \to \infty$.}
    \label{fig:B0.5contoursform0m2m10andeikonal}
\end{figure}

We scan the entire KNdS parameter space using the `{\it spherical polar parametrization}' $\{y_+,\calr,\Theta\}$ of KNdS introduced in \eqref{PolarCoord}. Recall that the range of these parameters is $y_+ \in (0, 1]$, $\calr \in (0, 1]$ and $\Theta \in [0, \frac{\pi}{2}]$. We typically use a numerical grid of $100$ points along each axis to do our scans  
 (with additional points near the extremal and Schwarzschild-dS limits, or whenever necessary). A first outcome of our analysis is that, in general, the numerical PS modes are in (very) good agreement with the eikonal approximation, even for angular harmonic quantum numbers as low as $m = \ell = 10$ (it becomes increasingly harder to get numerical modes with higher $m$). We also find that the $m=\ell$ numerical data approaches the eikonal frequencies, as it should, as $m=\ell$ increases. These two properties are illustrated in Figs.~\ref{fig:eikonal_verification} and~\ref{fig:B0.5contoursform0m2m10andeikonal}.

In Fig.~\ref{fig:eikonal_verification} we give the contour or density plots of $\beta$ for $m = \ell =10$ PS modes that we obtain numerically (onwards, let us call it $\beta_{\rm PS}$) as a function of $\Theta$ and $\calr$ for three different values of $y_+$, namely $y_+ = 0.1$ (top-left panel), $y_+ = 0.5$ (top-right panel) and $y_+ = 0.9$ (bottom panel). The darker (blue) regions describe small values of $\beta_{\rm PS}$ while the lighter (yellow) regions describe large $\beta_{\rm PS}$ values. In the these plots, the red line (around the top left corner) describes solutions where we have $\beta_{\rm PS} = 1/2$ (for $m=\ell=10$). On the other hand, the dashed green line describes the eikonal $\beta_{\rm eik} \equiv -\text{Im}(\omega_{\hbox{\tiny eik}})/\kappa_- = 1/2$ curve with $\omega_{\hbox{\tiny eik}}$ given by~\eqref{eqn:omega_eik}. We see that the red $m = \ell =10$ critical curve is always below  the  $m = \ell \to \infty$ dashed curve but they are quite close to each other (with the approximation being typically better for larger $y_+$ and smaller $\Theta$ and $\calr$).  These series of plots in Fig.~\ref{fig:eikonal_verification} for $y_+=0.1,0.5,0.9$ also show that the SCC conclusions can depend significantly on  $y_+ \equiv r_+/r_c$. For example, in the near-Nariai limit $y_+ \sim 1$ the PS modes enforce SCC almost for all KNdS independently of their $\calr$ and $\Theta$ (bottom plot). For smaller $y_+$ (top plots) one finds that $\beta>1/2$ in a relatively wide region `centred' around $(\Theta=0,\calr=1)$. In particular, this means that PS modes do not enforce Christodoulou's formulation of SCC for extremal RNdS as is well established~\cite{Cardoso:2017soq} (neither for weakly rotating KNdS near-extremality).

Moving now to the left panel of Fig.~\ref{fig:B0.5contoursform0m2m10andeikonal}, this plot complements the understanding gathered in the previous figure. Indeed, this time we instead fix our attention on a KNdS family with constant $\calr=0.95$ (so near-extremality) and we provide the $\beta_{\rm PS}$  contour plot for $m=\ell=10$ PS modes as a function of $\Theta$ and $y_+$. Again we identify the $\beta_{\rm PS}=1/2$ boundary red curve of $m=\ell=10$ modes (left and right panels) and we see that it approaches from the right the eikonal $\beta_{\rm eik}=1/2$ green-dashed boundary. In the left plot of Fig.~\ref{fig:B0.5contoursform0m2m10andeikonal} we further see that, to the right of the $\beta_{\rm PS}=1/2$ red curve, one has  $\beta_{\rm PS}<1/2$ which means that SCC holds for KNdS black holes in that region. This includes the Kerr-dS family ($\Theta=\pi/2$), as is well established~\cite{Dias:2018ynt},  but also weakly (or moderately weakly) charged KNdS black holes.

\begin{figure}[t]
    \centering
    \includegraphics[width=0.54\textwidth]{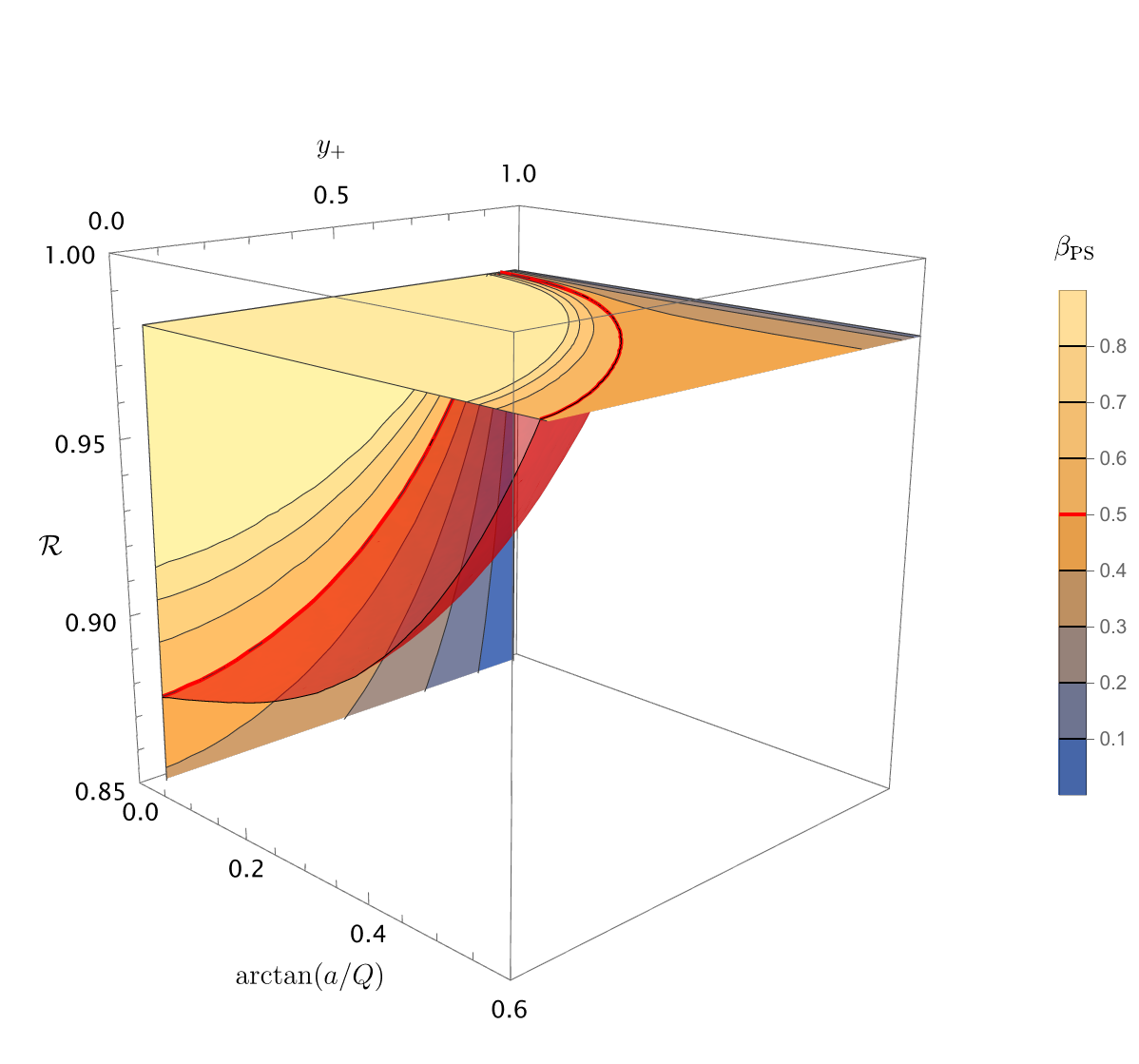}
  \includegraphics[width=0.45\textwidth]{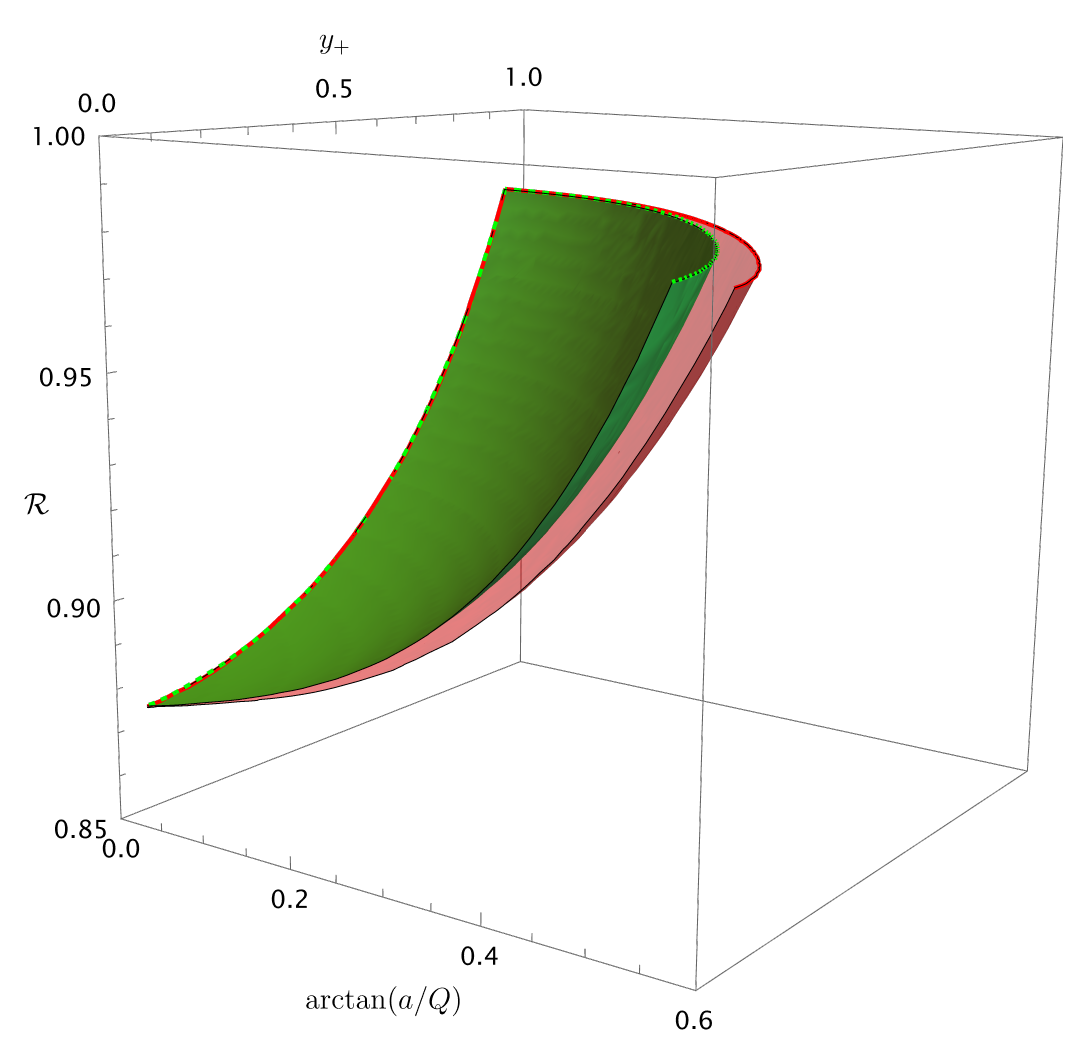}
    \caption{\textbf{Left panel:} Density plot of $\beta_{\rm PS}$ for exact numerical PS modes with $m=\ell=10$ across the  three-dimensional KNdS parameter space $\{y_+, \calr, \Theta=\arctan(a/Q)\}$ in the critical region where  $\beta_{\rm PS}$ crosses the critical value of $1/2$ (red `spherical-like' surface) for $m=\ell=10$. One has $\beta_{\rm PS}>1/2$ in the interior of this red surface. We display contour plots in the vertical plane with $\Theta = 0$ (RNdS black holes) and in the horizontal plane with $\calr = 0.98$ (very close to extremal KNdS).
    \textbf{Right panel:}  Same parameter space region and critical red surface with $\beta_{\rm PS}=1/2$ of $m=\ell=10$ as in the left panel but this time we also display the critical eikonal ($m=\ell\to \infty$) green `spherical-like' surface with $\beta_{\rm eik}=1/2$ (inside the red surface).
    \label{fig:eikonal_verification3D}     }
\end{figure}

 As the reader can infer from an inspection of Fig.~\ref{fig:B0.5contoursform0m2m10andeikonal}, it becomes significantly harder to study modes for small $y_+<0.1$. For this reason our scanning of the KNdS parameter in these regions was sparser. Consequently, our conclusions are less solid in these regions and thus we do not fill them in Fig.~\ref{fig:B0.5contoursform0m2m10andeikonal}. Nevertheless, we have done a detailed analyses of specific families of KNdS solutions in these regimes to be confident that we are not missing relevant physics as we now discuss. For very small $y_+$ (at least smaller than 0.1 to give a safe upper bound) it is often the case that, for a given $m=\ell$, de Sitter modes are the dominant modes  but only for very small $\Theta$. As $\Theta$ grows, dS modes start interacting with NH and PS modes, presenting splittings and mergers as illustrated in Fig. \ref{fig:m0l0RNdSR095slice}, so their distinction gets blurred. Still, there are very small regions where they can dominate and simultaneously have $\beta_{\rm dS}<1/2$. We do not have enough data to include these detailed regions (necessarily with $y_+<0.1$) with the required level of accuracy in Fig.~\ref{fig:B0.5contoursform0m2m10andeikonal} (neither in the later plot of Fig.~\ref{fig:eikonal_verification3D}).
 
  So far we have been focusing our attention on PS modes with $m=\ell=10$, as these values provide a middle ground. Indeed, these are large enough so that we can gather evidence that the eikonal $m=\ell\to \infty$ frequency is indeed a good approximation to the exact numerical data, but at the same time they present sufficiently small $m$ to allow for a numerically search of the exact modes. What happens with  $m = \ell  \neq 10$?
 For some value of $\Theta$ high enough (at fixed $y_+$ and $\calr$) we find that it is always PS modes with sufficiently high $m=\ell$ that end up enforcing the SCC. Thus, it is useful to identify the $\beta_{\rm PS} = 1/2$ boundary surface for each $m=\ell$ QNM family. This can be seen in the right panel of Fig.~\ref{fig:B0.5contoursform0m2m10andeikonal} where, for a near-extremal KNdS family with constant $\calr=0.95$, we plot the $\beta_{\rm PS} = 1/2$ boundary curve for $m=\ell=1$ (brown curve), $m = \ell = 2$ (blue curve), and $m=\ell=10$ (red line already displayed in the left panel). One has $\beta_{\rm PS}>1/2$ to the left of each of these $m=\ell$ boundary curves, which means that as we go through these boundary curves from the right one to the left one (\ie for increasingly higher $m=\ell$) we see that we reduce the area (volume if we also vary $\calr$) inside the shell, containing small extremal RNdS, where SCC can be violated. The right panel of Fig.~\ref{fig:B0.5contoursform0m2m10andeikonal}  provides further evidence for an important property: as we increase $m=\ell$ from $1,\cdots,10$ we see that the associated critical $\beta_{\rm PS}=1/2$ boundaries approach monotonically the $\beta_{\rm eik}=1/2$ boundary described by the green-dashed curve on the left. Although we have not attempted to explicitly confirm that critical curves indeed keep increasingly approaching the eikonal surface from the right for even larger $m=\ell$, we are confident they do so for an extra reason. This is because, for the particular Kerr-dS family, this exercise was completed up to $m=\ell=50$ and confirms the above expectation~\cite{Dias:2018ynt}. This should extend to all other KNdS black holes.

This means that eikonal PS modes with very large $m=\ell$ are the ones that end up enforcing SCC in the widest range of KNdS black holes (\eg all the region to the right of the black dashed curve in the right plot). The reader might question whether this statement might not be in contradiction with the properties observed \eg in Fig.~\ref{fig:eigenvalue_repulsion} for $m=\ell=1$ modes (the qualitative features are similar for other $m=\ell>0$ and, in particular, $m=\ell=10$). Indeed for moderately large $\calr$ (top panel of  Fig.~\ref{fig:eigenvalue_repulsion}) the PS modes are the least damped modes for sufficiently large $\Theta$. In contrast, the situation changes drastically as we approach further extremality, \ie for larger $\calr$ (middle and bottom plots in Fig.~\ref{fig:eigenvalue_repulsion}) due to {\it eigenvalue repulsions}: the blue diamond NH modes now become the least damped modes relevant for SCC. The key point here is that this is for a specific $m=\ell$ QNM family. But  arbitrarily large $m=\ell$ PS modes, \ie in the eikonal limit $m=\ell\to \infty$, always have, for sufficiently large $\Theta$,  the least damped frequency when compared with {\it any} frequency of {\it finite} $m=\ell$ modes (be they PS or NH modes). This is clearly illustrated in the plots of  Fig.~\ref{fig:eigenvalue_repulsion} where we see that the $m=\ell\to \infty$ green-dashed eikonal curve is indeed above the $m=\ell=1$ orange PS and blue NH curves. This is true only for sufficiently large $\Theta$; in particular, it is certainly true for $\Theta$'s larger than the one where the eikonal green-dashed curve crosses the critical red curve with ${\rm Im}(\omega r_c)=-\beta \kappa_-$ with $\beta=1/2$ also displayed in Fig.~\ref{fig:eigenvalue_repulsion}. Black holes with $\Theta$ larger than this special intersection point have $\beta_{\rm eik}<1/2$ and thus preserve SCC. For KNdS black holes with smaller $\Theta$, \ie to the left of this intersection point, the eikonal PS modes are no longer necessarily the least damped modes. For a small window of intermediate $\Theta$ the PS modes are still the dominant modes while for smaller $\Theta$ the NH modes become the dominant modes. But in both cases, one has $\beta>1/2$ and thus SCC is violated.

What remains is to establish that there are no other modes which could enforce SCC when the eikonal PS modes do not do so. We are particularly interested in KNdS solutions around small extremal RNdS as we know that extremal RNdS violates SCC when $y_+$ is not too large (\ie not close to $y_+=1$). We have already explained that within the PS family of modes, the eikonal critical $\beta=\frac{1}{2}$ boundary supersedes the ones of any finite $m=\ell> 0$ PS QNM family so we do not have to worry with these. On the other hand, the de Sitter (dS) QNM family of modes is typically very much more damped than the PS and NH modes, except for very small $y_+$  (at least smaller than 0.1 to give a safe upper bound) where it can often dominate when $\Theta$ is very small. 
There is even a very small region where they dominate and simultaneously have $\beta_{\rm dS}<1/2$. However, since our scanning of the $y_+<0.1$ region was much less sparser, the accuracy of our results and thus conclusions are less solid in this region and we do not fill it in Fig.~\ref{fig:B0.5contoursform0m2m10andeikonal}. To complement this discussion, we invite the reader to see Fig.~2 of~\cite{Casals:2020uxa} for a detailed analysis of a particular KNdS family with constant $\Lambda M^2$ that identifies the small region where dS modes can dominate and have $\beta>1/2$ (although~\cite{Casals:2020uxa} is for a particular massive scalar field, it illustrates the relevant features). 

We are left with the NH modes. Above, we have already discussed that although these modes can sometimes be less damped than PS modes for particular $m=\ell\neq 0$ modes, they are always more damped than the eikonal $m=\ell\to \infty$ modes. Hence, we do not have to worry with $m=\ell\neq 0$ NH modes. The only potentially dangerous modes that we have not yet discussed are the $m=\ell=0$ modes. As found in section~\ref{sect:QNM_spectrum_in_KNdS}, the QNM spectra for $m=\ell=0$ is qualitatively very distinct from the $m=\ell>0$ case since we found that, more often than not, $m=\ell=0$ NH modes are less damped modes than $m=\ell=0$ PS and dS modes. Moreover, for certain values of $(y_+, \mathcal{R})$ the NH modes can actually enforce SCC for all $\Theta$, as shown for $y_+ = 0.9$ and $\calr = 0.95$ in the right plot of Fig.~\ref{fig:PS_NH_ml0_R095_yp05_im}.  As scanning the entire KNdS parameter space for $m = \ell= 0$ modes is computationally costly due to the splittings/mergers between families of modes reported in section~\ref{sect:m=l=0slice}, we focused our attention on the dangerous region near extremality, \ie for $\calr\geq 0.90$. Then, we found the dominant mode for each of the PS, dS and NH families for $y_+ \in [0.1, 0.9]$ and $\Theta \in [0, \pi/2]$, by obtaining them first in RNdS limit and then marching along $\Theta$. Using the dominant eigenvalue at each point inside this slice, we identified the critical $\beta=1/2$ for the $m=\ell=0$ modes. For the particular case of $\calr = 0.95$,  this $\beta_{\rm NH}=1/2$ boundary is the blue curve displayed in the right panel of Fig.~\ref{fig:B0.5contoursform0m2m10andeikonal}, with $\beta_{\rm NH} < \frac{1}{2}$ above this curve.  It is well to the right/top of the eikonal PS green dashed curve. That is to say, the $m = \ell = 0$ NH modes do not enforce SCC in any region which was not already enforced by the eikonal PS modes. This is also the case for other values of $\calr\geq 0.9$.

The discussions so far were illustrated for particular families of KNdS where we were fixing either $y_+$ or $\calr$ to use reader-friendly 2-dimensional plots. But we can now extend the analysis to the full 3-dimensional parameter space of KNdS in the `spherical polar parametrization' $(\Theta,y_+,\calr)$. This is shown in Fig.~\ref{fig:eikonal_verification3D}. Recall that the off-extremality parameter $\calr \in [0,1]$ is such that the horizontal plane with $\calr=1$ in this figure describes extremal KNdS black holes, $y_+=r_+/r_c \in [0,1]$ is the off-Nariai parameter, and $\Theta=\arctan(a/Q)\in [0,\pi/2]$ runs along KNdS from the RNdS family with $\Theta=0$ and the Kerr-dS family with $\Theta=\pi/2$.  The left panel of Fig.~\ref{fig:eikonal_verification3D} shows the $\beta_{\rm PS}$ density plot for $m=\ell=10$ in the RNdS vertical plane with $\Theta=0$ and in the horizontal plane describing near-extremal KNdS with $\calr=0.98$.\footnote{Our scanning of the $\calr>0.98$ region was much sparser, and thus the accuracy of our results and  conclusions are less solid in this region and we do not fill it in Fig.~\ref{fig:eikonal_verification3D}; but we found strong evidence that the density plot of the $\calr=0.98$ plane extrapolates with the same features for higher $\calr$.} In each of these two planes with the contours, we identify the red curves with $\beta_{\rm PS}=1/2$ that are the intersections of these two planes with the critical $m=\ell=10$ $\beta_{\rm PS}=1/2$ red `spherical-like' surface also shown. Outside this `spherical-like' red surface one has $\beta_{\rm PS}<1/2$ and thus SCC certainly holds for such black holes (including in the Kerr-dS limit). In the right panel of Fig.~\ref{fig:eikonal_verification3D}, we display the very same plot on the left but this time we remove all information except the red spherical-like surface describing $m=\ell=10$, $\beta_{\rm PS}=1/2$ modes and we add the eikonal, $m=\ell\to \infty$, $\beta_{\rm eik}=1/2$ green surface. The latter green surface is slightly to the left of the former red surface (\ie as $m=\ell$ grows, the associated $\beta_{\rm PS}=1/2$ surface increasingly approaches this eikonal $\beta_{\rm eik}=1/2$ green surface), in agreement with the constant $\calr$ slice shown in the right panel of Fig.~\ref{fig:B0.5contoursform0m2m10andeikonal}. From exercises like the ones described in the analysis presented in previous paragraphs, we are confident that we have gathered strong numerical evidence to state that the eikonal $\beta_{\rm eik}=1/2$ green surface is ultimately the relevant surface (strictly speaking for $y_+\geq 0.1$) to discuss SCC in KNdS black holes and the identification of this surface is the main outcome of our manuscript. Namely, outside this eikonal green surface one has $\beta<\frac{1}{2}$ and thus SCC holds for KNdS in that region (including Kerr-dS~\cite{Dias:2018ynt}), while KNdS inside this eikonal surface have $\beta>\frac{1}{2}$ and thus SCC may be violated (this includes the case of near-extremal RNdS away from the Nariai $y_+=1$ limit~\cite{Dias:2018etb}). 
 
\section{Conclusions and Discussions\label{sec:Conclusions}} 

Christodoulou's formulation of Strong Cosmic Censorship (SCC) holds for Kerr-de Sitter black holes but is violated by Reissner–Nordström-de Sitter black holes. There must then be a boundary in the parameter space of Kerr-Newman-de Sitter (KNdS) black holes that marks the transition between solutions that respect and violate SCC.
For massless scalar field perturbations, we did a detailed full scan of the 3-dimensional parameter space of KNdS and identified this surface boundary. For $r_+/r_c\geq 0.1$, this is given by the eikonal photon sphere green surface in Fig.~\ref{fig:eikonal_verification3D}. This is roughly ``a quarter of a spherical-like boundary'' centred around small extremal RNdS black holes and Christodoulou's SCC is violated for KNdS solutions inside it and is preserved for KNdS black holes outside it.  Our study complements and extends the one of~\cite{Casals:2020uxa} which identified the transition curve (lying in our boundary surface) for a particular 2-parameter sub-family of Kerr-Newman de Sitter black holes with $\Lambda M^2=0.02$.
To complete our study of SCC in KNdS we necessarily had to perform a detailed investigation of QNMs in KNdS. The QNM spectrum of KNdS has a very rich structure with three main families (the photon sphere, near-horizon and de Sitter families) although it is not often easy to identify unambiguously each of them due to eigenvalue repulsion and splitting/merger phenomena that we also discussed with some detail.
 
In this paper we only considered scalar field perturbations. In our analytical discussions of section~\ref{sect:analytical_study} we included a mass $\mu$ and charge $q$ for the scalar field but all our numerical results reported in   section~\ref{sect:QNM_spectrum_in_KNdS} and in section~\ref{sect:SCC_in_KNdS} are only for $\mu=0$ and $q=0$. We now comment on how our numerical findings for the neutral massless scalar field and our analytical expressions for modes with generic $\mu,q$, combined with known past results in the RNdS and Kerr-dS cases and the data for the particular KNdS family  with constant $\Lambda M^2$ in~\cite{Casals:2020uxa}, can be used to produce educated expectations for the discussion of SCC in KNdS for other perturbations. 

Let us start with massive scalar perturbations. As mentioned in footnote \ref{foot:BTZ}, the behaviour of perturbations at the Cauchy horizon in 3-dimensional anti-de Sitter black holes (namely, the BTZ black hole) turns out to also be determined by the slowest decaying QNM (with their exponential decay), very much like in de Sitter black holes. In particular, this means that the least damped QNM (spectral gap) divided by the Cauchy horizon surface gravity is also the relevant quantity to discuss SCC in BTZ~\cite{diasBTZBlackHole2019}. Since~\cite{diasBTZBlackHole2019} did a detailed discussion of SCC for massive scalar fields in BTZ and their qualitative findings concerning the dependence on the mass of the scalar field should extend to de Sitter, we borrow their relevant conclusions. Combining them with our analytical MAE analysis of massive modes in KNdS we can produce reasonable expectations. Ref. \cite{diasBTZBlackHole2019} found that scalar fields in BTZ can violate SCC and this violation can be made arbitrarily worse by increasing the mass of the scalar field $\mu$. In the sense that massive scalar perturbations are extendible across the  Cauchy horizon with arbitrarily large degree of differentiability (\ie as $C^k$ with large $k$) if we are sufficiently close to extremality and we increase $\mu$ to have $\beta> k$. Our matched asymptotic expansion~\eqref{eqn:beta_MAE} for KNdS modes indeed displays such a mass-dependence: by increasing the mass $\mu$ we can make $\delta^2$ (defined in~\eqref{eqn:delta_definition}) arbitrarily negative, and therefore $\beta_{\MAE} > k$ if we have sufficiently large $\mu$. We thus expect that this also occurs for  massive scalar perturbations in KNdS away from the near-extremal MAE regime we studied in section~\ref{sect:NHmodesandNHgeometry} but only an actual numerical computation of the modes can confirm this is indeed the case. The expectation is that the eikonal green surface with  $\beta_{\rm eik}=1/2$ in Fig.~\ref{fig:eikonal_verification3D} also provides the transition boundary relevant for the discussion of SCC in the massive sector. This is because: 1) this eikonal transition surface, in the region near to extremal KNdS black holes, is also described by the MAE analysis for $m=\ell\gg 1$ modes,  and 2)~\cite{Casals:2020uxa} reports (for a particular scalar mass) that the photon sphere frequencies of massive scalar QNMs, like the massless ones,  also approach the eikonal limit (recall that the latter uses null circular orbits so it could have been irrelevant for massive fields).

Consider now charged scalar fields. For RNdS, it was shown that although charge tends to weaken the SCC violation, the fact is that there is always a neighbourhood of RNdS near extremality in which SCC is violated by perturbations arising from smooth initial data even for arbitrarily large scalar field charge $q$ \cite{Dias:2018ufh}. This violation was missed in the analysis of~\cite{Cardoso:2018nvb,Mo:2018nnu}  because these references did not consider sufficiently near-extremal black holes where important wiggles around $\beta=1/2$, so relevant for the final discussion, do appear~\cite{Dias:2018ufh}. These wiggles are also observed in the particular KNdS family with constant $\Lambda M^2$ analysed in~\cite{Casals:2020uxa}.  Altogether, we expect that charged scalar fields in KNdS will behave very much like in RNdS at least for very small rotation and most possibly beyond this (but only an actual computation can confirm this). Hence, charged scalar fields should  not rescue SCC for KNdS black holes: even when the charge of the field is large, there should always be a tiny neighbourhood of extremality in which SCC is violated.

Finally, consider gravito-electromagnetic perturbations. In Kerr-dS, this sector of perturbations, like the scalar field perturbations, also preserves Christodoulou's SCC~\cite{Dias:2018ynt}. However, in RNdS it was found that in this gravito-electromagnetic sector not only is the Christodoulou formulation of SCC violated (like in the scalar field sector) but so is the $C^2$ formulation of SCC~\cite{Dias:2018etb} (in contrast to the scalar field sector). Recall that  a violation of the $C^2$ formulation of SCC is even more worrying than a violation of the Christodoulou formulation since it implies that the solution can be extended across the Cauchy horizon as a twice differentiable solution: tidal forces and  curvature invariants are thus both finite at the Cauchy horizon. In fact, the gravito-electromagnetic sector of perturbations in RNdS have an even more dramatic violation of SCC. Indeed, there are solutions  that can  be extended beyond the Cauchy horizon $\mathcal{CH}^+_R$ as a solution of class $C^k$ (the metric and Maxwell potential perturbations are $C^k$), where $k$ can be made arbitrarily large, for sufficiently near-extremal and large RNdS (in units of the cosmological radius)~\cite{Dias:2018etb}. This is because there are near-extremal RNdS black holes with gravito-electromagnetic perturbations with $\beta>k$ and arbitrarily large $k$. All these features are expected to be present when we turn on a small rotation, \ie for weakly rotating KNdS black holes. The question, that only an exact computation could address, is how far these features extend relative to the eikonal green surface in Fig.~\ref{fig:eikonal_verification3D}.
On the other hand, the eikonal approximation that describes the eikonal surface is the leading order contribution of a WKB expansion in $1/m$ that is independent of the spin of the perturbations. That is to say, it is the same for scalar field perturbations and for gravito-electromagnetic perturbations and accurate (when compared with high $m=\ell$ numerical data) in both cases  as explicitly verified in the RNdS~\cite{Dias:2018etb} and Kerr-dS~\cite{Dias:2018ynt} cases. Thus, it seems natural to expect that it is also a good approximation for gravito-electromagnetic perturbations in KNdS. If this is indeed the case,  the eikonal $\beta_{\rm eik}=1/2$ green boundary surface in Fig.~\ref{fig:eikonal_verification3D} should also provide the critical $\beta=1/2$ boundary to discuss the validity of Christodoulou’s SCC in the context of gravito-electromagnetic extendibility beyond the Cauchy horizon.

\appendix
\section{Further details and comparisons with previous work}\label{sect:Appendix}

\subsection{The eikonal approximation and comparison with~\cite{Casals:2020uxa}}\label{sect:eikonal_analysis}

To confirm the accuracy of our eikonal approximation for photon sphere modes with $m=\ell\to \infty$, we performed an in-depth study, throughout the parameter space of KNdS, of the analytic approximation $\omega_{\hbox{\tiny eik}}$ provided in \eqref{eqn:omega_eik}. Additionally, we compared our eikonal approximation for the PS modes with the one provided in~\cite{Casals:2020uxa}. 

Before diving into the numerical results, we take a detour to compare the explicit expressions. In~\cite{Casals:2020uxa}, the eikonal PS modes are analytically captured by the expression 
\begin{align}\label{eqn:CasalsandMarinhoeikonal}
    \omega_{\rm [1]}^\pm = E^\pm - i \left(n + \frac{1}{2}\right)|\lambda_{\rm L}|, \quad \text{for } n = 0,1,2 \dots
\end{align}
where 
\begin{align}\label{eqn:CasalsandMarinhoeikonal2}
    E^\pm = \frac{L_\phi}{b_s^\pm} =\frac{a \pm \sqrt{\Delta_r(r_s^\pm)}}{(r_s^\pm)^2 + a^2 \pm a \sqrt{\Delta_r(r_s^\pm)}} L_\phi \to \frac{a m \pm \text{sign}(m)(\ell + 1/2) \sqrt{\Delta_r(r_s^\pm)}}{(r_s^\pm)^2 + a^2 \pm a \sqrt{\Delta_r(r_s^\pm)}}.
\end{align}

A first observation is that the Lyapunov exponent from~\cite{Casals:2020uxa} coincides with ours in \eqref{eqn:omega_eik}, hence we use the same symbol $|\lambda_{\rm L}|$. This imaginary part of the frequency is the only contribution relevant for the discussion of SCC. In particular, this fact and the agreement of the imaginary contribution of  \eqref{eqn:omega_eik}  and \eqref{eqn:CasalsandMarinhoeikonal}-\eqref{eqn:CasalsandMarinhoeikonal2} of~\cite{Casals:2020uxa} means that our SCC conclusions agree with those of~\cite{Casals:2020uxa}  for the range of parameters where there is overlap.

\begin{figure}[t]
    \centering
    \includegraphics[width=0.505\textwidth]{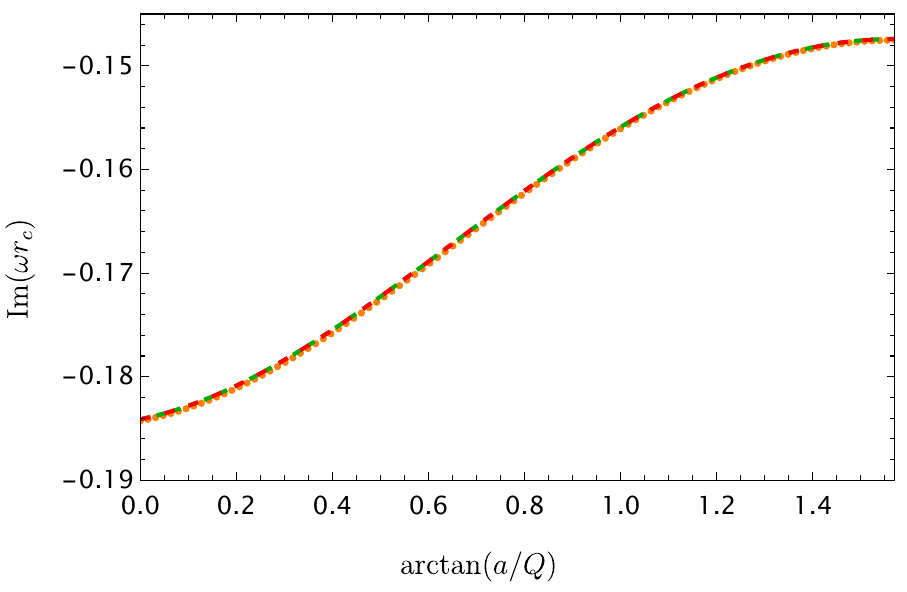}
    \includegraphics[width=0.475\textwidth]{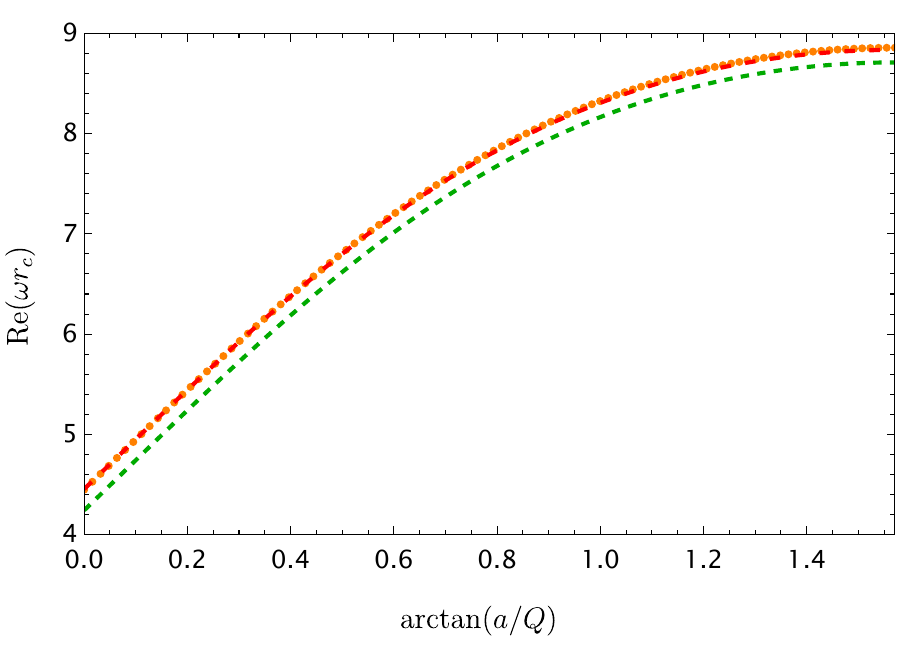}
     \includegraphics[width=0.505\textwidth]{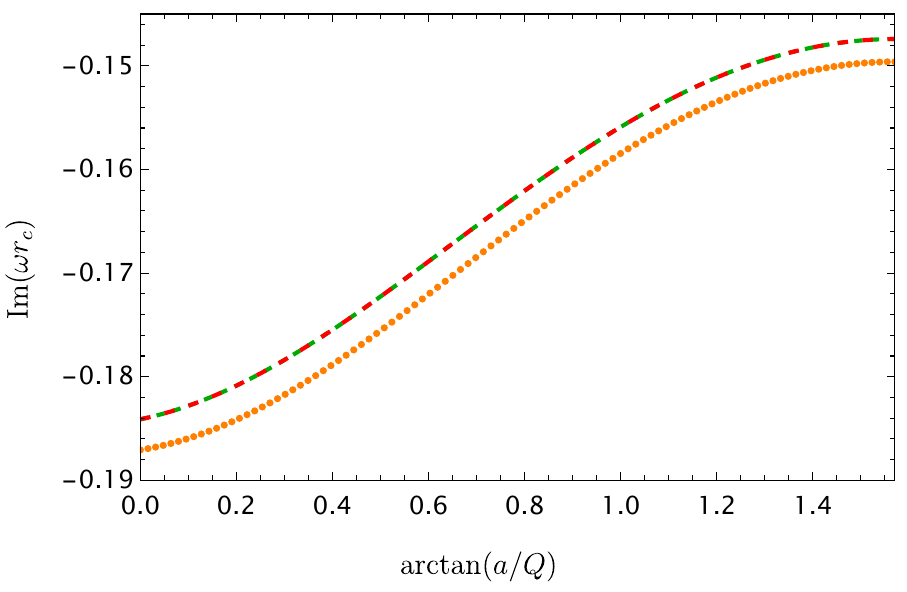}
    \includegraphics[width=0.48\textwidth]{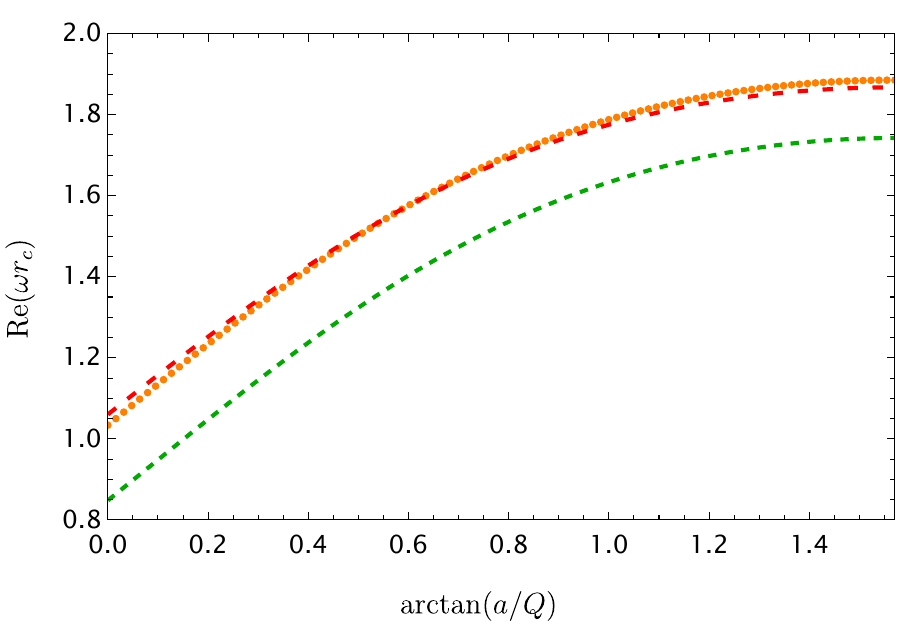}
    \caption{Comparison at $\calr = 0.5$, $y_+ = 0.5$, $m = \ell $ fixed and along $\Theta = \arctan(a/Q)$ of the dominant PS family frequencies (orange circles) against analytic approximations. We have the eikonal approximation $\omega_{\hbox{\tiny eik}}$ from \eqref{eqn:omega_eik} (green dashed line) and the analytical approximation of PS modes (red dashed line) from~\cite{Casals:2020uxa}. \textbf{Left column}: imaginary part of the frequencies for $m = \ell   = 10$ (top) and $ m = \ell  =  2$ (bottom). \textbf{Right column}: real part of the frequencies for $m = \ell   = 10$ (top) and $ m = \ell  =  2$ (bottom). These all correspond to $q =\mu = 0$ linear scalar perturbations on KNdS.}\label{fig:KNdSPSmodeVSWKBandCasals}
\end{figure}

Conversely, for the real parts of \eqref{eqn:omega_eik}  and \eqref{eqn:CasalsandMarinhoeikonal}-\eqref{eqn:CasalsandMarinhoeikonal2}, one concludes that ${\rm Re}( \omega_{\hbox{\tiny eik}}) \neq {\rm Re}(\omega_{\rm [1]})$. For a good reason. In \eqref{eqn:omega_eik},  ${\rm Re}( \omega_{\hbox{\tiny eik}})$ is obtained from a first principles derivation detailed in section~\ref{sect:PSmodes}. On the other, hand ${\rm Re}(\omega_{\rm [1]})$ is obtained from the very same first principles derivation followed by the phenomenological replacement/correction \eqref{eqn:CasalsandMarinhoeikonal2} designed precisely to enhance the accuracy of the improved eikonal approximation \eqref{eqn:CasalsandMarinhoeikonal}-\eqref{eqn:CasalsandMarinhoeikonal2} . Here, we do not lose the opportunity to further assert the accuracy of the improved \eqref{eqn:CasalsandMarinhoeikonal}-\eqref{eqn:CasalsandMarinhoeikonal2} of~\cite{Casals:2020uxa}.

First, let us briefly review the phenomenological reason that motivates the replacement in \eqref{eqn:CasalsandMarinhoeikonal2}. When using the photon sphere correspondence (see section~\ref{sect:PSmodes}), we map $L_\phi \to m$, as these agree in the eikonal approximation up to leading order. But this is just the leading term of a formal WKB analysis for $|m|=\ell\gg 1$.
To effectively account for the next-to-leading order WKB correction (without computing it),~\cite{Casals:2020uxa} proposes the transformation \eqref{eqn:CasalsandMarinhoeikonal2} whereby $L_\phi$ is mapped to $m$ in the contribution where it  does not multiply $a$, but it is mapped to $\pm (\ell + 1/2)$ in the term where it multiplies $a$. The motivation is to phenomenologically match the mappings of $L_\phi$ from~\cite{Dias:2018etb} when $Q = 0$ and from~\cite{Dias:2018ynt,Glampedakis:2019dqh} when $a = 0$. The computation of~\cite{Glampedakis:2019dqh} (but not the one of~\cite{Dias:2018etb}) includes higher order WKB corrections and the latter amount to do the transformation \eqref{eqn:CasalsandMarinhoeikonal2} when $a=0$. Consequently, the real parts of \eqref{eqn:omega_eik}  and \eqref{eqn:CasalsandMarinhoeikonal}-\eqref{eqn:CasalsandMarinhoeikonal2} differ by  
\begin{align}
    \text{Re}(\omega_{\rm [1]}^\pm - \omega_{\hbox{\tiny eik}}^\pm) = \frac{\pm\sqrt{\Delta_r(r_s^\pm)}}{2((r_s^\pm)^2 \pm a \sqrt{\Delta r (r_s^\pm)} + a^2)},
\end{align}
which does not depend on $m$, only on the point of the KNdS parameter space we are interested in. 

Let us now explicitly quantify the improvement in the accuracy of $\text{Re}(\omega_{\rm [1]}^\pm)$ with respect to $\text{Re}(\omega_{\hbox{\tiny eik}}^\pm)$. This is done in Fig.~\ref{fig:KNdSPSmodeVSWKBandCasals} for a KNdS family with $\calr=0.5$ and $y_+=0.5$ ranging from the RNdS limit ($\Theta=0$) till the Kerr-dS solution ($\Theta=\pi/2$). We display the exact numerical co-rotating ($+$) PS modes (orange disks) for $m=\ell=10$ (top panel) and for $m=\ell=2$ (bottom). Of course, the $1/m$ WKB corrections should be significant in the latter. In this figure, the left (right) panel plots the imaginary (real) part of the frequency. Recall that 
${\rm Im}(\omega_{\hbox{\tiny eik}}^+) ={\rm Im}(\omega_{\rm [1]}^+)$ and this is represented by the green dashed curve in the left plots. On the other hand, ${\rm Re}(\omega_{\hbox{\tiny eik}}^+)$ (green dashed curve) differs from  ${\rm Re}(\omega_{\rm [1]}^+)$ (red dashed curve) in the right plots. The plots speak for themselves. The eikonal approximation for the imaginary part of the frequency (that we use heavily in our SCC discussion of section~\ref{sect:SCC_in_KNdS} and associated plots) is accurate for values as low as $m=\ell=10$ (top left plot) but is is remarkably good even for $m=\ell=2$ (bottom-left plot). On the other hand, by design, $\text{Re}(\omega_{\rm [1]}^\pm)$ (dashed red curve) is a better approximation than $\text{Re}(\omega_{\hbox{\tiny eik}}^\pm)$ (green dashed curve) already for $m=\ell=10$ (top right plot) and even more so for $m=\ell=2$ (bottom-right plot). 
The accuracy of both approximations improves as $\Theta \to \pi/2$. For reference, for $m = \ell = 10$ both estimations yield an excellent result, with maximum relative errors of 0.02\% for $\omega_{\rm [1]}$ and of 4\% for $\omega_{\hbox{\tiny eik}}$.  These findings are in agreement with previous computations in the literature, see \eg~\cite{Davey:2022vyx, Davey:2023fin, Dias:2018etb, Cardoso:2017soq, Casals:2020uxa, Dias:2018ynt, Dias:2018ufh} and references therein.
Our computations also show that the eikonal approximation for the imaginary part remains a good approximation for generic values  of $\calr$  and $y_+$. But, as Fig.~\ref{fig:B0.5contoursform0m2m10andeikonal} illustrates it worsens for smaller values of $y_+$. Also note that very near extremality, the eikonal approximation stops describing PS modes and starts describing instead NH modes as illustrated in Fig. \ref{fig:eigenvalue_repulsion} and its discussion.

It would be interesting to do a WKB computation in $1/m$ that would give a first principles derivation of the next-to-leading order WKB correction to the eikonal approximation along the lines of those done in~\cite{Glampedakis:2019dqh,Davey:2023fin} (see also \cite{Dias:2022oqm}).

\subsection{MAE analysis and comparison with~\cite{Cardoso:2017soq}}\label{sect:MAEanalysis}

In this section, we further discuss the accuracy of the MAE approximation $\omega_{\hbox{\tiny MAE}}$ in \eqref{eqn:omega_MAE} and \eqref{eqn:omega_MAE_positive} derived in section~\ref{sect:MAEexpansion}. We are particularly interested on the RNdS case (and $\mu=q=0$), and we compare our $\omega_{\hbox{\tiny MAE}}$  with a similar analytical approximation $\omega_{\rm [2]}$ presented  in~\cite{Cardoso:2017soq}.

\begin{figure}[t]
    \centering
    \includegraphics[width=0.495\textwidth]{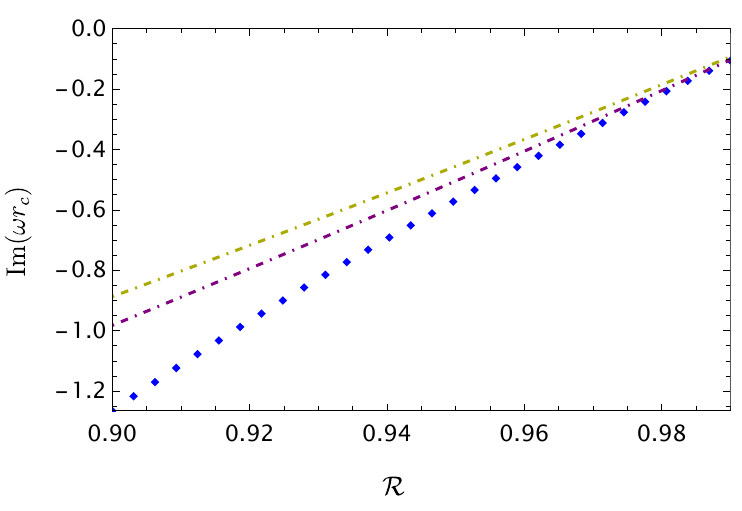}
    \includegraphics[width=0.495\textwidth]{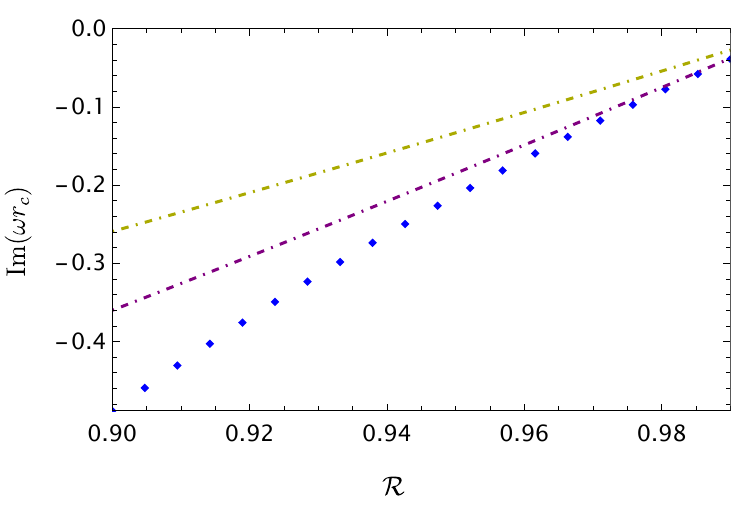}
    \includegraphics[width=0.50\textwidth]{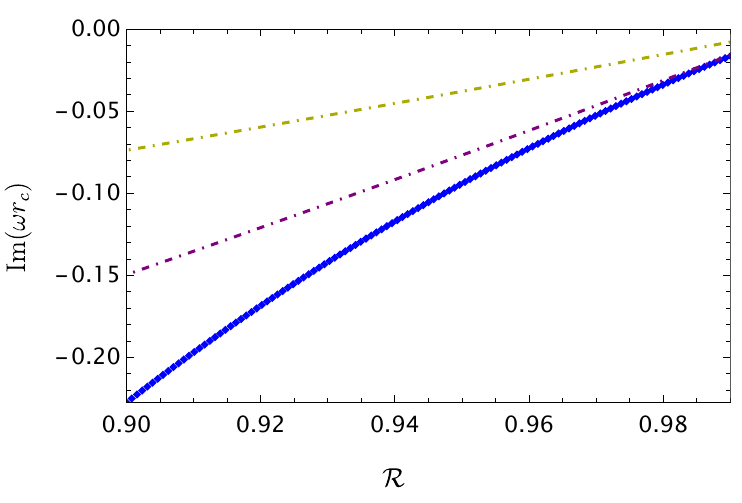}
    \caption{Imaginary part of the frequency of near-horizon (NH) modes for scalar field QNMs with $q =\mu = 0$ and $m = \ell = 2, n=0$ for near-extremal  RNdS as a function of $\calr \in [0.9, 0.99]$  for three distinct RNdS families with fixed $y_+$, namely $y_+ = 0.25$ (top left), $y_+ = 0.5$ (top right) and $y_+ = 0.75$ (bottom).  The numerical exact NH modes are represented by the blue diamonds, the $n = 0$ MAE approximation $\omega_{\hbox{\tiny MAE}}$ is described by the purple dot-dashed line and the near-horizon approximation $\omega_{[2]}$ of~\cite{Cardoso:2017soq} by the yellow dot-dashed line.}\label{fig:cardosoNHapproximationVSMAE}
\end{figure}

Consider RNdS black holes ($\Theta = 0$, \ie $a = 0$).
Ref.~\cite{Cardoso:2017soq} presented an analytical approximation for the frequency of NH modes. It is given by~\cite{Cardoso:2017soq}
\begin{align}\label{eqn:NHcardosoapproximation}
    \omega_{\rm [2]} = -i (\ell +  n + 1) \kappa_+.
\end{align}
In Fig.~\ref{fig:cardosoNHapproximationVSMAE} we display the imaginary part of the frequency for  $m=\ell=2$ NH modes for RNdS in the near-extremal region $\calr \in [0.9, 0.99]$  for three different values of $y_+$, namely for  $y_+ = 0.25$ (top left plot),  $y_+ = 0.5$ (top right plot) and $y_+ = 0.75$ (bottom plot). In all these plots, the blue diamonds describe the numerical exact frequencies, the purple dot-dashed line is given by $\omega_{\hbox{\tiny MAE}}$ in \eqref{eqn:omega_MAE}  and \eqref{eqn:omega_MAE_positive} and the yellow line is  $\omega_{\rm [2]}$ in \eqref{eqn:NHcardosoapproximation}.  
Fig.~\ref{fig:cardosoNHapproximationVSMAE} shows that both our $\omega_{\hbox{\tiny MAE}}$  and  $\omega_{\rm [2]}$ get increasingly more  accurate as $\calr \to 1$. But, overall, $\omega_{\hbox{\tiny MAE}}$ is visibly a better approximation, in particular, off-extremality and also for large values of $y_+$. Nevertheless,  the accuracy of $\omega_{\hbox{\tiny MAE}}$ also improves as we decrease $y_+$, with the better approximation obtained for $y_+ = 0.25$ (top plot). 

As detailed in the discussion of Fig.~\ref{fig:eigenvalue_repulsion}, the MAE frequency $\omega_{\hbox{\tiny MAE}}$ of \eqref{eqn:omega_MAE} and \eqref{eqn:omega_MAE_positive} remains an excellent approximation in KNdS, as long as we are sufficiently close to extremality. So we do not discuss the accuracy of $\omega_{\hbox{\tiny MAE}}$ further for KNdS. 
MAE approximations similar to ours (or exactly the same as ours in appropriate limits) have also been  discussed in detail in~\cite{Davey:2022vyx, Davey:2023fin, Yang:2012pj, Yang:2013uba, Zimmerman:2015trm}.

\subsection{NH modes for $m = \ell = 0$  at extremality}\label{sect:MAEvalueforvanishingm}

In section~\ref{sect:MAEexpansion} we used a matching asymptotic expansion to find an analytic approximation for the near-horizon frequencies of KNdS black holes close to extremality. It is valid for $m=\ell\geq 0$ modes. 
Recall that in order to compute $\omega_{\MAE}$ in~\eqref{eqn:omega_MAE} we first need to solve the angular equation~\eqref{eqn:KG_angular}, evaluated at extremality, to find the separation constant $\lambda$. This is straightforward to find numerically. However, in the case of $m = \ell = 0$ we simply have $\lambda = 0$, as we show analytically in this appendix. Furthermore, using this, we will derive a zeroth-order approximation for $\beta_{\MAE}^{m = \ell = 0}$ that is valid exactly at extremality (and only there). This allows us to conclude that the dominant $m = \ell = 0$ NH modes, if captured by the matching asymptotic expansion, have $\beta_{\MAE}^{m = \ell = 0} \to 1$ at extremality and so do not enforce SCC there.

Firstly, we analytically show that for massless scalar perturbations of extremal KNdS black holes with $m = \ell = 0$ we must have separation constant $\lambda = 0$, regardless of the value of $\Theta$ (\ie not only in the RNdS limit). Evaluating \eqref{eqn:KG_angular} at extremality and with $m = 0$ we obtain
\begin{align}\label{eqn:KG_angular_m=0}
     (1-x^2)(1 + \alpha x^2)S''(x) -2x\big[1+\alpha(2x^2-1)\big]S'(x) + \lambda S(x) = 0,
\end{align}
where we defined $\alpha = a^2/L^2$ for convenience and both $a$ and $L$ are evaluated at extremality. In the RNdS limit $\alpha \to 0$,  (\ref{eqn:KG_angular_m=0}) becomes Legendre's equation, with $\lambda = \ell(\ell+1)$ being required by regularity. Setting $\ell = 0 \Rightarrow \lambda = 0$, we find that $S(x) = c$ a constant is the unique solution regular at $x = \pm 1$. As we are free to rescale the eigenfunctions, we will set $S(x) = 1$ for the solution to (\ref{eqn:KG_angular_m=0}) in the RNdS limit. 

We perturbatively solve (\ref{eqn:KG_angular_m=0}) in powers of $\alpha$ with $m = \ell = 0$. Our aim is to show that $\lambda = 0$ to arbitrary order in $\alpha$. We expand $\lambda$ and $S(x)$ as 
\begin{align}\label{eqn:perturbative_ansatz_for_m=l=0_eqn}
   \lambda = \sum_{k = 1}^\infty \lambda_k \alpha^k, \qquad\quad S(x) = 1 + \sum_{j = 1}^{\infty} S_j(x)\alpha^j.
\end{align}
In these expansions we have used that $\lambda_0 = \ell(\ell+1) = 0$ and $S_0(x) = 1$. Additionally, the $\lambda_k$ are constants and we will impose regularity on $S_j(x)$ at $x = \pm 1$. After defining $S_{-1}(x) = 0$, substitution of the ansatz (\ref{eqn:perturbative_ansatz_for_m=l=0_eqn}) in (\ref{eqn:KG_angular_m=0}) yields
\begin{multline}\label{eqn:KG_with_perturbative_ansatz}
 \sum_{n=0}^\infty \alpha^n \Bigg[\sum_{l=0}^{n} \lambda_l S_{n-l} - 2x S_n'(x) + (1-x^2) S_n''(x) - \\ 
 - 2x(2x^2 -1)S_{n-1}'(x) + x^2(1-x^2)S_{n-1}''(x) \Bigg] = 0
\end{multline}
We now employ strong induction to show that $S_n(x) = 1$ and $\lambda_n = 0$, for all $n$ orders in $\alpha$. The base case is provided by the solution to Legendre's equation with $m = \ell  =0$ in the RNdS limit. Now let $p$ be a positive integer and suppose that  $S_p(x) = 1$ and $\lambda_p = 0$ for all $p \in \{1, 2, \dots, n-1\}$. Our aim is to show that $S_{n} = 1$ and $\lambda_n = 0$. Looking at the coefficient of $\alpha^n$ in (\ref{eqn:KG_with_perturbative_ansatz}) and using our induction hypothesis, we obtain 
\begin{align*}
    \mathcal{O}(\alpha^n): \lambda_{n} - 2x S_{n}'(x) + (1-x^2)S_{n}''(x) = 0. 
\end{align*}
This is solved by 
\begin{align*}
    S_{n}(x) = c_1 + \frac{1}{2}(\lambda_n + c_2)\log|1-x| + \frac{1}{2}(\lambda_n - c_2)\log|1+x|.
\end{align*}
Imposing regularity at $x = \pm 1$ implies that $c_2 = \lambda_n = 0$  and normalising the eigenfunction $S_n(x)$ to 1 yields $c_1 = 1$. Hence, for $m = \ell = 0$ at extremality, we have proven that $\lambda = 0$ regardless of the value of $a/Q$, \ie independent of $\Theta$.

We compute $\beta$ for NH modes with $m=\ell=0$ exactly at extremality using our matching asymptotic approximation. At extremality with $m = \ell = 0$, using the definition of $\delta$ from (\ref{eqn:delta_definition}), we have that
\begin{align}\label{eqn:beta_mae_ml0}
    \beta_{\MAE}^{m = \ell = 0} = 1 + n.
\end{align}
\begin{figure}[t]
    \centering
    \includegraphics[width=0.75\textwidth]{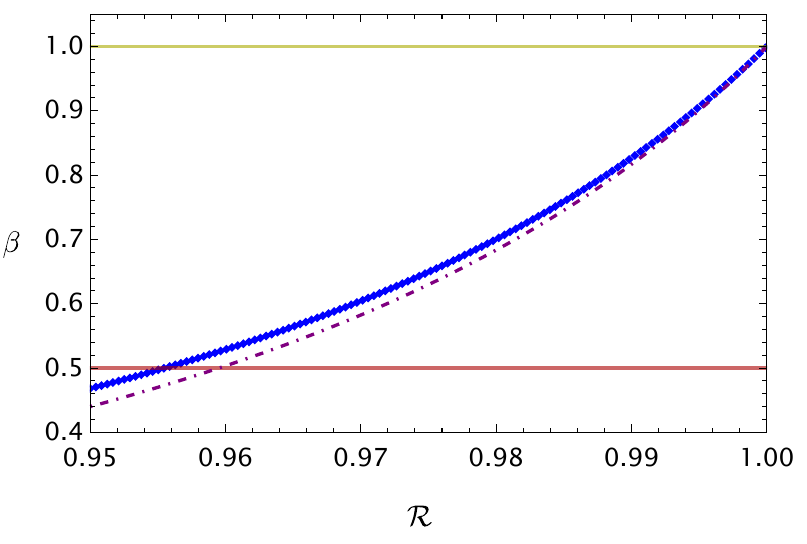}
    \caption{Value of $\beta$ for the dominant NH mode (blue diamonds) for $\mathcal{R} \in [0.95, 0.9999]$ for $q =\mu = 0$ linear scalar perturbations with $m = \ell = 0$ on KNdS with $\Theta = \pi/4 \sim 0.79,\, y_+ = 0.9$.
    The $n = 0$ ($n=1$) MAE approximation $\omega_{\MAE}$ to the NH modes is given by a dot-dashed purple curve. The red (yellow) horizontal line indicates $\beta = \frac{1}{2}$ ($\beta = 1$).}\label{fig:NH_ml0_R095_to_R09999_yp09_beta}
\end{figure}
This result is in good agreement with our exact numerical NH modes, and an example of this is given in Fig.~\ref{fig:NH_ml0_R095_to_R09999_yp09_beta}, where we plot the dominant NH mode (blue diamonds) with $m = \ell = 0$ at a fixed $\Theta=\pi/4\sim 0.79$ and $y_+ = 0.9$, as we approach extremality $\mathcal{R} \in [0.95, 0.9999]$. We also display the MAE approximation~\eqref{eqn:omega_MAE} (which necessarily has separation constant $\lambda = 0$ as just shown) as a dot-dashed purple curve. The red (yellow) horizontal line indicates $\beta = \frac{1}{2}$ ($\beta = 1$). Starting at the left-hand side of the plot, $\mathcal{R} \sim 0.95$, some of the NH modes have $\beta_{\rm NH} < \frac{1}{2}$, in agreement with the blue line in the right plot of  Fig.~\ref{fig:B0.5contoursform0m2m10andeikonal} which describes the dominant $m = \ell = 0$ modes with $\beta = \frac{1}{2}$ (and also in agreement with the right plot of Fig.~\ref{fig:PS_NH_ml0_R095_yp05_im}). At the other end, towards the right hand side of Fig.~\ref{fig:NH_ml0_R095_to_R09999_yp09_beta}, $\mathcal{R} \sim 1$, the NH mode rather rapidly approaches $\beta \to 1$, in agreement with~\eqref{eqn:beta_mae_ml0}. In RNdS, it has previously been shown that the dominant $m = \ell = 0$ NH modes approach $\beta = 1$ in the extremal limit~\cite{Cardoso:2017soq}. Our result~\eqref{eqn:beta_mae_ml0}, in conjunction with our numerical results, show that this feature of the $m = \ell = 0$ NH modes extends away from the RNdS limit, for all $\Theta$ even up to the Kerr-dS limit.

\subsection{Numerical convergence tests}\label{sect:numerical_convergence}

\begin{figure}[t]
    \centering
    \includegraphics[width=0.9\textwidth]{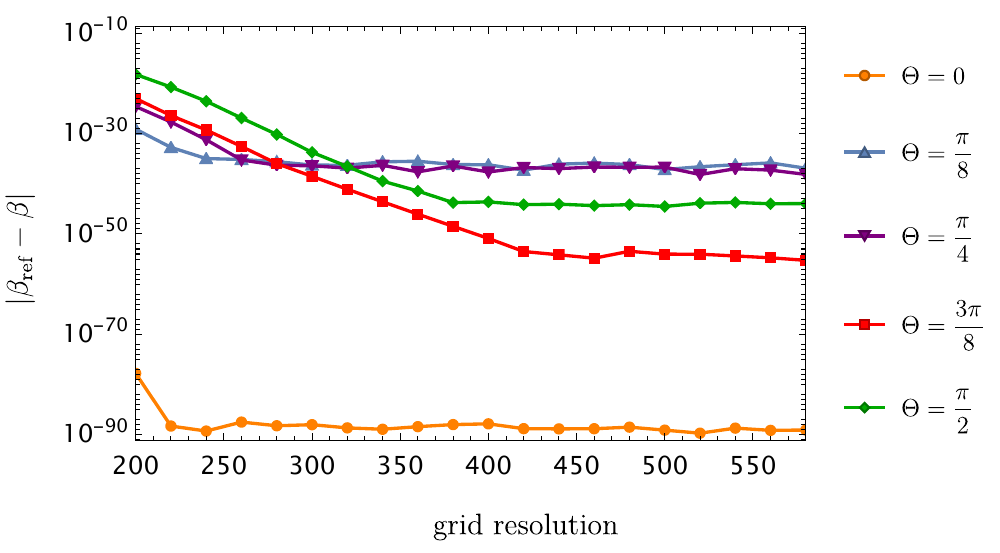}
    \caption{Numerical converge tests of $\beta$ for KNdS at $y_+ = 0.9$, $R = 0.95$, with $m = \ell = 10$ and distinct $\Theta$ values. These modes were converged for a fixed numerical precision of 100 digits and a range of resolutions, starting at a minimum of 200 grid points and then linearly scaling up to a 600 grid points. The maximum values were used to compute the reference value $\beta_{\text{ref}}$.}
    \label{fig:convergenceplot}
\end{figure}

All of our numerical results converge with an error less than or equal to $10^{-12}$. In Fig.~\ref{fig:convergenceplot}, we present the results of a numerical convergence test for KNdS with $y_+=0.9$, $\calr = 0.95$ and $m = \ell = 10$. To test convergence, we repeat the computation of the dominant PS mode at each value of $\Theta$ at fixed precision of 100 digits and increasing grid resolution, following a linear increase up to $600$ grid points. The value of $\beta$ at the maximum resolution defines $\beta_{\text{ref}}$. When using pseudospectral collocation methods, we expect to observe exponential convergence as the grid resolution increases~\cite{Dias:2015nua}. Indeed, this is what we find in Fig. \ref{fig:convergenceplot}, with the maximum error computed in these tests being $|\beta_{\text{ref} }- \beta| \sim 10^{-19}$, given by the first point of the convergence test with $\Theta = \pi/2$.

\acknowledgments

O.D. acknowledges financial support from the STFC ``Particle Physics Grants Panel (PPGP) 2018'' Grant No.~ST/T000775/1 and PPGP 2020 grant No.~ST/X000583/1.
The authors acknowledge enlightening discussions with Jorge Santos.
OD acknowledges the organizers of the YITP-ExU long-term workshop
{\it Quantum Information, Quantum Matter and Quantum Gravity (QIMG2023)}, Yukawa Institute for Theoretical Physics, Kyoto University, during which part of this work was completed.
The authors acknowledge the use of the IRIDIS High Performance Computing Facility and associated support services at the University of Southampton in the completion of this work.

\bibliographystyle{JHEP}
\bibliography{SCC_KNdS}
\end{document}